\documentclass[
  journal=pasa,
  manuscript=research-paper, 
  year=2024,
  volume=00,
]{cup-journal}

\usepackage{graphicx}
\usepackage{lscape}
\usepackage{color}
\usepackage{booktabs}
\usepackage{amsmath} 
\usepackage{caption}
\usepackage{subcaption}
\usepackage{aas_macros}
\usepackage{hyperref}
\usepackage{pdflscape}
\hypersetup{colorlinks,citecolor=blue,linkcolor=blue,urlcolor=blue}

\newcommand{\hi}{\mbox{H\,{\sc i}} } 
\newcommand{\hix}{\mbox{H\,{\sc i}}} 
\newcommand{\mgii}{\mbox{Mg\,{\sc ii}} } 

\title{\color{MidnightBlue} The First Large Absorption Survey in \hi (FLASH): II. Pilot Survey data release and first results}

\author{Hyein Yoon}
\affiliation{Sydney Institute for Astronomy, School of Physics A28, University of Sydney, NSW 2006, Australia}
\alsoaffiliation{ARC Centre of Excellence for All-Sky Astrophysics in 3 Dimensions (ASTRO 3D)}
\alsoaffiliation{Institute for Data Innovation in Science, Seoul National University, 1 Gwanak-ro, Gwanak-gu, Seoul 08826, Republic of Korea}
\alsoaffiliation{Astronomy Program, Department of Physics and Astronomy, Seoul National University, 1 Gwanak-ro, Gwanak-gu, Seoul 08826, Republic of Korea}
\email[Hyein Yoon]{hiyoon.astro@gmail.com}

\author{Elaine M. Sadler}
\affiliation{Sydney Institute for Astronomy, School of Physics A28, University of Sydney, NSW 2006, Australia}
\alsoaffiliation{ARC Centre of Excellence for All-Sky Astrophysics in 3 Dimensions (ASTRO 3D)}
\alsoaffiliation{ATNF, CSIRO, Space and Astronomy, PO Box 76, Epping, NSW 1710, Australia}

\author{Elizabeth K. Mahony}
\affiliation{ATNF, CSIRO, Space and Astronomy, PO Box 76, Epping, NSW 1710, Australia}

\author{J.N.H.S. Aditya} 
\affiliation{Sydney Institute for Astronomy, School of Physics A28, University of Sydney, NSW 2006, Australia}
\alsoaffiliation{ARC Centre of Excellence for All-Sky Astrophysics in 3 Dimensions (ASTRO 3D)}

\author{James R. Allison} 
\affiliation{First Light Fusion Ltd., Unit 9/10 Oxford Pioneer Park, Mead Road, Yarnton, Kidlington OX5 1QU, UK}

\author{Marcin Glowacki}
\affiliation{Institute for Astronomy, University of Edinburgh, Royal Observatory, Edinburgh, EH9 3HJ, United Kingdom}
\alsoaffiliation{International Centre for Radio Astronomy Research (ICRAR), Curtin University, Bentley, WA 6102, Australia}
\alsoaffiliation{Inter-University Institute for Data Intensive Astronomy, Department of Astronomy, University of Cape Town, Cape Town, South Africa}

\author{Emily F. Kerrison}
\affiliation{Sydney Institute for Astronomy, School of Physics A28, University of Sydney, NSW 2006, Australia}
\alsoaffiliation{ARC Centre of Excellence for All-Sky Astrophysics in 3 Dimensions (ASTRO 3D)}
\alsoaffiliation{ATNF, CSIRO, Space and Astronomy, PO Box 76, Epping, NSW 1710, Australia}

\author{Vanessa A. Moss}  
\affiliation{ATNF, CSIRO, Space and Astronomy, PO Box 76, Epping, NSW 1710, Australia}

\author{Renzhi Su}
\affiliation{Research Center for Astronomical Computing, Zhejiang Laboratory, Hangzhou 311100, China}
\alsoaffiliation{Key Laboratory for Research in Galaxies and Cosmology, Shanghai Astronomical Observatory, 80 Nandan Road, Shanghai 200030, China }
\alsoaffiliation{University of Chinese Academy of Sciences, 19A Yuquan Road, Beijing 100049, China}
\alsoaffiliation{ATNF, CSIRO, Space and Astronomy, PO Box 76, Epping, NSW 1710, Australia}

\author{Simon Weng}
\affiliation{Sydney Institute for Astronomy, School of Physics A28, University of Sydney, NSW 2006, Australia}
\alsoaffiliation{ARC Centre of Excellence for All-Sky Astrophysics in 3 Dimensions (ASTRO 3D)}
\alsoaffiliation{European Southern Observatory, Karl-Schwarzschild-Strasse 2, Garching bei M\"unchen, 85748, Germany}

\author{Matthew Whiting}
\affiliation{ATNF, CSIRO, Space and Astronomy, PO Box 76, Epping, NSW 1710, Australia}

\author{O. Ivy Wong}
\affiliation{ATNF, CSIRO, Space and Astronomy, PO Box 1130, Bentley, WA 6102, Australia}
\alsoaffiliation{International Centre for Radio Astronomy Research (ICRAR), The University of Western Australia, 35 Stirling Hwy, Crawley, WA 6009, Australia}
\alsoaffiliation{ARC Centre of Excellence for All-Sky Astrophysics in 3 Dimensions (ASTRO 3D)}

\author{Joseph R. Callingham}
\affiliation{ASTRON, the Netherlands Institute for Radio Astronomy, Oude Hoogeveensedijk 4, NL-7991 PD Dwingeloo, The Netherlands} 
\alsoaffiliation{Anton Pannenkoek Institute for Astronomy, University of Amsterdam, Science Park 904, Amsterdam, The Netherlands} 

\author{Stephen J. Curran} 
\affiliation{School of Chemical and Physical Sciences, Victoria University of Wellington, PO Box 600, Wellington 6140, New Zealand}

\author{Jeremy Darling}
\affiliation{Center for Astrophysics and Space Astronomy, Department of Astrophysical and Planetary Sciences, University of Colorado, 389 UCB, Boulder, CO 80309-0389, USA }

\author{Alastair C. Edge}
\affiliation{Centre for Extragalactic Astronomy, Department of Physics, Durham University, South Road, Durham, DH1 3LE, UK}

\author{Sara L. Ellison}
\affiliation{Department of Physics \& Astronomy, University of Victoria, Finnerty Road, Victoria, British Columbia, V8P 1A1, Canada)}

\author{Kimberly L. Emig}
\affiliation{National Radio Astronomy Observatory, 520 Edgemont Road, Charlottesville, VA 22903, USA}

\author{Lilian Garratt-Smithson}
\affiliation{International Centre for Radio Astronomy Research (ICRAR), The University of Western Australia, 35 Stirling Hwy, Crawley, WA 6009, Australia} 
\alsoaffiliation{ARC Centre of Excellence for All-Sky Astrophysics in 3 Dimensions (ASTRO 3D)}

\author{Gordon German}
\affiliation{Australian SKA Regional Centre (AusSRC), CSIRO, Space and Astronomy, PO Box 1130, Bentley, WA 6102, Australia} 

\author{Kathryn Grasha}
\affiliation{Research School of Astronomy and Astrophysics, Australian National University, Canberra, ACT 2611, Australia}
\alsoaffiliation{ARC Centre of Excellence for All-Sky Astrophysics in 3 Dimensions (ASTRO 3D)}

\author{B\"{a}rbel S. Koribalski}
\affiliation{ATNF, CSIRO, Space and Astronomy, PO Box 76, Epping, NSW 1710, Australia}
\alsoaffiliation{School of Science, Western Sydney University, Locked Bag 1797, Penrith, NSW 2751, Australia}

\author{Raffaella Morganti} 
\affiliation{ASTRON, the Netherlands Institute for Radio Astronomy, Oude Hoogeveensedijk 4, NL-7991 PD Dwingeloo, The Netherlands}
\alsoaffiliation{Kapteyn Astronomical Institute, University of Groningen, Postbus 800, NL-9700 AV Groningen, The Netherlands}

\author{Tom Oosterloo} 
\affiliation{ASTRON, the Netherlands Institute for Radio Astronomy, Oude Hoogeveensedijk 4, NL-7991 PD Dwingeloo, The Netherlands}
\alsoaffiliation{Kapteyn Astronomical Institute, University of Groningen, Postbus 800, NL-9700 AV Groningen, The Netherlands}

\author{C\'eline P\'eroux} 
\affiliation{European Southern Observatory, Karl-Schwarzschild-Strasse 2, Garching bei M\"unchen, 85748, Germany}
\alsoaffiliation{Aix Marseille Universit\'e, CNRS, LAM (Laboratoire d'Astrophysique de Marseille) UMR 7326, F-13388, Marseille, France} 

\author{Max Pettini} 
\affiliation{Institute of Astronomy, University of Cambridge, Madingley Road, Cambridge CB3 0HA, UK}

\author{Kevin A. Pimbblet} 
\affiliation{Centre of Excellence for Data Science, AI, and Modelling (DAIM), University of Hull, Cottingham Road, Kingston-upon-Hull, UK. HU6 7RX}

\author{Zheng Zheng}
\affiliation{National Astronomical Observatories, Chinese Academy of Sciences, 20A Datun Road, Chaoyang, District, Beijing, 100101, China}

\author{Martin Zwaan}
\affiliation{European Southern Observatory, Karl-Schwarzschild-Strasse 2, Garching bei M\"unchen, 85748, Germany}

\author{Lewis Ball}
\affiliation{SKA Observatory, Jodrell Bank, Lower Withington, Cheshire, SK11 9FT, UK}

\author{Douglas C.-J. Bock}
\affiliation{ATNF, CSIRO, Space and Astronomy, PO Box 76, Epping, NSW 1710, Australia}

\author{David Brodrick}
\affiliation{Research School of Astronomy and Astrophysics, Australian National University, Canberra, ACT 2611, Australia}

\author{John D. Bunton}
\affiliation{ATNF, CSIRO, Space and Astronomy, PO Box 76, Epping, NSW 1710, Australia}

\author{F. R. Cooray}
\affiliation{ATNF, CSIRO, Space and Astronomy, PO Box 76, Epping, NSW 1710, Australia}

\author{Philip G. Edwards}
\affiliation{ATNF, CSIRO, Space and Astronomy, PO Box 76, Epping, NSW 1710, Australia}

\author{Douglas B. Hayman}
\affiliation{ATNF, CSIRO, Space and Astronomy, PO Box 76, Epping, NSW 1710, Australia}

\author{Aidan W. Hotan}
\affiliation{ATNF, CSIRO, Space and Astronomy, PO Box 1130, Bentley, WA 6102, Australia}

\author{K. Lee-Waddell}
\affiliation{Australian SKA Regional Centre (AusSRC), The University of Western Australia, 35 Stirling Hwy, Crawley, WA 6009, Australia}
\alsoaffiliation{ATNF, CSIRO, Space and Astronomy, PO Box 1130, Bentley, WA 6102, Australia}
\alsoaffiliation{International Centre for Radio Astronomy Research (ICRAR), Curtin University, Bentley, WA 6102, Australia}

\author{N. M. McClure-Griffiths}
\affiliation{Research School of Astronomy and Astrophysics, Australian National University, Canberra, ACT 2611, Australia}

\author{A. Ng}
\affiliation{ATNF, CSIRO, Space and Astronomy, PO Box 76, Epping, NSW 1710, Australia}

\author{Chris J. Phillips}
\affiliation{ATNF, CSIRO, Space and Astronomy, PO Box 76, Epping, NSW 1710, Australia}

\author{Wasim Raja}
\affiliation{ATNF, CSIRO, Space and Astronomy, PO Box 76, Epping, NSW 1710, Australia}

\author{Maxim A. Voronkov}
\affiliation{ATNF, CSIRO, Space and Astronomy, PO Box 76, Epping, NSW 1710, Australia}

\author{Tobias Westmeier}
\affiliation{International Centre for Radio Astronomy Research (ICRAR), The University of Western Australia, 35 Stirling Hwy, Crawley, WA 6009, Australia}

\doi{10.1017/pas.\the\year.xxx}

\received {dd Mmm YYYY}
\revised  {dd Mmm YYYY}
\accepted {dd Mmm YYYY}
\published{dd Mmm YYYY}

\keywords{galaxies: active -- galaxies: ISM -- methods: observational -- radio lines: galaxies -- radio continuum: general -- surveys}

\begin{document}

\begin{abstract} 
The First Large Absorption Survey in \hi (FLASH) is a large-area radio survey for neutral hydrogen in and around galaxies in the intermediate redshift range $0.4<z<1.0$, using the 21-cm \hi absorption line as a probe of cold neutral gas. The survey uses 
the ASKAP radio telescope and will cover 24,000 deg$^2$ of sky over the next five years. FLASH breaks new ground in two ways – it is the first large \hi absorption survey to be carried out without any optical preselection of targets, and we use an automated Bayesian line-finding tool to search through large datasets and assign a statistical significance to potential line detections. 
Two Pilot Surveys, covering around 3000\,deg$^2$ of sky, were carried out in 2019-22 to test and verify the strategy for the full FLASH survey. The processed data products from these Pilot Surveys (spectral-line cubes, continuum images, and catalogues) are public and available online. In this paper, we describe the FLASH spectral-line and continuum data products and discuss the quality of the \hi spectra and the completeness of our automated line search. Finally, we present a set of 30 new \hi absorption lines that were robustly detected in the Pilot Surveys, almost doubling the number of known \hi absorption systems at $0.4<z<1$.
The detected lines span a wide range in \hi optical depth, including three lines with a peak optical depth $\tau>1$, and appear to be a mixture of intervening and associated systems. Interestingly, around two-thirds of the lines found in this untargeted sample are detected against sources with a peaked-spectrum radio continuum, which are only a minor (5-20\%) fraction of the overall radio-source population.
The detection rate for \hi absorption lines in the Pilot Surveys (0.3 to 0.5 lines per 40\,deg$^2$ ASKAP field) is a factor of two below the expected value. One possible reason for this is the presence of a range of spectral-line artefacts in the Pilot Survey data that have now been mitigated and are not expected to recur in the full FLASH survey. A future paper in this series will discuss the host galaxies of the \hi absorption systems identified here.
\end{abstract}

\section{Introduction}
\label{sec:intro}

Neutral atomic hydrogen (\hix) is a fundamental ingredient in cosmic star formation and galaxy evolution and is crucial to understanding the cosmic baryon cycle \citep{peroux20}. 

For nearby galaxies, large-area surveys for 21\,cm \hi emission like HIPASS \citep{koribalski2004,meyer2004}, ALFALFA \citep{haynes2011}, and WALLABY \citep{koribalski2012,koribalski2020,westmeier2022}, combined with targeted surveys \citep[e.g.][]{walter2008,chung09,deblok24} provide a comprehensive picture of the typical \hi content of galaxies, the physical state of the neutral gas, and its relationship to the galaxy environment and star formation rate. 

For galaxies beyond the local ($z < 0.1$) Universe, the faintness of the 21\,cm line means that very long integration times are needed to detect \hi emission from even the most gas-rich galaxies. A further limitation for \hi studies at $0.09 < z < 0.2$ is the presence of strong radio frequency interference (RFI) from the Global Navigation Satellite Systems (GNSS). In spite of these challenges, direct detections of \hi emission at $z>0.2$ have been made in several surveys, including BUDHIES \citep{jaffe2013}, HIGHz \citep{catinella15}, CHILES \citep{fernandez2016,hess2019} and FUDS \citep{xi2022,xi2024}. In addition, \cite{chakraborty2023} have detected \hi emission from a star-forming galaxy at $z\sim1.3$ that was lensed by a foreground galaxy along the line of sight.

\subsection{Neutral hydrogen in the distant Universe} 

\begin{figure}[!t]
\centering
\includegraphics[width=\textwidth]{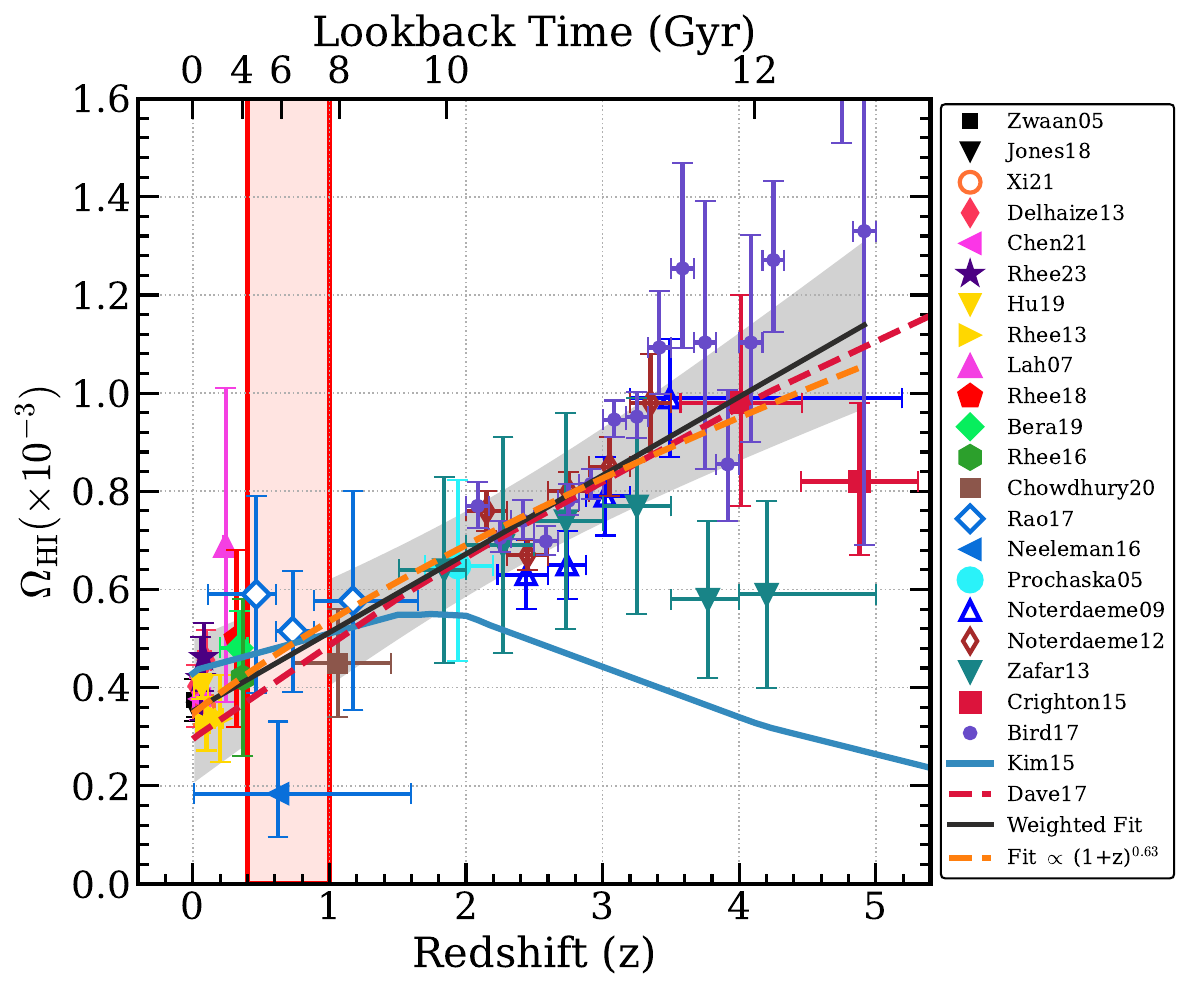}
\caption{Cosmic \hi gas density $\Omega_{\rm HI}$ as a function of redshift (bottom axis) and lookback time (top axis), adapted from \cite{rhee18,rhee23}. This plot uses damped Ly$\alpha$ absorption ($0.1 < z < 5$) and \hi stacking (up to $z \sim 1$) and direct detection of \hi emission at lower-$z$. The redshift range observed in the FLASH survey is shown by light red shading. The black solid line and background grey shade show a weighted linear fit of all $\Omega_{\rm HI}$ measurements and its 95 per cent confidence interval. The orange dashed line is a power-law fit. The predictions from semi-analytic models \citep{kim2015} and cosmological hydrodynamical simulations \citep{dave2017} are denoted with blue solid line and red dashed line, respectively. Further details on the observational data points used for this plot, as well as relevant references, are provided in Figure\,14 of \cite{rhee18} and Figure\,25 of \cite{rhee23}.}
\label{fig:rhee18}
\end{figure}

Most of our current knowledge about \hi in the distant Universe comes from studies of the redshifted 1215.7\,\AA\ Lyman-alpha (Ly$\alpha$) line detected in absorption against background quasars \citep{wolfe1986, wolfe2005}. 
Large ground-based quasar absorption-line surveys \citep[e.g.][]{noterdaeme12,zafar13,sanchez-ramirez2016} have enabled measurements of the cosmic \hi mass density ($\Omega_{\rm HI}$) over a wide redshift range at $z\geq1.7$, as shown in Figure \ref{fig:rhee18}. 

At redshift $z<1.7$ however, the Ly$\alpha$ line lies in the ultraviolet part of the spectrum where it is only observable with space-based telescopes. 
As a result, samples of Ly$\alpha$ absorbers at $z<1.7$ remain small \citep{neeleman16, rao17} and (as can be seen in Figure \ref{fig:rhee18}), our knowledge of the amount and distribution of \hi in the redshift range $0.3<z<1.7$, a timespan of 6.5 Gyr, or almost half the age of the Universe, is patchy and highly incomplete.

21\,cm emission-line stacking techniques can also provide estimates of $\Omega_{\rm HI}$ beyond the local Universe \citep{lah2007, delhaize2013, rhee18, chowdhury2020,chowdhury22,sinigaglia2022,bera23}. Such measurements can give valuable insights into the \hi properties of galaxies at different cosmic epochs but they provide statistical measures rather than information about the properties of individual galaxies.

As discussed by recent studies \citep[e.g.][]{sadler20,allison22,deka24}, the 21\,cm \hi absorption line provides an alternative tool for probing the cold gas content of galaxies beyond the local Universe, since the strength of the absorption line is independent of the distance to the absorber. The 21\,cm line is also unaffected by dust extinction, and surveys for 21\,cm \hi absorption can be carried out over a wide redshift range if a sufficiently large sample of bright background radio sources is used. 
The optical depth of the 21\,cm \hi absorption line is inversely related to the gas spin temperature T$_{\rm S}$, so radio \hi absorption surveys are most sensitive to gas within the cold neutral medium of galaxies with 
T$_{\rm S} \sim100$\,K \citep{braun12}. This cold component is likely to be an effective tracer of star formation in galaxies across cosmic time \citep[e.g.][]{kanekar2014a}. 

\subsection{Surveys for intervening 21\,cm \mbox{H\,{\sc i}} absorption}
Until recently, almost all 21\,cm \hi absorption searches were targeted in redshift, using some form of optical pre-selection such as the detection of \mgii absorption lines \citep{briggs1983} or the presence of a known damped Ly$\alpha$ (DLA) system \citep{york07,kanekar2014a}. 

As a result, existing samples of intervening \hi absorption systems (which trace cold gas in `normal' galaxies along the line of sight to the background source) are relatively small and may suffer from 
selection effects that are difficult to quantify. In particular, the equivalent width of the MgII absorption line mainly measures kinematic spread rather than gas density \citep{rao17} so there is no simple relationship between \hi column density and MgII absorption-line strength in DLA systems \citep{ellison06,ellison09,dessauges-zavadsky09}. The use of optical pre-selection may also introduce a bias against sightlines through dusty, gas-rich galaxies \citep[e.g.][]{fall1993,carilli1998}.

In a targeted search with the Arecibo telescope using pre-selection based on \mgii absorption, \cite{briggs1983} detected two \hi lines from their sample of 18 \mgii systems at $0.36<z<1.94$. They found no correlation between \mgii equivalent width and \hi optical depth, and suggested that the lack of an observed correlation between the optical and radio absorption properties was related to the multi-component nature of the absorbing gas. 

\cite{kanekar2014a} used the Green Bank Telescope (GBT) to search for 21\,cm \hi absorption at the redshift of 22 known quasar DLA systems with $0.8<z<3.5$ and made three detections. By combining the 21\,cm and optical DLA data, these authors were able to derive spin temperatures (or lower limits) for each system. They found statistically-significant evidence for an increase in the spin temperature of DLA systems at higher redshift. Other targeted searches, mainly using \mgii pre-selection \citep{lane98,lane2001,gupta2012,kanekar2014a,dutta2020} have also produced a small number of intervening 21\,cm detections. 

A few intervening \hi absorption detections have been made in untargeted surveys with the GBT in the redshift range $0.3<z<0.8$ \citep{brown73,brown79,brown83,darling2004}. More recently, \cite{grasha2020} used the GBT to search for \hi absorption against 260 radio sources in the redshift range $0<z<2.74$. They re-detected ten known absorption systems, but made no new detections. From their results, \cite{grasha2020} derived measurements of $\Omega_{\rm HI}$ consistent with other methods, inferring a relatively mild evolution in \hi mass density over the redshift range $0<z<2.74$.

\subsection{Surveys for associated 21\,cm \hi absorption}
The detection of cold gas within the host galaxies of radio-loud AGN (``associated \hi absorption'') can provide unique insights into the distribution and kinematics of gas in the nuclear regions of active galaxies and the role gas may play in the evolution of AGN \citep{morganti2018}. In particular, the radio jets in these systems may drive rapid outflows of neutral gas \citep[e.g.][]{Morganti2005,morganti2016,scnulz2021}.  
Previous searches for associated 21\,cm absorption lines have generally used a known optical redshift to target a specific radio frequency. Most of these searches have been carried out at low redshift, though some higher-redshift lines have also been detected. 

Large and sensitive targeted searches for associated \hi absorption in nearby radio galaxies have been carried out by \cite{gereb2014} and \cite{maccagni2017}, with detection rates of up to 30\%. Recent studies have expand these searches to investigate gas in the environments of radio-loud AGN at higher redshifts, reaching beyond $z \sim 3$ \citep[e.g.][and references therein]{kanekar2007,srianand2012,aditya18a,aditya18b,aditya2019,murthy2022}.

The results so far suggest that the detection rate of associated \hi absorption lines is higher in compact radio sources than in extended sources \citep{vermeulen2003, gupta16}, and appears to be lower at redshift $z>0.5$ than in nearby galaxies \citep{aditya18b,su2022,aditya2024}. However, the detection rate at high redshift appears to be influenced by the morphology of the sources. Most \hi detections at high-z are associated with peaked-spectrum sources \citep{aditya18a,glowacki19,aditya2024}, whereas compact, flat-spectrum quasars exhibit a lower detection rate \citep{aditya18b}.

\cite{curran2012} postulate that there is a critical UV continuum luminosity of L$_{\rm 1216} \sim 10^{23}$ W\,Hz$^{-1}$ above which neutral hydrogen is completely ionised and 21\,cm absorption is no longer detected. They suggest that this may partly account for the lower \hi detection rate at higher redshifts where powerful radio-loud quasars are more common. 

More recently, \cite{murthy2022} investigated selected targets where the UV luminosity fell below the threshold of 10$^{23}$ W\,Hz$^{-1}$, finding that neither UV nor radio luminosity of the AGNs is likely to cause the lower detection rate. This work suggests that the lower detection rate of associated \hi 21\,cm absorption lines at high redshifts may be due to evolution in the physical conditions of \hix, possibly either lower \hi column densities or higher spin temperatures in high-$z$ AGN environment. There have also been studies of associated absorption in Ly$\alpha$, i.e. the proximate DLAs (PDLAs), which do indeed seem to be more common towards radio selected QSOs \citep[e.g.][]{ellison2002,ellison2010,russell2006}.

\subsection{\hi absorption studies with SKA precursor telescopes}
A new parameter space for \hi absorption studies has recently been opened up by the development of wide-band correlators for radio interferometers that provide instantaneous redshift coverage approaching that of optical spectrographs, and by the construction of the SKA precursor telescopes ASKAP \citep{johnston2008,hotan21} and MeerKAT \citep{jonas2009} on radio-quiet sites where RFI contamination is minimised. Importantly, these radio-quiet sites uniquely allow near-continuous coverage of the 21\,cm line at frequencies between 0.5 and 1\,GHz, enabling spectral -line studies of \hi at $0.4<z<1.8$ \citep[e.g.][]{allison2017,deka2023}. 

The MeerKAT Absorption-Line Survey \citep[MALS;][]{gupta16} is currently conducting a search for \hi absorption lines at redshift $0<z<1.45$ in several hundred fields (each of area $\sim1\,{\rm deg}^2$, the total sky coverage of $\sim1000\,{\rm deg}^2$) centred on bright radio-loud quasars with a known optical redshift \citep{gupta2022}. This survey is optimised for lines with low \hi column densities, reaching limits as low as N$_{\rm HI} \sim 10^{19}$\,cm$^{-2}$ for the brightest background sources and N$_{\rm HI} \sim  10^{21}$\,cm$^{-2}$ for the other sources in general.

The wide (36\,deg$^2$) field of view, broad spectral coverage and radio-quiet site of the Australian SKA Pathfinder (ASKAP) radio telescope \citep{hotan21} makes it possible to carry out the first `all-sky' survey for 21\,cm \hi absorption across a wide redshift range ($0<z<1.0$) without any optical pre-selection of targets. The design and science goals for such a survey, the ASKAP First Large Absorption Survey in \hi (FLASH) are described in detail in the survey design paper by \cite{allison22}. In contrast to the Meerkat MALS survey, FLASH is optimised for the detection of high column-density \hi absorption lines (typically with N$_{\rm HI} > 10^{21}$\,cm$^{-2}$) against relatively faint radio sources.

The first \hi absorption-line searches with ASKAP \citep[][]{allison15,moss17,glowacki19,sadler20,mahony2022} used a smaller array of 6--12 ASKAP dishes with a single bright continuum source placed near the field centre. These observations detected a variety of associated and intervening \hi absorption lines, and showed that ASKAP can produce high-quality radio spectra that are essentially free of terrestrial RFI in the 700--1000\,MHz band. 

\cite{allison20} presented the first ASKAP detection of \hi absorption against a source away from the field centre in a wide-field observation with 12-14 dishes (the GAMA\,23 Early Science field). Like the earlier ASKAP observations, the GAMA23 data were processed with a custom data pipeline based around the Miriad software package \citep{sault95}. 
The much larger datasets produced by the full 36-dish ASKAP array require the use of a dedicated observatory-built data pipeline, ASKAPsoft \citep{wieringa20}, and so a key aim of the ASKAP Pilot Survey program was to test and verify this pipeline.

The huge size of the ASKAP spectral-line datasets means that spectral-line visibilities cannot be stored once the data have been processed through ASKAPsoft \citep{hotan21}, so re-processing of the raw data is not possible. Testing and verification of the full ASKAPsoft data pipeline for FLASH data is therefore essential, and this was achieved through the ASKAP Pilot Survey program described in this paper. 

\subsection{Outline of this paper} 
Section~\ref{sec:pilots} of this paper describes the two ASKAP FLASH pilot surveys carried out from 2019--22, which together covered around 3,000\,deg$^2$ of sky. Section~\ref{sec:obs} gives details of the fields observed, Section~\ref{sec:data} describes the data processing and Section~\ref{sec:casda} discusses the data products released through the CSIRO ASKAP Data Archive (CASDA)\footnote{\url{https://research.csiro.au/casda/}}. 
Section~\ref{sec:cont} briefly describes the continuum images and catalogues released from the pilot surveys, while Section~\ref{sec:quality} discusses the overall quality of the processed spectral-line data released in CASDA. 

Section~\ref{sec:lines} presents the results from the FLASHfinder, an automated tool to search for \hi absorption and discusses machine learning classifications of the detected lines. A Discussion section, an outline of future work, and a brief summary can be found in Sections~\ref{sec:discussion} and \ref{sec:summary}. Some further technical details are included in the Appendices, and plots of the detected \hi lines are shown in \ref{sec:plots}.

\section{The ASKAP-FLASH Pilot Surveys}
\label{sec:pilots} 

A program of ASKAP pilot surveys was carried out before the start of the full set of five-year ASKAP surveys. Along with other ASKAP Survey Science teams, the FLASH team was allocated 100~hours of observing time for a first round of pilot surveys in 2019--20 and a further 100 hours for the second pilot survey round in 2021--22. 

\subsection{FLASH survey strategy and data pipeline} 

The FLASH pilot surveys had two main goals: to test and validate the observing and data processing strategies for the full all-sky FLASH survey, and to provide a first look at the population of \hi absorption systems detected in a wide-field radio absorption-line survey conducted without any optical pre-selection. 

\begin{figure*}[!ht]
\centering
\includegraphics[width=\textwidth]{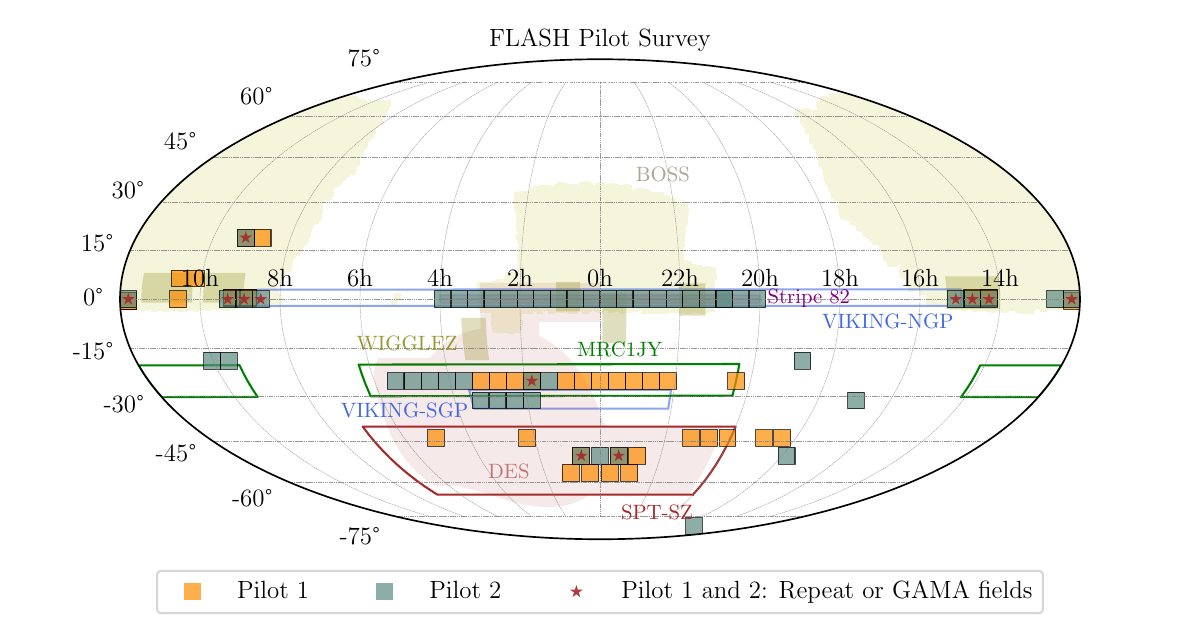}
\caption{FLASH Pilot Survey Sky Coverage. Orange squares show the Pilot Survey 1 fields, and green squares the Pilot Survey 2 fields. Fields observed in both Pilot~1 and 2 are marked with star symbols. The coordinates are equatorial, with right ascension versus declination. The footprints of multi-wavelength surveys, such as BOSS, DES, MRC~1Jy, Stripe~82, WiggleZ, VIKING, and SPT-SZ, are included.}
\label{fig:footprint}
\end{figure*}

In the pilot surveys, we carried out end-to-end tests and validation of the ASKAPsoft data pipeline to ensure that it was able to produce the required spectra, continuum images and catalogues to the quality specified in the original FLASH survey proposal as outlined by \cite{allison22}. We also assessed the variation in sensitivity across the full ASKAP field of view at 711.5-999.5~MHz and measured the system performance and noise properties over a wide range in declination (from at least $+20$\,deg to $-50$\,deg declination) and for both daytime and nighttime observations. Figure~\ref{fig:footprint} shows the sky coverage of the survey.

\subsection{Choice of observing parameters}
The choice of observing frequency for FLASH is a trade-off between the frequency-dependent sensitivity of the ASKAP telescope and our desire to optimize the redshift path-length for detecting \hi absorption. This trade-off is discussed in detail by \cite{allison22}, who show that (i) a frequency range of \ 711.5--999.5\,MHz maximises the total absorption path length sampled, and (ii) an integration time of around two hours per field maximises the sky area that can be covered to reasonable depth in a fixed time, which in turn maximises the number of absorption lines detected and the discovery potential of the survey. 

\begin{figure*}[!ht]
\centering
\includegraphics[width=0.48\columnwidth]{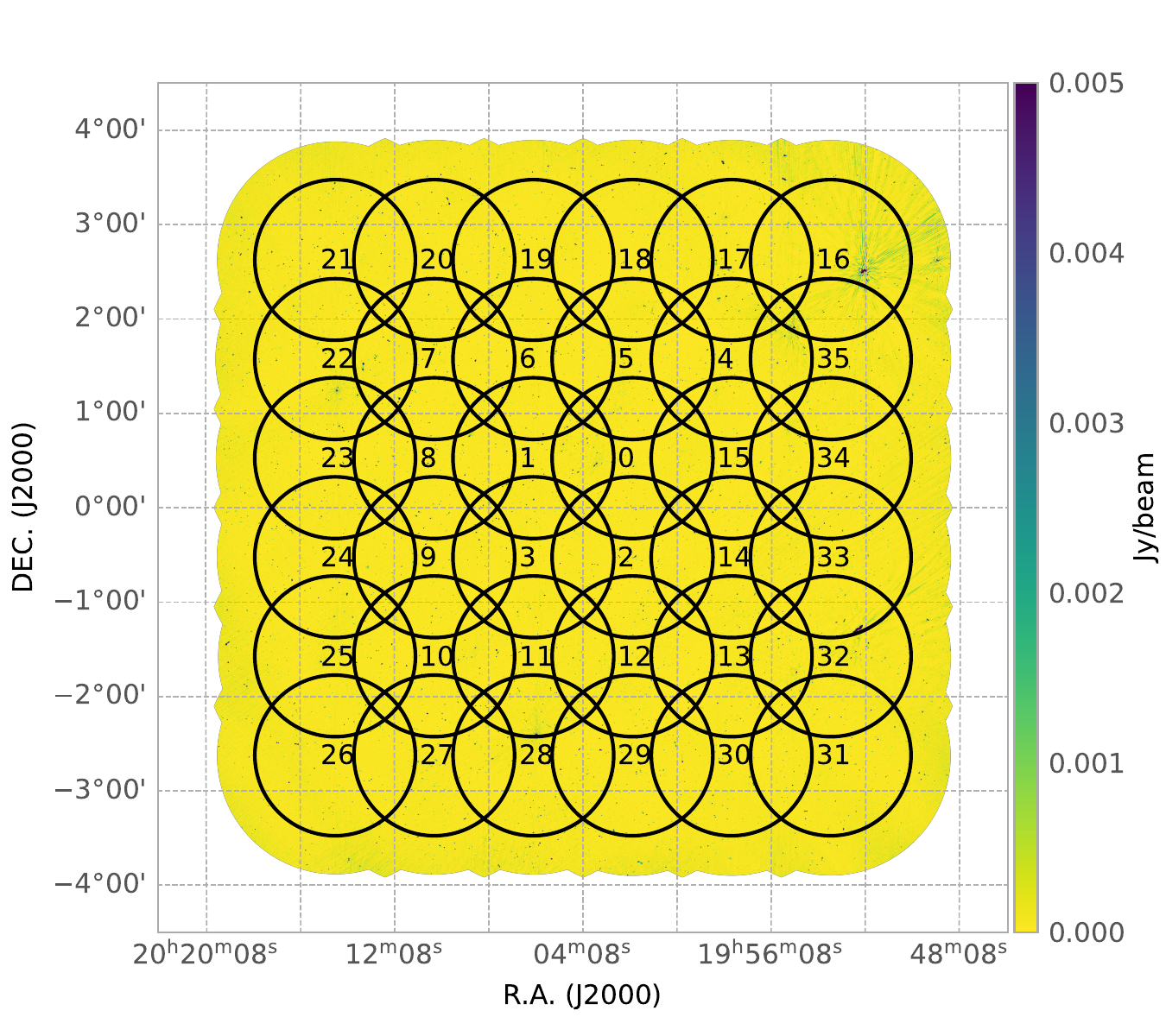}
\includegraphics[width=0.48\columnwidth]{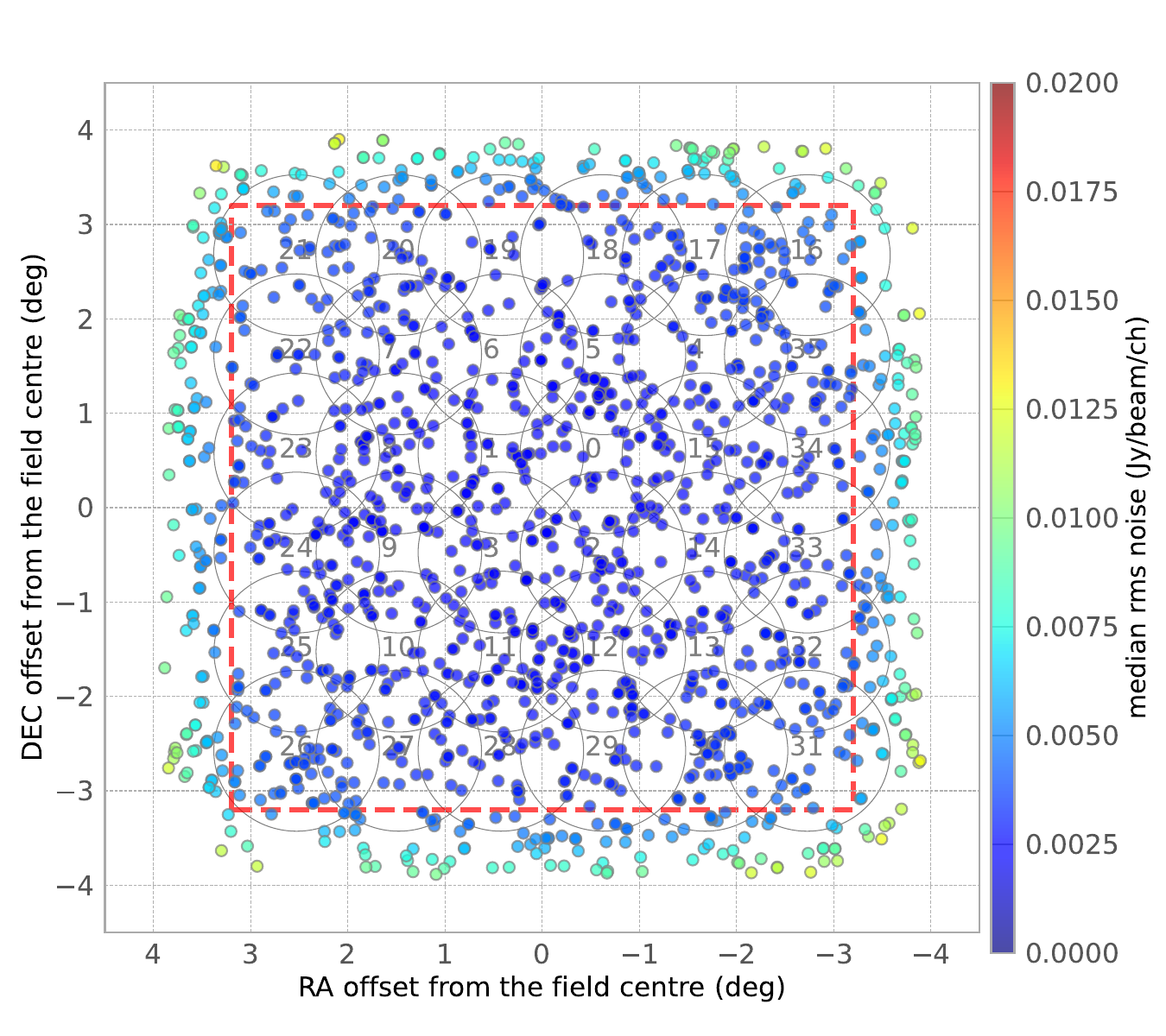}
\caption{FLASH field of view and noise properties: (left) a FLASH continuum image showing the position of the 36~ASKAP beams in the {\tt square\_6$\times$6} foootprint used to form the image, (right) measurements of spectral-line noise for extracted spectra in a 6-hour Pilot 1 observation (SBID~13293), showing that the noise is roughly uniform within the central red dashed square, covering 6.4 deg$\times$6.4~deg on the sky (total area 40.9~deg$^2$).}
\label{fig:beam}
\end{figure*}

Table~\ref{tab_obs_param} lists the observing parameters used for the FLASH pilot survey. Individual observations used the grid of fields and pointing centres listed in Table 2 of \cite{allison22}, which is designed to provide uniform sensitivity across the survey area and is the same grid used by the first RACS low survey \citep[The Rapid ASKAP Continuum Survey;][]{mcconnell2020}. We used the {\tt square\_6$\times$6} ASKAP beam footprint \citep{hotan21} shown in Figure \ref{fig:beam}. In the frequency range used for FLASH, this provides an rms spectral-line noise that is uniform across the central 6.4\,deg $\times$6.4 deg of the ASKAP field of view, as also shown in Figure \ref{fig:beam}.

\begin{table}[t!]
\caption{Observing parameters for the FLASH Pilot Surveys. For a detailed description of the ASKAP array, see \citet{hotan21}.}
\begin{tabular}{lll}
\hline
\hline
Number of antennas & 36 (diameter = 12\,m) \\
Frequency range & 711.5 -- 999.5\,MHz \\
Central frequency & 855.5\,MHz \\
Redshift coverage & $z=0.42-1.00$\ (\hi line) \\
& $z=0.67-1.34$\ (OH 1665-67 lines) \\
Footprint & {\tt square\_6$\times$6} \\
Interleaving & None \\
Field of view & $6.4\times6.4$\,deg$^2$ \\
Integration time & 2\,hr (standard) \\
                  & 6\,hr (deep GAMA fields) \\ 
Channel width & 18.5~kHz \\
Velocity width & 5.5 -- 7.8\,km\,s$^{-1}$\\
Noise level of spectra & 5.5~mJy~b$^{-1}$(see Figure~\ref{fig:spec_noise_all_sbids}) \\           
Spatial resolution of continuum & 15~arcsec\\
\hline\hline
\end{tabular}
\label{tab_obs_param}
\end{table}

\subsection{Spectral-line beamforming intervals with ASKAP} 
\label{sec:bfinterval}
ASKAP's Phased Array Feeds allow us to digitally form up to 36 beams across the field of view. By default, the beam-forming weights are calculated in 1 MHz intervals across the full 288 MHz bandwidth (see \cite{hotan21} for further details), which is prone to produce bandpass imperfections at the same frequency interval. At the redshifts covered by the FLASH survey this corresponds to velocity widths of 235--420 km\,s$^{-1}$ which would severely limit our ability to detect broader \hi absorption lines. 

To overcome this, fixing the beam-weights over larger bandwidths is required. FLASH pilot survey observations carried out observations using both 5 and 9\,MHz beam-forming intervals to assess which mode was most suitable moving forward. The majority of Pilot 1 fields used 9~MHz beamforming intervals with the exception of SBIDs (Scheduling block IDs) 10849, 10850, 11051, 11052, 11053, 11068 which used 5\,MHz intervals. Observations using 9\,MHz intervals showed larger phase jumps between neighbouring intervals and required higher order polynomials for the bandpass smoothing so it was concluded that 5\,MHz intervals provided the optimum observing set-up for all Pilot Survey~2 observations and future FLASH observations.

\section{Observations}
\label{sec:obs} 

\subsection{Pilot Survey 1: 2019-20}
The first FLASH pilot survey, covering $\sim1510$\,deg$^2$\ of sky was carried out with the full 36-dish ASKAP telescope between December 2019 and September 2020. 
We observed 40~fields with an integration time of two hours each (the 2-hr fields). For six of these fields, we also carried out a longer (6\,hr) integration (the 6-hr fields). 
The observed fields were chosen to span a wide range in declination, but also to include some areas of the sky with good quality optical data. 

In Pilot Survey~1, 37 of the 2-hr fields and four 6-hr fields were successfully processed through the ASKAPsoft data pipeline and released in CASDA. One of these fields (listed as J2022-2507P in Table~\ref{tab:obs}) was observed at the wrong RA because of a scheduling error\footnote{The cooordinates for some of the earliest Pilot Survey observations were entered by hand, and in this case were entered incorrectly. The telescope now has an automated scheduler that should ensure correct pointing.},  but still contains useful data. The fields released are shown in Figure~\ref{fig:footprint} and are listed in \ref{sec:obslogs}. 

\subsection{Pilot Survey 2: 2021-22}

After some further refinements to the data processing pipeline, a second pilot survey, was carried out from November 2021 to August 2022. The fields observed in this second pilot survey are also shown in Figure~\ref{fig:footprint} and are listed in \ref{sec:obslogs}. 

In Pilot Survey 2, we observed 50 fields with an integration time of two hours each. 
Twelve of these were repeats of fields observed in Pilot 1, and were used to test the reliability and reproducibility of the absorption lines detected by ASKAP as discussed in Section \ref{sec:repeat}. 
The other 38 fields were observed for the first time and and cover $\sim1550$\,deg$^2$ of sky.
 
\subsection{Sky areas covered by the two Pilot Surveys}

The data released in CASDA from the two FLASH pilot surveys include 75 unique ASKAP fields with a total sky coverage of 3000\,deg$^2$ (see Figure \ref{fig:footprint}).

The Pilot Survey fields include several well-observed regions with multi-wavelength photometric and spectroscopic information. These include: 
\begin{itemize}
    \item 
The three equatorial fields (at 9h, 12h and 15h RA) from the GAMA galaxy survey \citep[Galaxy And Mass Assembly;][]{driver11}, each covering 48\,deg$^2$ of sky. 
The GAMA fields were observed on non-standard field centres in the first pilot survey, so that each field was completely covered by two ASKAP pointings \citep{su2022}. In the second pilot survey, the GAMA fields were re-observed on the standard FLASH pointing centres listed by \cite{allison22}. 

\item
All of {\bf SDSS Stripe 82} \citep{annis14}, a well-studied 300 deg$^2$ region of sky with RA between 20h and 04h (300 to 60 deg) and declination between -1.26 and +1.26 deg. 
The available multi-wavelength data products include deep optical \citep[][$r < 24.6$~mag]{jiang14} and 1.4~GHz radio imaging \citep{hodge11} as well as extensive optical spectroscopy. 

\item 
Part of the {\bf WiggleZ }\  \citep{drinkwater18} redshift survey area. WiggleZ targeted a UV-selected galaxy sample of candidate star-forming galaxies at redshift $0.2<z<1.0$ across several fields covering a total of $\sim1000$\,deg$^2$ of sky, so is well-matched to the redshift range covered by FLASH. Further analysis of the FLASH WiggleZ data is provided by \cite{eden2025}. 

\item 
Part of the {\bf MRC\,1Jy} \citep{mccarthy1996} survey area, which covers much of the southern extragalactic sky between declination $-20$ and $-30$\,degrees. An early analysis of FLASH data from the MRC\,1Jy region is given by \cite{aditya2024}.
\end{itemize}

\section{Data Processing}
\label{sec:data}

\subsection{Pipeline processing}
The observations were processed through the standard ASKAPsoft pipeline \cite{whiting2020}. This is a scripted workflow that, for FLASH pilot processing, ran on the \textit{galaxy} supercomputer at the Pawsey Supercomputing Centre\footnote{\url{https://pawsey.org.au}}. It performs all the necessary calibration, imaging and source-extraction tasks required to produce science-ready data products. A detailed description of the pipeline can be found elsewhere \citep[e.g.][]{hotan21}, but we give specific details relevant for the FLASH processing here.

The bandpass and primary flux calibration were done using PKS~B1934$-$638, which was observed at the centre of each beam for $\sim200$~sec. The bandpass solutions were derived for each beam, antenna, polarisation combination and smoothed over intervals of 1~MHz to improve the signal-to-noise of the calibration solutions. Each beam of the science observation was calibrated with the solutions from the corresponding beam of the bandpass observation.

All imaging of the science observations was done independently for each beam. A continuum image (using visibilities averaged to 1~MHz resolution) was made first, through an iterative self-calibration approach. This involves imaging, then calibrating the time-dependent complex gains against a model derived from the image, then re-imaging (with more cleaning) after applying those gain solutions. The imaging used multi-scale (scales of 0, 6, 15, 30, 45, 60 pixels) multi-frequency (2 Taylor-terms) synthesis, with Wiener preconditioning applied using robustness=0.0. This gives an average PSF size (averaged over all Pilot observations) of $18\times13$~arcsec. A second image was also made with a robustness=1.0 (giving an average resolution of $28\times20$~arcsec) to more closely match the spatial resolution of the spectral line cube.

The full-spectral resolution data were then imaged over the full band. First, the continuum was subtracted from the visibilities by forming a model from the continuum image and transforming to the ($u,v$) plane. After this, the data were imaged in each channel to form a spectral cube. 
Two cuts on the ($u,v$) distance were applied, by removing baselines $<70$m to limit solar interference, and $>4000$m to limit the size of the full image. Multi-scale imaging (scales 0, 3, 10, 30~pixels) was done, using Wiener preconditioning with robustness=1.0 -- this, in addition to the ($u,v$) cuts, provides an average PSF resolution in the middle of the band of $29\times22$~arcsec, ranging from $34\times36$~arcsec to $25\times19$ arcsec across the band. Following imaging, a further round of continuum-subtraction was then applied, to further remove any continuum residuals. Each spectrum in the cube (that is, each spatial pixel) had the continuum level fitted within each beam-forming interval (or, in the case of Pilot Phase 1, each 1~MHz interval) and subtracted. Breaking at the edge of each beamforming interval allows the removal of any residual discontinuities not completely removed by the bandpass calibration (this is particularly important for brighter sources that might have a higher signal-to-noise than we get on B1934$-$638 in 200~sec).

The continuum (1~MHz-resolution) data were also imaged in each channel to form continuum cubes. For Pilot 2 these were done with the same ($u,v$) cuts and preconditioning as the spectral cubes, to provide matching-resolution data to enable more accurate extraction of continuum spectra in post-processing.

Once imaging was complete, the individual beams were mosaicked together to form the full field images in each of continuum, continuum-cube and spectral-cube modes. Source-finding was run with Selavy \citep{whiting2012} to create catalogues of continuum components and islands, and the full-spectral-resolution spectra at the location of each component brighter than 45~mJy (20~mJy for the 6-hour observations) were extracted and stored as 1-dimensional spectra.

\subsection{Pilot 1 data and post-processing}
\label{sec:post}

Observations for the first FLASH pilot survey used 5 or 9\,MHz beam-forming intervals instead of the standard 1~MHz (see Sec~\ref{sec:bfinterval}). However, during data processing we found that these larger intervals were not functioning as anticipated, still causing amplitude and phase jumps at 1\,MHz intervals. As these jumps occurred precisely at the 1\,MHz boundaries, they could be corrected for in the pipeline processing by smoothing the bandpass solutions on 1\,MHz intervals as well as removing any residual artefacts by using 1\,MHz intervals for the image-based continuum subtraction. This procedure corrects for the large amplitude jumps and leads to a clean, smooth bandpass across the entire band. However, 1\,MHz corresponds to a line width of approximately 300\,km\,s$^{-1}$ at these redshifts, meaning that processing the data using 1\,MHz intervals risks removing any real \hi absorption signatures of similar line widths. This could mean that some \hi absorption lines are either completely subtracted out or (if detected) the shape of the line is significantly altered because of the limited line-free channels used for the continuum subtraction in each 1\,MHz interval. 

To address this issue, we carried out some further post-processing outside of the ASKAPsoft pipeline. This involved downloading the spectral cubes from CASDA that were made prior to the image-based continuum subtraction, and extracting the spectra towards all bright radio sources. For the majority of the fields observed the flux limit for spectrum extraction was set at 30\,mJy, which is deeper than the pipeline limit of 45~mJy. For the longer 6\,hr integrations, a flux density threshold of 10\,mJy was used. 
A spectrum was extracted at the peak pixel of each catalogued continuum component. 

After extracting the spectra from the spectral cube, some further continuum subtraction was carried out to remove any residual continuum not subtracted in the visibility domain. This involved fitting a second-order polynomial within each 5 or 9\,MHz beam-forming interval, while masking any strong lines (either due to real absorption or RFI) that were more than 3-$\sigma$ above the noise. 

\begin{figure}[!t]
\begin{center}
\includegraphics[width=1.1\textwidth]{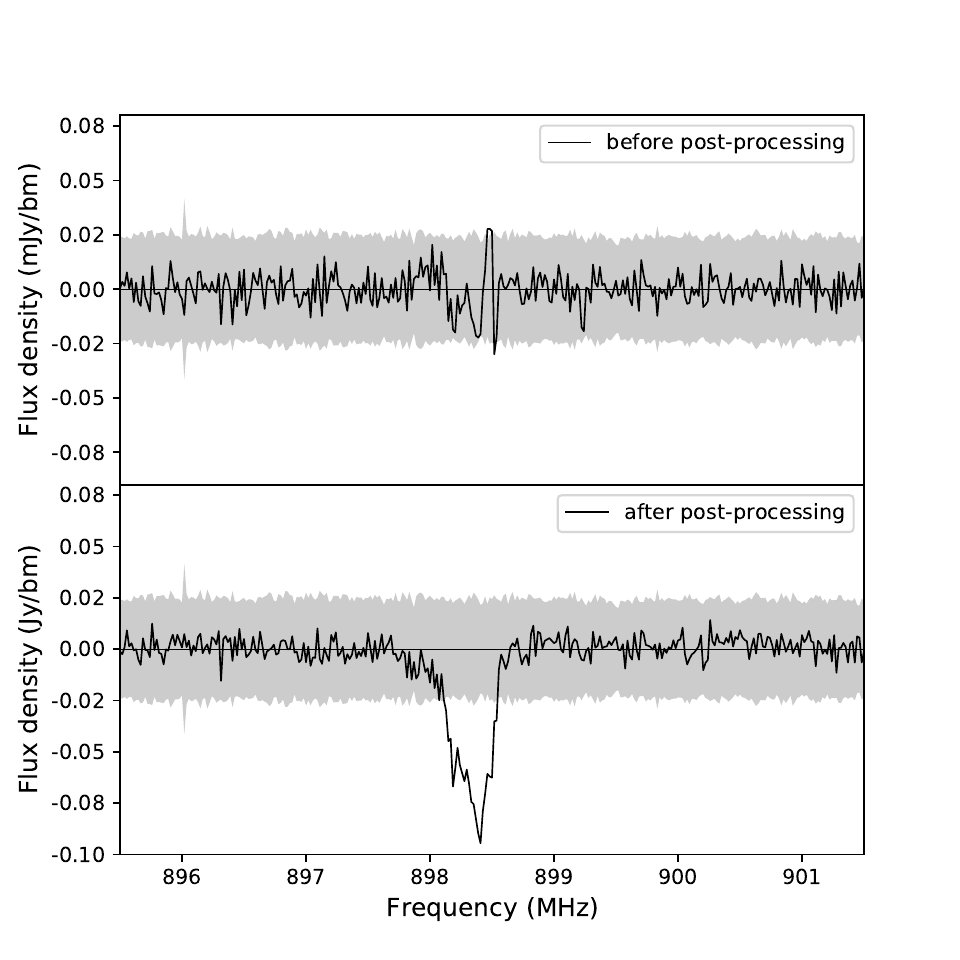}
\caption{An example of a broad \hi absorption line detected after post-processing of the FLASH pilot survey data. The spectrum was extracted toward PKS~2311-477 in FLASH~160P field (SB~15873 component~5a, $S = 1.04$\,Jy). The grey shaded region is set at five times the rms spectral-line noise. The absorption is found at $z = 0.5811$, and the FWHM linewidth from a single Gaussian fit is 112~km~s$^{-1}$.
}
\label{fig:post-processing}
\end{center}
\end{figure}

This post-processing procedure allowed us to recover some \hi absorption lines that were inadvertently subtracted out by using the default pipeline processing parameters, but did lead to some broader spectral ripples in some frequency ranges, particularly towards the brighter continuum sources. Figure~\ref{fig:post-processing} shows an example of a broad \hi absorption line that was subtracted out in the initial processing, but was recovered after re-doing the continuum subtraction over wider frequency intervals. The post-processed data will be publicly released in CASDA alongside this publication.

\subsection{Pilot 2 data} 
The overall data processing method for Pilot Survey 2 was similar to that for Pilot 1. Improvements in correctly implementing larger beam-forming intervals were made prior to the Pilot 2 survey, meaning that 5-MHz intervals were used for all Pilot-2 survey observations and post-processing of the data was not required.

Unfortunately, over half of the Pilot 2 fields show significant ripples in their spectra, caused by a processing error in the bandpass calibration for a large batch of observations made between 17 December 2021 and 15 January 2022 (see Section \ref{sec:artefacts} for more details). These ripples are particularly noticeable in the spectra of bright sources, and the features occur across the whole frequency band. 

Although 24 of the SBIDs with spectral ripples were re-observed after February 2022, more than half of these repeated fields were affected by tropospheric ducting of RFI (see Section \ref{sec:artefacts}) across at least 20\% of the frequency range. These fields will be repeated in the full FLASH survey.  

We are confident that the problems seen in the Pilot 2 spectra are unlikely to recur for the full FLASH survey. In particular, we now monitor for the presence of tropospheric ducting of RFI (which mainly affects the 700-1000\,MHz ASKAP band) and take steps to avoid observing FLASH fields when ducting is present.

\begin{figure*}[!ht]
\begin{center}
\includegraphics[width=\textwidth]{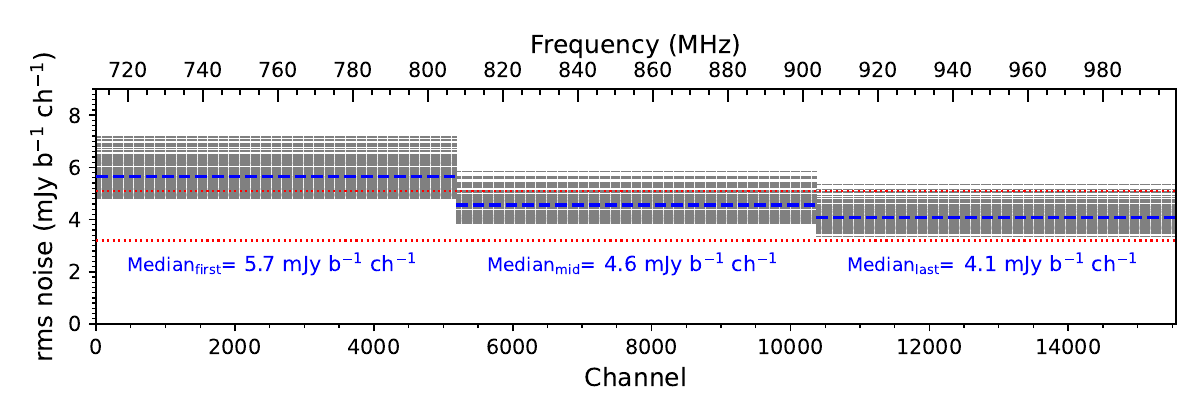}
\caption{Measurements of median spectral-line noise (in mJy beam$^{-1}$ ch$^{-1}$; channel width: 18.5\,kHz) in three sub-bands of FLASH field 307 (SBID~13268), showing the values in each sub-band within the 712-1000\,MHz observing band. Grey lines show the values for individual sources within 3.2 degrees of the field centre, and the dashed blue line shows the median for the field as a whole. The dotted red lines indicate the range of predicted noise across the entire band (3.2--5.1\,mJy beam$^{-1}$ ch$^{-1}$), as estimated by \cite{allison22}.
}
\label{fig:noise_13268}
\end{center}
\end{figure*}

\section{Data products released in CASDA}
\label{sec:casda}

\subsection{Processed data products}
Processed data products from the FLASH pilot surveys have been released through the CSIRO ASKAP Science Data Archive (CASDA) at \url{https://research.csiro.au/casda/} under project code AS~109. To download data, users need to register and create a CASDA account.

The output data products loaded into CASDA for each observation are listed with the examples of the filenames as follows:
\begin{enumerate}
\item
{\bf Continuum catalogues:}\ A component catalogue and a separate island catalogue are produced by the Selavy source finder \citep{whiting2012}. These catalogues are used for data validation, and the component catalogue positions are also used to extract spectra for the absorption-line search. The island catalogues provide a useful resource for optical cross-matching.\\ 
{\it $-$ File examples:} 
``selavy-image.i.SB15873.cont.taylor.0.\\~restored.conv.components.xml'', 
``selavy-image.i.SB15873.\\~cont.taylor.0.restored.conv.islands.xml''
\item
{\bf Continuum images and cubes:}\ We produce both a Stokes I continuum image at $\sim15$\,arcsec resolution (rms noise $\sim90\,\mu$Jy/beam) 
and a continuum cube. The continum cubes are  matched to the 30\,arcsec resolution of the spectral-line cube for the later Pilot fields.\\ 
{\it $-$ File examples:} ``image.i.SB15873.cont.taylor.0.restored.\\~conv.fits'', ``image.restored.i.SB15873.contcube.conv.fits'' 
\item
{\bf Spectral-line cubes:} Two full spectral-line cubes are produced for each SBID. An initial spectral cube is created after continuum subtraction in the visibility domain, and a final spectral cube after the image-based continuum subtraction. This second cube is used to extract the individual \hi\ spectra. Both cubes cover the full 288\,MHz bandwidth at 18.5\,kHz spectral resolution with all 36 beams mosaiced together to produce a single cube of the full ASKAP field.\\
{\it $-$ File examples:} ``image.restored.i.SB15873.cube.contsub.fits''
\item
{\bf Individual spectra at positions of continuum sources:} The ASKAPsoft pipeline extracts individual source and noise spectra for all radio components with flux above 45\,mJy (20\,mJy for 6-hr fields), which is the predicted detection limit by \cite{allison22}. The spectra can be downloaded individually, or as a bulk .tar file for each SBID.\\
$-$ {\it File examples:} 
``spec$\_$SB15873$\_$component\_1a.fits"
\item
{\bf Validation reports:} These reports include metrics and general notes on data quality. 
\end{enumerate}

\subsection{Data validation and quality assessment} 
The processed data products from the ASKAPsoft pipeline are assessed and validated by the FLASH science team before their public release. This data validation uses a checklist that takes into account the metrics provided by the ASKAPsoft pipeline (such as measurements of the rms noise in line and continuum), as well as a visual inspection of the images and a representative sample of spectra by members of the science team. 

The completed checklist provides a numerical score, which the team use to classify each SBID as `Good', `Uncertain' or `Bad'. SBIDs classified as Good or Uncertain are released, but those classified as Bad remain unreleased and are flagged for reobservation. Among the released Pilot Survey data, 17 and 99~SBIDs are classified as `Good' and `Uncertain', respectively. The high fraction of `Uncertain' is from the early observations requiring post-processing (Pilot~1; see Section~\ref{sec:post}) and the presence of artefacts (Pilot~2; see Section~\ref{sec:artefacts}). The overall quality of the data are good for Pilot~1, and the artefacts observed in Pilot~2 have been mostly resolved for the Full Survey. The team also provide some brief Release Notes to accompany the CASDA data. 

\subsubsection{Continuum data validation}
The validation data provided in CASDA for the continuum images include beam diagnostics, the spatial distribution of identified sources, an measurement of the rms noise in the continuum, and a flux density comparison with other surveys (NVSS, SUMSS or RACS). Astrometric checks are also carried out by measuring the median position offset from published surveys. 

As part of the data validation process, the FLASH team checked that the continuum catalogues were present and contained a reasonable number of sources, typically between 10,000 and 20,000. We also visually inspected each continuum image to ensure that there were no obvious problems such as large-scale patterns or artefacts around strong sources. 

\subsubsection{Spectral-line data validation}
In addition to the diagnostics provided by ASKAPsoft, the FLASH team verifies the presence of the image cubes, and the presence of several hundred spectra that are expected from each of our observations. We also measured the rms noise in the spectral line data both for the whole frequency range and across three sub-bands as shown in Figure \ref{fig:noise_13268}, and checked that this was in line with expectations. Further details will be described in Section~\ref{sec:quality}.

\begin{figure*}[!ht]
\begin{center}
\includegraphics[width=0.85\textwidth]{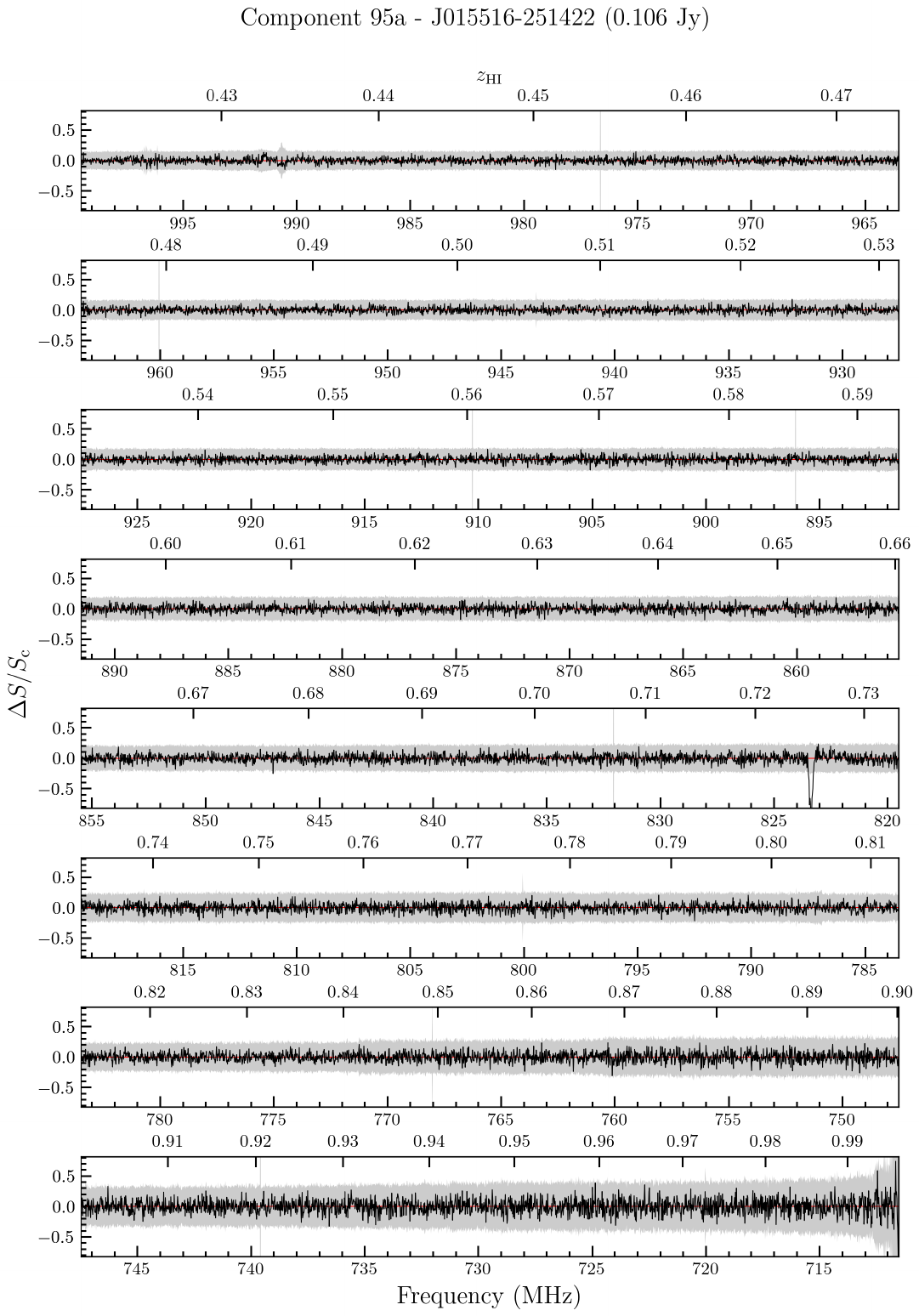}
\caption{An example of a spectral-line plot produced by the FLASH team during post-processing and used in data validation. In this figure, the gray band is expanded to five times the RMS spectral-line noise to clearly visualize the noise level across the band. The source shown here (NVSS\,J015516-251423, SB~37475 component~95a) has an \hi absorption line at $z=0.7251$~($f=823.4$~MHz), which extends beyond the gray band, indicating a peak S/N of ~> 5. (see Table \ref{tab:finder2}).}
\label{fig:spectrum}
\end{center}
\end{figure*}

Finally, as part of the post-processing workflow, we produce spectral-line plots like that shown in Figure \ref{fig:spectrum}. The extracted spectra of the ten brightest sources in each field are then inspected to check for any artefacts or non-uniformity of the bandpass. 

\section{Continuum images and catalogues}
\label{sec:cont}

\subsection{Wide field continuum images}
FLASH observations provide high-quality continuum products, including wide field images and source catalogues, in addition to the spectral-line data. 
As can be seen from Table \ref{tab_cont_comp}, the FLASH continuum images are intermediate in sensitivity between the first epoch of RACS-Low \citep{mcconnell2020} and the EMU Pilot Survey \citep{norris2021}.

\begin{table*}[t]
\small
\caption{Comparison of continuum properties for images from three ASKAP surveys:  FLASH, RACS-Low, and EMU. }
\centering
\begin{tabular}{@{}lcrrccc@{}}
\toprule
Survey & Central & Integration & 1\,$\sigma$ rms  & Resolution & \\
       & \multicolumn{1}{c}{freq. (MHz)} & \multicolumn{1}{c}{time} & \multicolumn{1}{c}{($\mu$Jy/beam)} & \multicolumn{1}{c}{arcsec} & \\
\hline       
RACS-Low & 888 & 15 minutes & $\sim250$ & 15 &  \\
FLASH    & 856 &  2 hours & $\sim90$ & 15  &  \\ 
EMU      & 944 &  10 hours & $\sim30$ & 12-15 & \\
\bottomrule
\end{tabular}
\label{tab_cont_comp}
\end{table*}

The restored total-intensity (Stokes I) continuum images are accumulated over the entire bandwidth being processed. Within 3.2~deg from the field centre, the rms noise in the continuum images is uniform. Most continuum images are of excellent quality and revealed several radio sources with complex asymmetric morphologies.

The rms noise level in the 2-hour continuum images is typically $\sim90\,\mu$Jy beam$^{-1}$, i.e. roughly five times deeper than NVSS (450$\mu$Jy beam$^{-1}$) and an order of magnitude deeper than SUMSS ($\sim1$\,mJy beam$^{-1}$). The initially expected median image noise, estimated from the results of RACS-Low \citep{mcconnell2020}, is $\sim93\,\mu$Jy beam$^{-1}$ \citep{allison22}, which is close to our observed values.

\subsection{Selavy radio source catalogues}
As described in Section~5.1, the ASKAPsoft pipeline generates two different continuum source catalogues via the ASKAP source-finder Selavy:\\
(i) An {\bf island}\ catalogue is generated first - this is a catalogue of groups of contiguous pixels that are above some detection threshold. \\
(ii) A {\bf component}\ catalogue is then generated, where each component is a two-dimensional Gaussian, parameterised by a location, flux, size and orientation. Each island has one or more components fitted to it, so there is a one-to-many relationship between the island and component catalogues. \\
Both catalogues are available for download through CASDA. 

\subsection{Astrometric accuracy}
The astrometric accuracy of the listed ASKAP continuum catalogues is determined by the combination of a statistical component (set by the S/N of the source and the angular resolution of the image), and a systematic component (set by the accuracy to which the measured source positions can be aligned with a standard reference frame) \citep{heywood2016}. For the FLASH continuum sources of interest for this project, the statistical component is small ($<0.1$\,arcsec for a 10\,mJy source) and the position uncertainty is dominated by the systematic component. The size of this systematic component is estimated at the data validation stage through a cross-comparison with other large-area source catalogues. For the pilot surveys, these catalogues were NVSS \citep{Condon1998} and SUMSS \citep{Mauch2003}. The continuum data validation reports, available via CASDA, present astrometric offsets between the sky positions of FLASH and NVSS/SUMSS for individual SBIDs.

The validation reports show systematic offsets of up to 1\,arcsec (and occasionally larger) in both RA and Dec for individual ASKAP fields. In this paper we adopt an indicative astrometric error of 1\,arcsec for the FLASH positions, noting that \cite{hale2021} found offsets of a similar size for the ASKAP RACS-Low fields. Although we could improve the astrometric accuracy of the pilot survey positions by correcting for the known RA and Dec offsets of each SBID, we chose not to do so in this paper because the pipeline positions are accurate enough for us to be able to match our bright ($>30$\,mJy) sources with published radio catalogues. 
We plan to revisit the ASKAP astrometry in future, in a paper that identifies and discusses the host galaxies of the \hi absorption lines found in the FLASH pilot surveys. 

\subsection{Flux-density scale}
To check the ASKAP flux density scale in the FLASH band (712--1000\,MHz), we cross-matched sources in our southern (Dec $<-30^\circ$) pilot fields with the SUMSS catalogue \citep{Mauch2003}. SUMSS was chosen because its 843\,MHz survey frequency is close enough to the FLASH band centre at 856\,MHz that we expect the flux densities to be directly comparable. If the field is outside the SUMSS coverage, NVSS is used instead. The flux-density ratio as function of S/N and its variation by sky position for individual SBIDs can be found in the continuum data validation reports on CASDA.

From a cross-matched sample of over 5000 bright ($>20$\,mJy) and spatially unresolved FLASH continuum components, we find that the FLASH and SUMSS sources are on the same flux density scale to within 2--3\%. Since SUMSS itself is consistent with the 1.4 GHz NVSS flux density scale \citep{Mauch2003}, this gives us confidence in the reliability of the ASKAP flux density measurements. 

\section{Spectral-line data} 
\label{sec:quality}

\begin{figure*}[!ht]
\begin{center}
\includegraphics[width=0.92\textwidth]{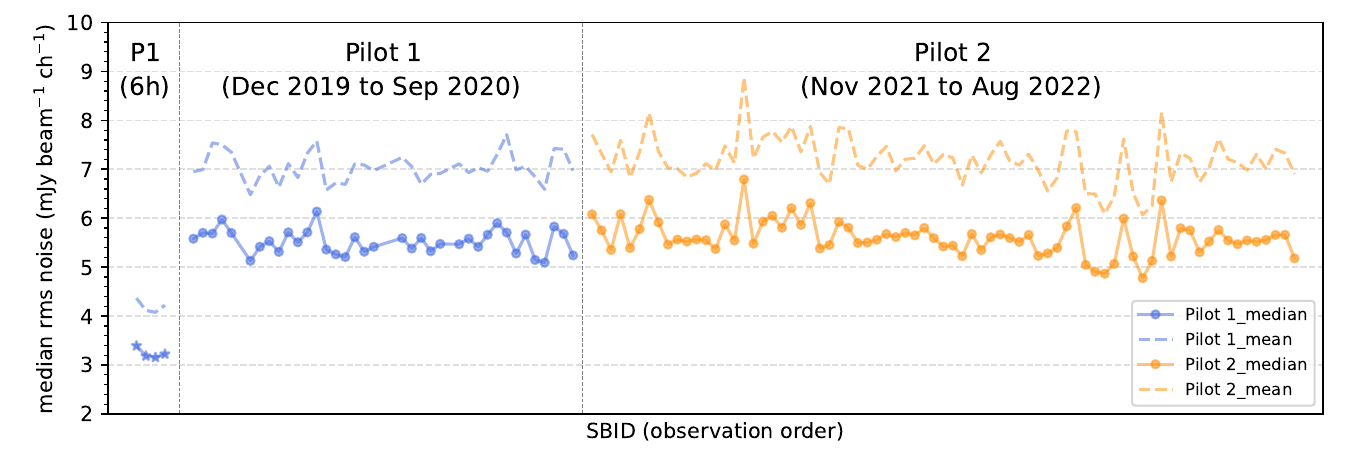}
\caption{RMS spectral-line noise estimates measured from the noise spectra of all FLASH Pilot fields. The estimates account for all sources located at the edge of the field. Blue and orange symbols represent Pilot 1 and Pilot 2 fields, respectively. Individual dots on the dashed and solid lines indicate the mean and median noise from the mid-5184 channels of all noise spectra in each field. The 6-hr fields with lower rms levels are marked as stars. The x-axis is ordered by SBID, following the observation sequence except for the 6-hr fields.}
\label{fig:spec_noise_all_sbids}
\end{center}
\end{figure*}

\begin{figure*}[!ht]
\begin{center}
\includegraphics[width=0.32\textwidth]{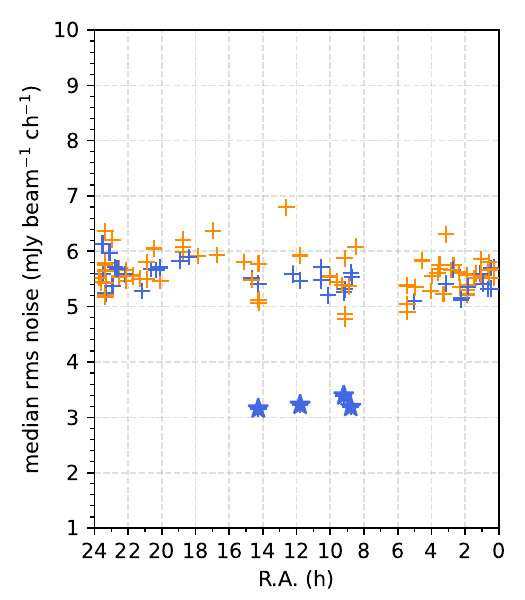}
\includegraphics[width=0.32\textwidth]{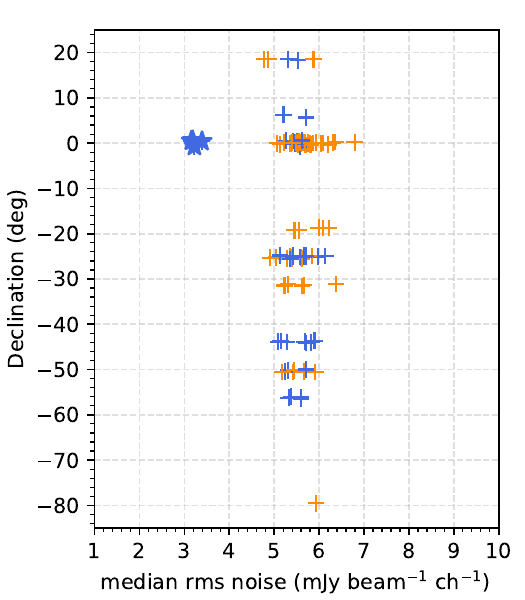}
\includegraphics[width=0.34\textwidth]{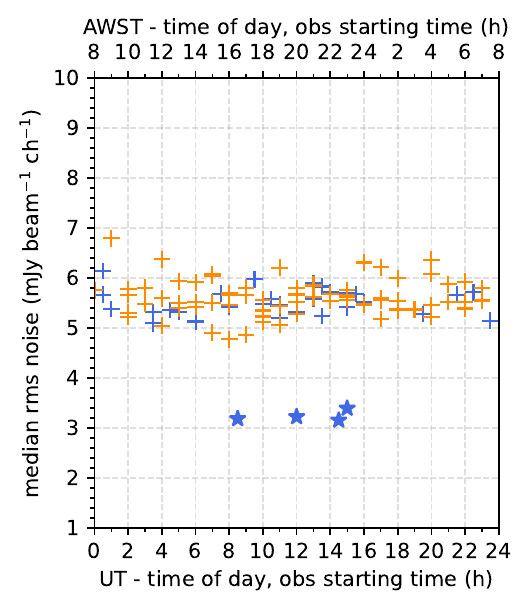}
\caption{RMS spectral noise comparison across different right ascension (left), declination (middle), and day/night observation (right) demonstrating uniform noise. Blue and orange crosses are from Pilot 1 and Pilot 2, respectively. 6-hr fields with lower rms levels are marked as stars. The median rms of each field is derived from the middle 5184 channels of the individual noise spectra for all sources in that field. Each mean noise level spans 3.16--3.39\,mJy\,beam$^{-1}$\,ch$^{-1}$ (Pilot~1, 6-hr fields), 5.10--6.13\,mJy\,beam$^{-1}$\,ch$^{-1}$ (Pilot~1), and 4.78 to 6.80 (Pilot~2) \,mJy\,beam$^{-1}$\,ch$^{-1}$.}
\label{fig:noise_comparison}
\end{center}
\end{figure*}

\subsection{Uniform sensitivity of the data}
One aim of the Pilot Surveys was to check the uniformity and reproducibility of the ASKAP data. To do this, we tested for variations in sensitivity either with declination or between day-time and night-time observations. 

\subsubsection{Spectral-line sensitivity}
Figure \ref{fig:spec_noise_all_sbids} shows the overall rms noise in the spectral-line, measured 
across all FLASH fields in date order. The rms noise in both continuum and spectral-line data remained roughly uniform over the period of almost three years during which the observations were carried out. 
 
The rms noise in the extracted spectra from the 2-hour fields has a median value of 4.6~mJy per beam per channel in the central third of the frequency band (808--904\,MHz) and ranges from 4.2 to 5.3\,mJy beam$^{-1}$ ch$^{-1}$ in this sub-band (Figure~\ref{fig:noise_13268}). 
This is consistent with the range of 3.2--5.1 \,mJy beam$^{-1}$ ch$^{-1}$ quoted in Table~1 of \cite{allison22}. The rms noise for the 6-hour fields was around 3.5~mJy per beam per channel, which is slightly higher but still close to the value expected for the longer observing time. The expected noise level for a 6-hour observation, reduced by a factor of $\sqrt{3}$ compared to the 2-hour estimate from \cite{allison22}, ranges from 1.8 to 2.9 \,mJy beam$^{-1}$ ch$^{-1}$.

The spectral-line sensitivity varies by 25--30\% across the observing band because of changes in the ASKAP effective system temperature T$_{\rm sys}$ with frequency \citep[see Figure 22 of][for details]{hotan21}. Figure~\ref{fig:noise_13268} shows this variation for a typical SBID. 

\subsubsection{Sensitivity versus sky position}
FLASH fields are widely spread over the sky, so we were able to compare the noise across a wide range in R.A. and declination, and between daytime and nightime. Figure~\ref{fig:noise_comparison} shows that the noise properties are roughly uniform with respect to both the right ascension and declination.

\subsubsection{Daytime versus night-time observing}
The right-hand plot in Figure~\ref{fig:noise_comparison} shows the rms noise as a function of local observing time (AWST). We see no significant difference in sensitivity between daytime observations and observations made at night. 

\subsection{Spectral-line artefacts} 
\label{sec:artefacts}
\begin{figure}[t!]
     \centering
        \includegraphics[width=1.0\columnwidth]{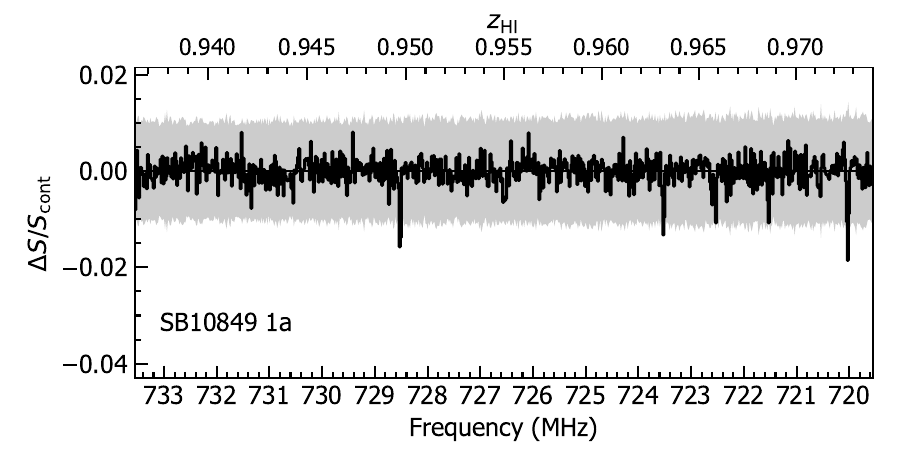}
        \vspace{-1mm}          
        \includegraphics[width=1.0\columnwidth]{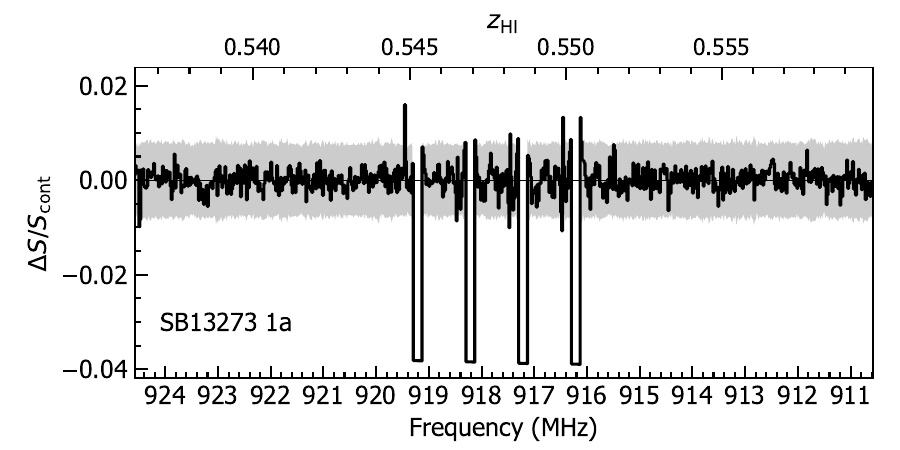}
        \vspace{-1mm}     
        \includegraphics[width=1.0\columnwidth]{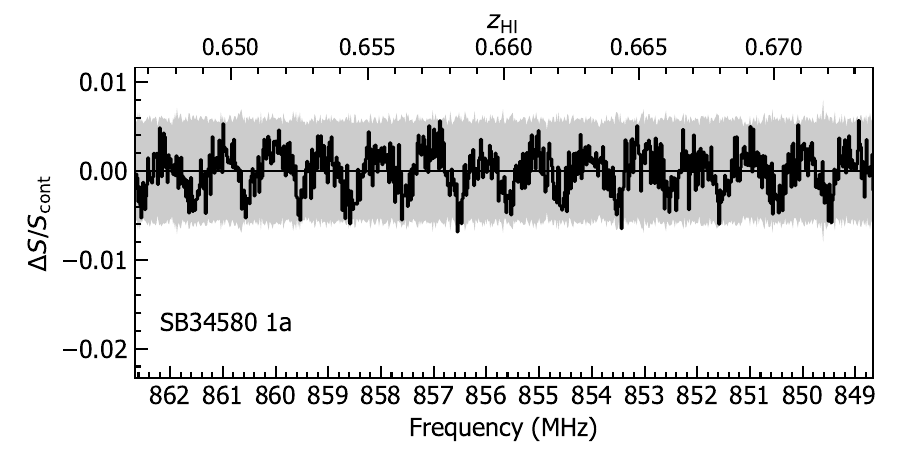}
        \vspace{-1mm} 
        \includegraphics[width=1.0\columnwidth]{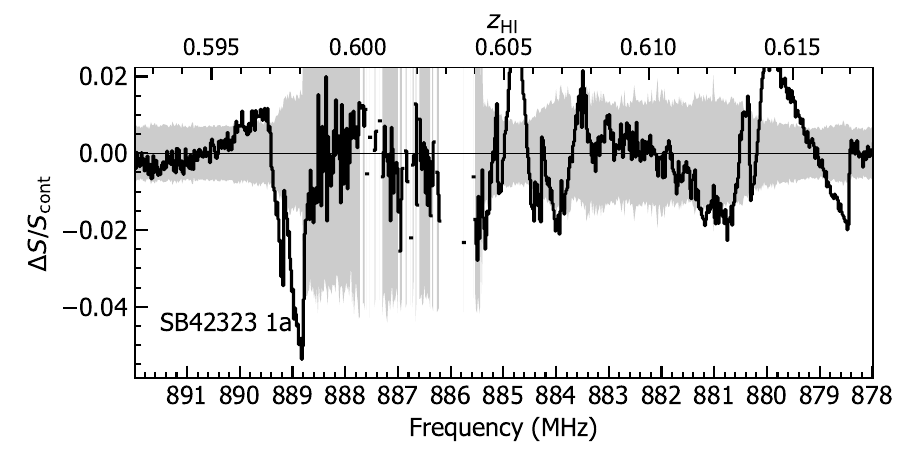}
        \vspace{-1mm} 
        \caption{Examples of spectral-line artefacts seen from the Pilot data. From top to bottom, each panels presents glitches, correlator dropouts, ripples, and tropospheric ducting of RFI in spectra. The grey background shade represents five times the rms spectral-line noise.} 
        \label{fig:artefacts}
\end{figure}

We identified several artefacts affecting the data quality of the spectra as shown in Figure~\ref{fig:artefacts}. The underlying cause of three of the artefacts seen in the pilot survey data  has now been addressed, and these artefacts are not expected to occur in future FLASH observations. These artefacts were:

\begin{enumerate}
    \item {\bf Glitches at the edges of beam-forming intervals}. Data from the first 18 Pilot~1 fields randomly showed glitches on 1~MHz interval due to the issues with bandpass smoothing parameterisation. This problem has been fixed, and is not seen in data taken after April 2020. 
    \item {\bf Correlator dropouts}. These are artefacts in specific frequency ranges, caused by data dropouts in an ASKAP correlator block that were not properly accounted for in processing.  
    The main frequency ranges affected are 915--920\,MHz (corresponding to \hi redshifts in the range $0.543<z<0.552$) and 987--992\,MHz (\hi at $0.431<z<0.439$.  This problem has largely been fixed, and mainly affected data taken before 2021.    
    \item {\bf Ripples}.  Wobble-like features are seen in the processed spectra of bright sources in almost all Pilot~2 fields observed between 17 December 2021 and 15 January 2022.  These ripples were produced during processing by the use of incorrect parameters in fitting the spectral bandpass, and affect data across the whole spectral band. Delays in uploading the processed data to the CASDA archive for data validation meant that the problem was not identified immediately, and because the ASKAP spectral-line visibilities are deleted immediately after processing it was not possible to reprocess the data with the correct parameters. The problem is not expected to recur.
\end{enumerate}

There is also one ongoing effect that can produce significant spectral-line artefacts: 

\begin{enumerate}
\setcounter{enumi}{3}
\item  {\bf Tropospheric ducting of RFI}. This occurs under particular atmospheric conditions, when radio signals from distant transmitters are refracted in the troposphere and can propagate over large distances. 
Ducting can occur in the ASKAP frequency range as well as at lower frequencies \citep{sokolowski2017}. 
For FLASH observations, tropospheric ducting can result in the telescope seeing RFI from mobile phone bands used in Perth and other communities well beyond the radio-quiet zone in which ASKAP is located. 
When ducting is present, ducted RFI signals can occupy up to 20\% of the FLASH frequency band. The frequencies most commonly affected are  $758-788$\,MHz, $870-890$\,MHz, and $943-960$\,MHz. 
\end{enumerate}

Ducted RFI was rarely seen in FLASH pilot survey data taken before January 2022, but became much more common, after that time. The reason for this is not yet understood, but well over half the FLASH pilot survey observations between May and August 2022 (17/24 fields) showed ducted RFI in the spectra of bright sources (see Figure \ref{fig:artefacts} for an example). 

If this high rate of ducting occurs again in future, we can minimize its effects in two ways. The weather conditions that give rise to tropospheric ducting can often be predicted, and we can avoid scheduling FLASH observations when these conditions are present. Alternatively, if ducting conditions are detected at the telescope while a FLASH observation is underway, the observation will be stopped, the data discarded, and the observation rescheduled for a later time. A more detailed discussion of ducting effects at the ASKAP site is given by \cite{indermuehle2018} and \cite{lourenco2024}. 

\section{FLASHfinder - an automated search for \hi absorption lines}
\label{sec:lines}

\subsection{An automated absorption line-finder for ASKAP}
The FLASHfinder\footnote{\url{https://github.com/drjamesallison/flash_finder}} Bayesian line-finding tool \citep{allison12} was designed to identify and parameterise \hi absorption lines in large ASKAP spectral-line datasets in an automated and efficient way, and to assign a statistical significance to individual line detections. The line finder has been tested on simulated data by \cite{allison12}, and its application to real data is discussed in detail by various recent studies \citep[e.g.][]{allison2012b,glowacki19,allison20,allison21,mahony2022,su2022,aditya2024}.

\begin{figure*}[!ht]
    \includegraphics[width=0.48\textwidth]{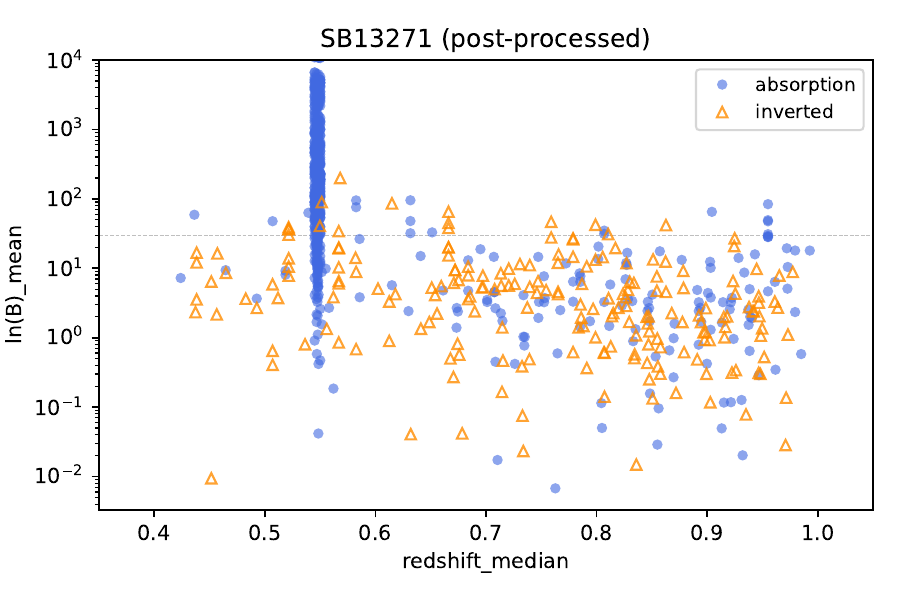}
    \includegraphics[width=0.48\textwidth]{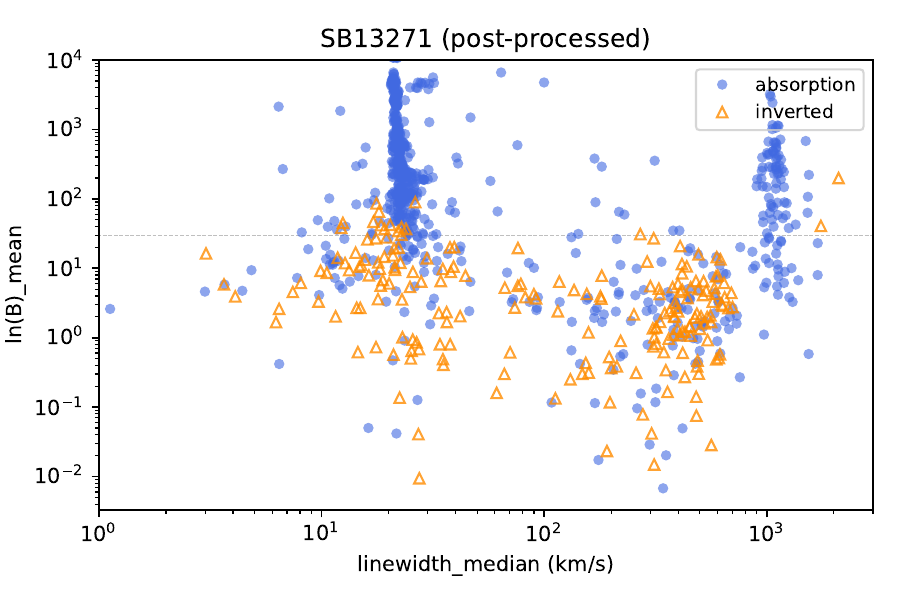}
    \includegraphics[width=0.48\textwidth]{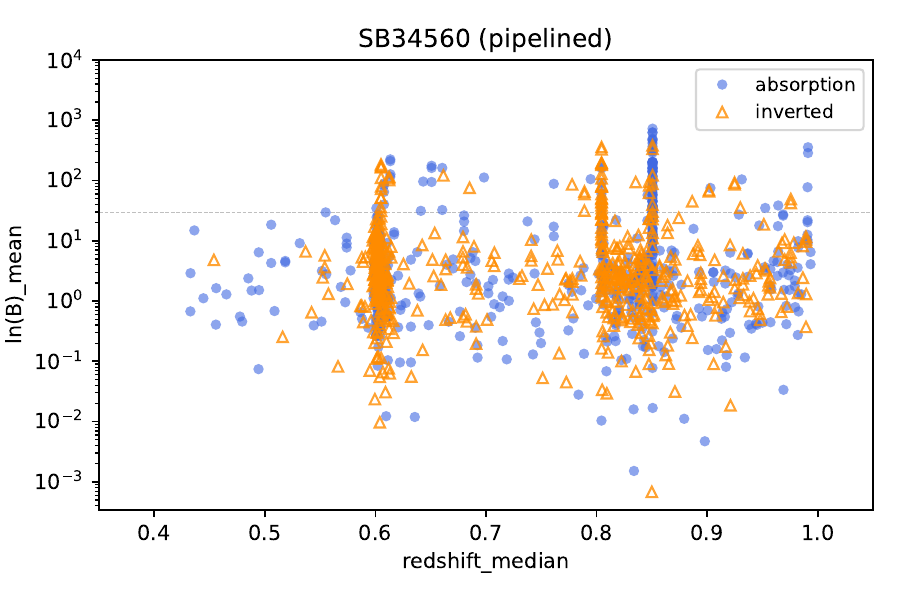}
    \includegraphics[width=0.48\textwidth]{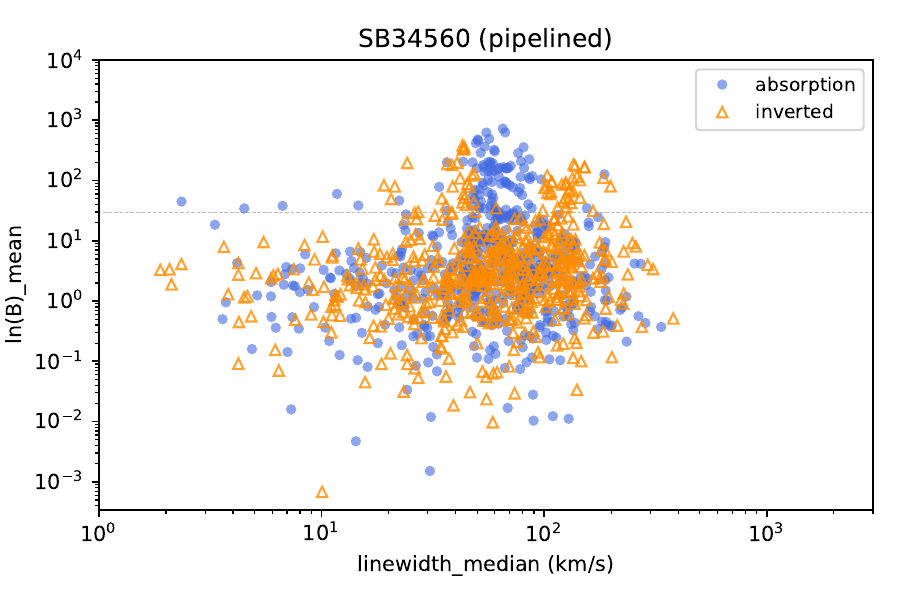}    
    \caption{Plot of Bayesian evidence with ln\,(B) versus redshift (left) and linewidth (right) for extracted spectra from FLASH field 719, with the Pilot 1 linefinder results (top) and the Pilot 2 results (bottom). In both plots the lines found from the original absorption spectra are plotted in blue and those from the inverted spectra in orange.}
    \label{fig:abs_inv}
\end{figure*}

\subsection{Line identification in the Pilot Survey spectra}
The FLASHfinder was initially run across the full 711.5--999.5 MHz range for each of the (several hundred) extracted spectra from each SBID. 
We use a single Gaussian model with nlive = 1000, which is the number of live points used for sampling \citep{allison12}. The output data file produced by the line finder includes measurements of the following parameters and their 1$\sigma$\ uncertainties, based on fitting a single Gaussian component to each candidate line: 
\begin{itemize}
    \item 
$z_{\rm HI}$, the redshift at the peak of the candidate line.
\item
$\tau_{\rm pk}$, the peak optical depth of the line.
\item 
$\tau_{\rm int}$, the integrated optical depth of the line, in km\,s$^{-1}$.
\item 
The FWHM linewidth with a single Gaussian fit of the line, in km\,s$^{-1}$.
\item 
The Bayesian evidence value, ln\,(B), which reflects the extent to which the line detection is favoured over the null hypothesis in which no line is present (for a discussion of the relationship between ln\,(B) and the signal-to-noise ratio, see \ref{sec:bayes}).  
\end{itemize}

\begin{figure*}[!ht]
\includegraphics[width=0.95\textwidth]{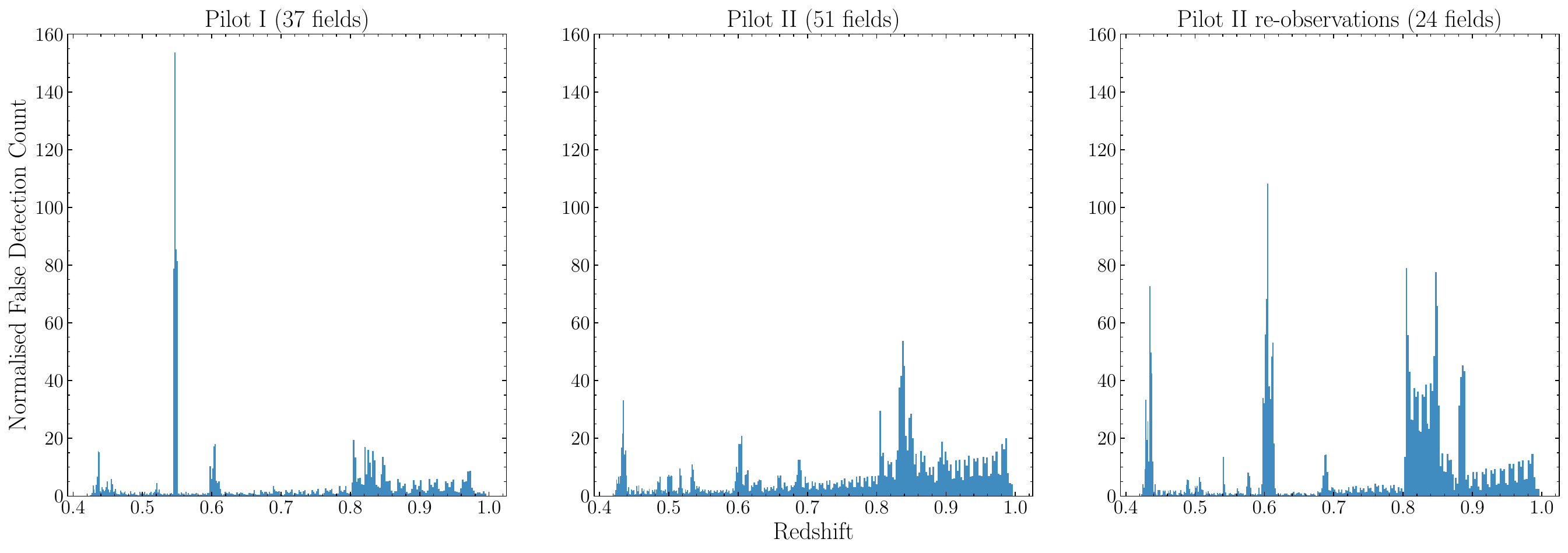}
\caption{The false detection (artefact) count normalised by the number of fields in the absorption and inverted spectra as a function of redshift across the FLASH pilot surveys. The bin sizes correspond to the redshift range covered by 1~MHz interval. The highest peaks for the first pilot survey are mainly caused by correlator dropouts, while those from the second pilot survey are mainly due to ducted RFI.}
\label{fig:bad_intervals}
\end{figure*}

In this initial run the linefinder was set to search at all possible redshifts in each spectrum, with no prior constraints on where a line might be. 
The output files for each SBID contain a set of candidate detections that may be genuine astronomical signals, spectral-line artefacts, or noise peaks.  

\subsection{Identifying genuine absorption lines }
To distinguish astronomical signals from noise peaks, we took a conservative approach by considering only lines for which ln\,(B)$\geq30$. 
We chose this value by inverting the original spectra and re-running the line finder on the inverted spectra (which we expect to contain only spectral artefacts and noise, since the \hi emission line is too weak to be seen in the FLASH redshift range). As can be seen from Figure \ref{fig:abs_inv}, a value of ln\,(B)$\sim30$ represents an upper envelope to most of the linefinder points from the inverted spectrum, with the exception of several strong spectral artefacts seen as vertical lines in both plots. 

We next aimed to distinguish astronomical signals from spectral artefacts like those shown in Figure \ref{fig:artefacts}. In general, we expect that an \hi absorption line at any given redshift should occur in only one or two spectra in an SBID, while the spectral artefacts occur in the same position in many different spectra. 

To identify repeated artefacts and spectral-line glitches at specific frequencies across many SBIDs, 
we aggregated the total number of detections returned by the FLASHfinder algorithm using a log-likelihood cut of ln\,(B)$>0$.  
From the histograms in Fig~\ref{fig:bad_intervals}, the sharp spikes with counts above $\sim$40 correspond to regions with correlator dropouts or ducting. 
From these plots, we compiled a list of frequencies at which the most common spectral artefacts occurred. 
In general, the Pilot 2 data more often contains artefacts (see Section~7.2) than the Pilot 1 data after normalising by the number of fields observed. 

\subsection{Properties of the detected lines}
In all, we detected 30 lines with ln\,(B)\,$> 30$ that we are confident are robust detections of genuine \hi absorption lines. Additionally, three lines from well-known sources have been identified but are not included in this paper as they have been previously studied (see \ref{sec:notes}). As we will show in section \ref{sec:completeness}, it is likely that other \hi lines with smaller ln\,(B) values are present in the Pilot survey data, but we have not attempted to identify them at this stage and this will be the subject of a future paper. 

\begin{figure*}[h!]
     \centering
     \begin{subfigure}[b]{0.495\textwidth}
         \includegraphics[width=\textwidth]{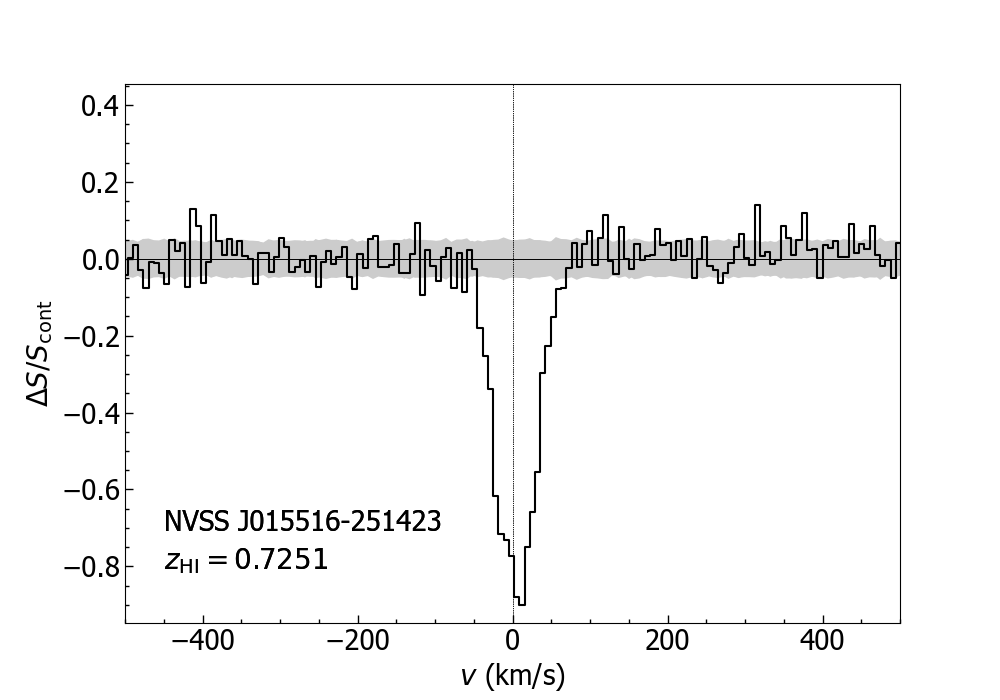}
         \caption{NVSS~J015516-251423 (SB~13281~103a)}
     \end{subfigure}
     \begin{subfigure}[b]{0.495\textwidth}
        \includegraphics[width=\textwidth]{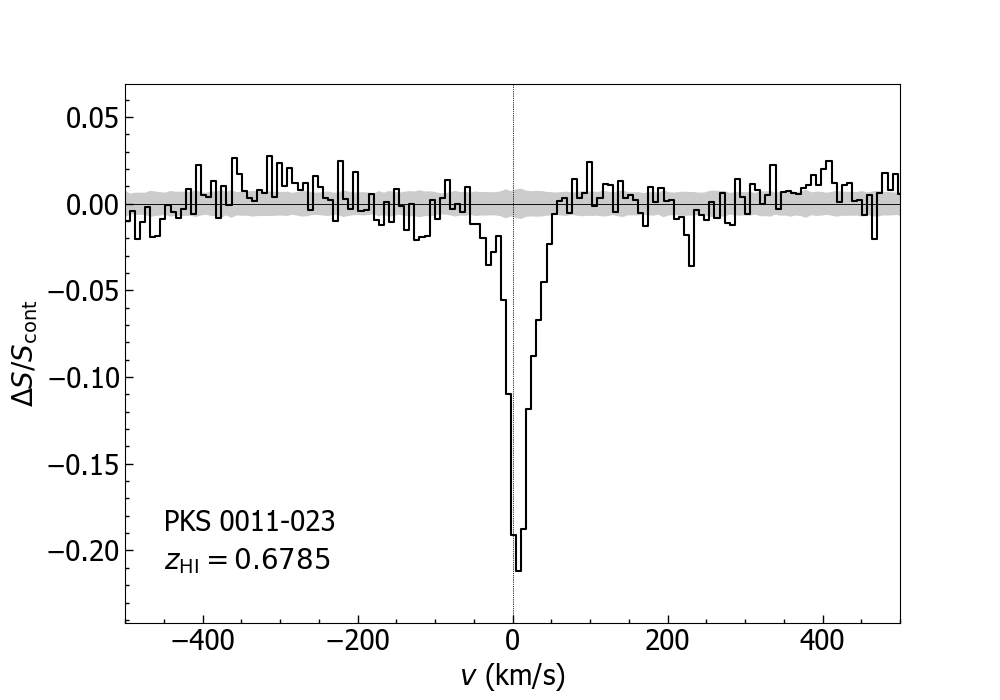}
         \caption{PKS 0011-023 (SB~34581~9a)}
     \end{subfigure}
     \hfill
        \vspace{-2mm}          
        \caption{Two strong \hi absorption lines identified in Pilot Survey Phase~1 and 2. The line profiles of all the detections are added in \ref{sec:plots}. The grey shaded region in this figure and the figures in \ref{sec:plots} represents 1$\sigma$ level of the rms spectral-line noise.}
        \label{fig:spectra}
\end{figure*}

\begin{figure}
    \centering
    \includegraphics[width=1.0\textwidth]{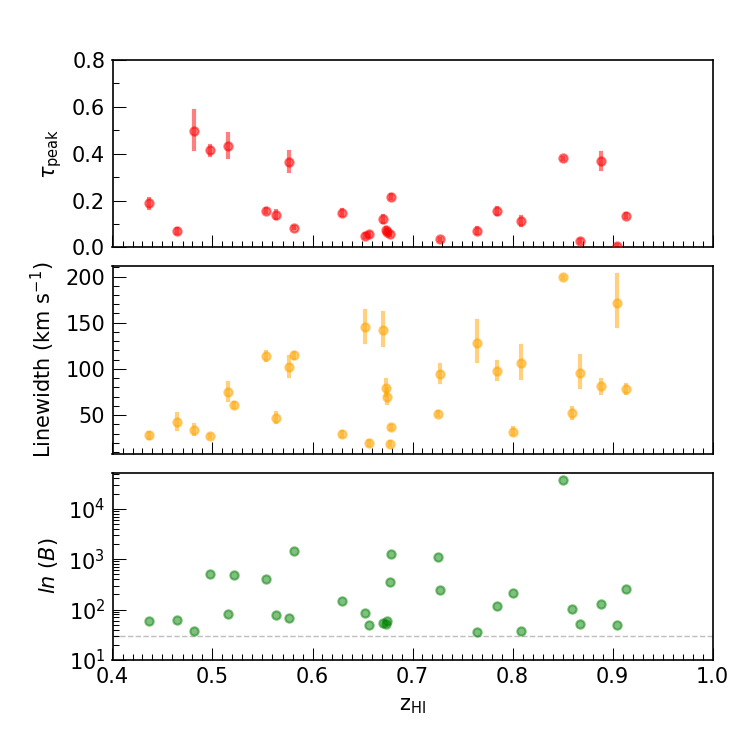}
   \caption{Peak optical depth (red) and linewidth (yellow) with 1$\sigma$ error as a function of redshift. ln~(B) (green) is presented in the bottom panel with the dotted line of ln (B) = 30, the cutoff used in the linefinder search. 
}
    \label{fig:linefinder_summary}
\end{figure}

Table \ref{tab:finder} in \ref{sec:linefinder} lists these lines, along with details of the continuum sources against which they were detected. We next extracted spectra in a smaller region around each line, and plotted the optical depth against velocity in the region around the line peak, as shown in Figure \ref{fig:spectra}. We then re-ran the line finder on these extracted spectra, after first co-adding the spectra of sources with two or more observations in Table \ref{tab:finder}. The final results for these co-added spectra are listed in Table \ref{tab:finder2} and plots of the individual spectra are shown in Figure~\ref{spectra1}. Fig~\ref{fig:linefinder_summary} shows the optical depth, linewidth, and logarithmic Bayes number as a function of redshift. The 33~detections, including the three known detections, were found across the entire redshift range covered by the survey with linewidths ranging from $\sim$20 to 218~km~s$^{-1}$.

Currently, genuine absorption lines are differentiated from spectral artefacts through visual inspection and investigation by the team. This is highly inefficient and not sustainable for the entire FLASH survey. As such, we are undertaking a separate project to investigate the use of tree-based machine learning methods to improve source-identification efficiency (Liu et al. in prep.) for future FLASH observations.

\subsection{Machine Learning classification of detected absorption lines} 
\label{sec:ML}
We expect the lines identified from the observations (listed in Table~\ref{tab:finder} in \ref{sec:linefinder}) to be a mixture of intervening and associated \hi absorption systems. As a preliminary way of distinguishing these in the absence of optical spectroscopy for many of our objects, we used an automated machine learning (ML) methodology \citep{curran2016b,curran2021} to classify each of the absorbers detected in this survey. 

\begin{figure*}[t!]
    \centering
    \includegraphics[width=1.0\textwidth]{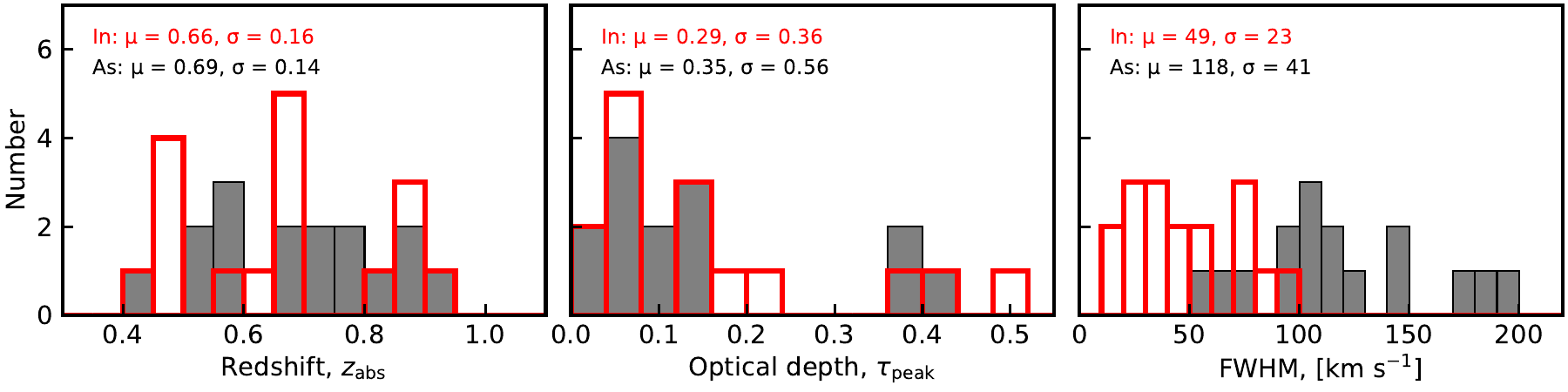}
   \caption{The distributions of the dominant model features for the putative intervening (In) and associated (As) absorbers. For the redshift and optical depth, t-tests give probabilities of p(t) = 0.64 and 0.70 of the means being the same, respectively. For the FWHM this is $p(t) = 1.68 \times 10^{-6}$.
}
    \label{fig:mltest_histo}
\end{figure*}

We use the \textit{logistic regression} (ML) classifier of \citet{curran2021}, which was the best performing of the other common classifiers and compile the results in the final column of Table \ref{tab:finder2}. To classify the lines, we used a training set of 116 out of a sample of 138 known absorbers from the literature, selected to have equal numbers of associated and intervening systems. 16 of the lines in Table \ref{tab:finder2} were classified as intervening and 14 as associated, with the line width (FWHM) being the main driver of the classification (Figure~\ref{fig:mltest_histo}). 

The broader associated profiles could be due to the additional nuclear component, hypothesised by unified schemes of active galactic nuclei \citep[AGN, e.g.][]{antonucci1993,urry1995,curran2016a} being preferably detected through the disk of the host \citep[cf.][]{curran2010}, whereas intervening absorption favours face-on systems, where the coverage of the background flux is maximised \citep{curran2016b}. 

In the Pilot Survey sample, we find an equal mix of associated and intervening systems. A potential limitation that we note here is that the training set used consists mainly of low-z systems and contain an optical pre-selection, which may not be fully representative of the sources that could be found from an untargeted survey such as FLASH. We will return to this question in a subsequent paper discussing the host galaxies of the Pilot Survey absorption systems. 
 
We also caution that a match between the redshift of a detected \hi line and the optical redshift of the galaxy in which the \hi absorption occurs does {\it not necessarily}\ imply 
that the line can be classified as associated. 
Figure 9 of \cite{allison20} shows one example of an intervening \hi detection in a galaxy (GAMA J224500.05-343031.7) where the radio and optical redshifts match. Here, the background radio source against which the line is detected is offset by 2.6\,arcsec from the centre of the galaxy in which the detected \hi is located.   
Although the background radio-source host is obscured by the foreground galaxy in this case, the astrometric offset shows that the line is intervening rather than associated. Section 4.1 of \cite{allison22} includes a more detailed discussion of the problem of distinguishing intervening and associated lines in an untargeted survey like FLASH. 

\begin{table*} 
\centering
\footnotesize
\tabcolsep 6pt
\caption{FLASHfinder results for co-added spectra; (1) ID; (2) Source name; (3) Number of spectra combined; (4) Field name; (5) Continuum flux density; (6) \hi redshift; (7) Peak optical depth and uncertainty; (8) Integrated optical depth and uncertainty; (9) Linewidth and uncertainty; (10) ln\,(B); (11) Machine Learning (ML) classification, using the methodology of \citet{curran2021}, where `As' indicates a probable associated line and `In' a probable intervening line, see Section~\ref{sec:ML}; (12) Notes (N = Individual Note in \ref{sec:notes}). The uncertainties listed in columns (7) to (9) represent the 68\% credible interval about the median. The individual spectra before combining are listed in Table~\ref{tab:finder} in \ref{sec:linefinder}.
} 
\label{tab:finder2}
\begin{tabular}{clclrrrrrrcc}
\hline													\multicolumn{1}{c}{ID}	&	\multicolumn{1}{c}{Source name}	&	\multicolumn{1}{c}{N$_{\rm spec}$}	& \multicolumn{1}{c}{FLASH}	& \multicolumn{1}{c}{S$_{\rm cont.}$} &	\multicolumn{1}{c}{$z_{\rm HI}$}	&	\multicolumn{1}{c}{$\tau_{\rm peak}$}  & \multicolumn{1}{c}{$\tau_{\rm int}$}	& \multicolumn{1}{c}{Linewidth}	& \multicolumn{1}{c}{ln\,(B)} & \multicolumn{1}{c}{ML} & \multicolumn{1}{c}{Note} \\
\multicolumn{1}{c}{}	&	\multicolumn{1}{c}{}		&	\multicolumn{1}{c}{ }	&	\multicolumn{1}{c}{field}	&	\multicolumn{1}{c}{(Jy)}	&	\multicolumn{1}{c}{}	&&& 	\multicolumn{1}{c}{(km s$^{-1}$)} &	&	\multicolumn{1}{c}{class} &	\\
\multicolumn{1}{c}{(1)}	&	\multicolumn{1}{c}{(2)}		&	\multicolumn{1}{c}{(3)}	&	\multicolumn{1}{c}{(4)}	&	\multicolumn{1}{c}{(5)}	&	\multicolumn{1}{c}{(6)}	&	\multicolumn{1}{c}{(7)}	&	\multicolumn{1}{c}{(8)}	&	\multicolumn{1}{c}{(9)}	&	\multicolumn{1}{c}{(10)} & \multicolumn{1}{c}{(11)} & \multicolumn{1}{c}{(12)}  \\
\hline

1	& PKS\,0011-023 & 1 & 525 & 0.694 & 0.6785 &	$0.216\substack{+0.007\\-0.006}$ &   $7.90\substack{+0.20\\-0.20}$ &  $36.6\substack{+1.2\\-1.2}$ 	&  1239.9 & In & N \\	
2 & NVSS\,J002331+010114 & 1 & 525	& 0.067	& 0.5159 & 
$0.431\substack{+0.052\\-0.047}$ &  $32.30\substack{+3.20\\-3.06}$ &  $74.8\substack{+9.6\\-8.4}$ & 80.4 & As 	& N \\	
3	&	MRC\,0023-482 &	2 &	123	& 0.379	& 0.6745 & $0.066\substack{+0.007\\-0.007}$ &	 $4.63\substack{+0.47\\-0.46}$ &  $70.0\substack{+7.8\\-7.1}$  & 58.7 & In & N \\		
4	& NVSS\,J014141-231511 & 2 & 306 &	0.137 &	0.6707 &	$0.122\substack{+0.014\\-0.014}$ & 	$17.42\substack{+1.91\\-1.92}$ &	 $142.2\substack{+18.7\\-16.1}$ &   55.2 &  As & - \\	
5	& NVSS\,J015516-251423 & 2 & 306 &	0.103 &	0.7251 &	$1.830\substack{+0.119\\-0.105}$ &  $94.53\substack{+4.31\\-4.08}$ &  $51.6\substack{+1.7\\-1.6}$ &  1121.3 & As & N \\
6	& PKS\,0253-259 & 2 & 308/309 &	0.705 &	0.6564	&	
$0.057\substack{+0.006\\-0.006}$ &  $1.12\substack{+0.13\\-0.12}$ & 	  $19.7\substack{+2.6\\-2.3}$ &  50.4 & In & N \\	
7	& PKS 0405-280 & 1 &	311	& 1.278 & 0.7280 & $0.035\substack{+0.002\\-0.002}$ &  $3.34\substack{+0.22\\-0.22}$ &  $94.6\substack{+9.4\\-9.1}$ & 242.4 & As & N \\
8 & SUMSS\,J045501-423858	& 1 & 170 &	0.288	& 0.6525 & $0.051\substack{+0.005\\-0.005}$ &  $7.38\substack{+0.70\\-0.67}$ & $145.7\substack{+17.5\\-16.4}$ &  86.2 & As & - \\
9 & NVSS J051806-245502 &	3 & 314	& 0.186 & 0.5538 &
$0.156\substack{+0.007\\-0.006}$ &  $17.85\substack{+0.68\\-0.70}$ &  $114.2\substack{+3.9\\-4.1}$ &	 411.4 & As & N \\
10	& MRC 0531-237 	& 2 & 314 &	1.746 &	0.8508 & $0.381\substack{+0.002\\-0.002}$ & $75.82\substack{+0.35\\-0.35}$  & $199.1\substack{+0.8\\-0.7}$ & 36724.9 & As& N \\
11 & NVSS\,J090331+010846 & 2 & G09B/546&	0.059 &	0.5218	&	$1.748\substack{+0.155\\-0.145}$ &  $106.74\substack{+6.93\\-6.68}$ &	 $61.1\substack{+2.4\\-2.4}$ & 483.8 & As & N \\	
12 & NVSS\,J090425+012015	& 3	&	G09B/546/547 & 0.087	& 0.8004 & $1.158\substack{+0.146\\-0.127}$ &  $37.42\substack{+2.79\\-2.74}$ &  $32.1\substack{+3.5\\-3.0}$ & 210.7 & In & N \\	
13 & NVSS\,J091256+030021 & 1 & G09B\_long	&	0.033	&	0.8592 	& $1.255\substack{+0.212\\-0.176}$	&  $64.93\substack{+7.09\\-6.64}$ &	$51.7\substack{+5.7\\-5.4}$ & 104.7 & In & N \\ 
14 &	PKS\,0917+18 & 2 &	719 &	1.791	& 0.9044 &	$0.008\substack{+0.001\\-0.001}$ &  $1.37\substack{+0.17\\-0.16}$ &  $171.5\substack{+30.4\\-25.4}$ & 48.7 & As & N \\		
15	&	NVSS\,J092012+161238 & 3 & 719 	&	0.171	&	0.4362	& 	$0.188\substack{+0.021\\-0.019}$ &  $5.28\substack{+0.55\\-0.53}$ & 	 $28.0\substack{+3.0\\-2.6}$ & 	 58.3 &  In & - \\	
16	& NVSS J094650-202044 & 1 & 377	&	0.550	&	0.9134	&	$0.135\substack{+0.007\\-0.007}$	&  $10.61\substack{+0.54\\-0.54}$	&  $78.2\substack{+4.2\\-4.0}$	& 252.4 & In & N \\
17	&	NVSS J100238-195917 & 1 & 378 &	0.057	&	0.4815	&	$0.499\substack{+0.082\\-0.077}$ &  $17.16\substack{+2.31\\-2.24}$ &  	$34.2\substack{+5.4\\-4.3}$ & 	 38.3 &  In & - \\
18	&	NVSS\,J113622+004850 &	3	&	G12A/553	&	0.154	&	0.5632 	&	$0.140\substack{+0.014\\-0.014}$ &  $6.57\substack{+0.61\\-0.61}$ & 		$46.9\substack{+5.3\\-4.5}$ & 76.8 & In & N \\
19 & NVSS J150506+022927 &	1 & 561	& 0.150	& 0.8085 & 
$0.112\substack{+0.017\\-0.015}$ &  $11.93\substack{+1.54\\-1.50}$  &  	$106.4\substack{+18.8\\-16.0}$ & 37.7 & As & N \\
20 & NVSS J170135-294918 & 1 &	287	&	0.403	&	0.6299	&	$0.149\substack{+0.011\\-0.011}$	&  $4.43\substack{+0.30\\-0.30}$ &  $29.8\substack{+2.1\\-2.1}$ & 146.5 & In & N \\	
21	&	PKS\,2007-245	& 1 & J2022-2507	&	1.827	&	0.6778	&	$0.056\substack{+0.003\\-0.003}$ &  $1.06\substack{+0.07\\-0.06}$ & 	 	$19.0\substack{+1.8\\-1.7}$ & 358.5 & In & N \\ 
22 & NVSS J205147+021740 &	2 & 575	&	0.111	&	0.8884	& $0.368\substack{+0.035\\-0.032}$	&  $29.91\substack{+2.35\\-2.35}$ &	 	$81.0\substack{+7.3\\-6.9}$ & 	  127.3 & In & - \\
23	& NVSS\,J215924-241752 & 1	& 351	&	0.806	&	0.8679	&	$0.028\substack{+0.004\\-0.004}$ &  $2.69\substack{+0.35\\-0.32}$ & 	 	$95.0\substack{+19.1\\-14.9}$ & 52.7 & In& N \\	
24 & NVSS J223317-015739 &	2	&	579	&	0.235 &	0.6734	&	$0.074\substack{+0.008\\-0.008}$	&  $5.91\substack{+0.62\\-0.63}$ & 	$79.0\substack{+9.4\\-7.8}$ &		51.9 &  In & - \\ 
25	& NVSS\,J223605-251919 &	1	& 352	&	0.215	& 0.4974	&	$0.414\substack{+0.021\\-0.019}$	&  $11.29\substack{+0.45\\-0.47}$ &  $27.2\substack{+1.3\\-1.3}$	&  515.0 & In & - \\	 
26	& NVSS\,J223620-222430 &	1 &	352	&	0.207 & 0.7846	& $0.157\substack{+0.013\\-0.012}$ &  $15.32\substack{+1.26\\-1.18}$ &		$97.7\substack{+9.6\\-8.3}$ & 	 115.9 & As & - \\			
27	& MRC\,2234-254	& 1	&	352	&	0.343	&	0.4641	& $0.071\substack{+0.010\\-0.009}$ &  $3.06\substack{+0.39\\-0.35}$ &  $42.9\substack{+8.1\\-7.6}$ &   60.9 & In & - \\	
28	& PKS\,2311-477 &	2 & 160	& 0.998 & 0.5811 & $0.084\substack{+0.002\\-0.002}$ &  $9.65\substack{+0.22\\-0.21}$ &  $114.6\substack{+2.8\\-2.7}$ & 1428.4 & As & N \\
29	& SUMSS\,J233432-585646 & 1 & 121 & 0.074 & 
0.5769 & $0.365\substack{+0.042\\-0.040}$ &  $37.38\substack{+3.60\\-3.77}$ &  $102.1\substack{+11.1\\-10.0}$ & 68.6 & As& N \\		
30	& NVSS J233702-015209 &	1 &	582	& 0.232	& 0.7645 &	
$0.072\substack{+0.010\\-0.010}$ &  $9.26\substack{+1.23\\-1.21}$ &  
$127.8\substack{+23.5\\-19.2}$ &  36.6 & As & - \\	
\hline
\end{tabular}
\end{table*}
\newpage

\subsection{Completeness tests}
\label{sec:completeness}

Next, we investigated the completeness of the FLASHfinder {\hi} 21-cm detections using the techniques described by \citet{allison20} and \citet{allison22}. The completeness is defined as the probability of recovering an absorption feature with a given peak signal-to-noise ratio (S/N; in an 18.5\,kHz channel) and line width (in $\rm km~s^{-1}$) from the spectrum, using the linefinder \citep[][]{allison12}.

This is estimated by selecting 100~spectra randomly from a sample of `test' spectra, and inducing a Gaussian absorption line with a specific peak signal-to-noise ratio (S/N) and line width in each spectrum at random locations. The linefinder is then run on these 100~spectra, after which we estimate the fraction of the recovered absorption lines. We made two separate runs. The first run used a detection limit of ln\,(B)~$> 20$, which is slightly deeper than the threshold described in the earlier sections, to test the depth of our threshold. The second run used a detection limit of ln\,(B)~$> 30$, consistent with our usual threshold. 

\begin{figure*}[ht!]
    \centering
    \includegraphics[width=0.475\textwidth]{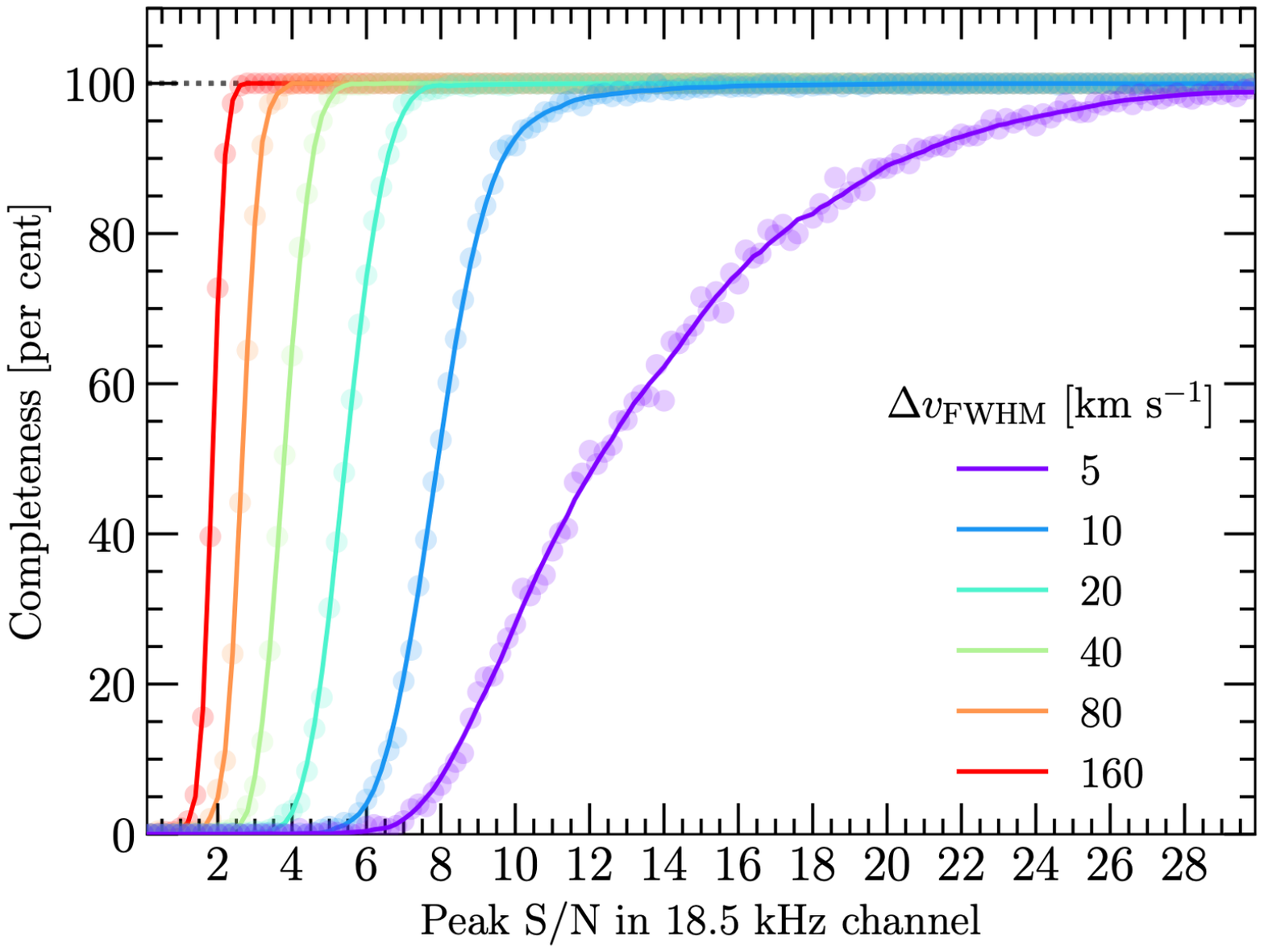}
    \includegraphics[width=0.5\textwidth]{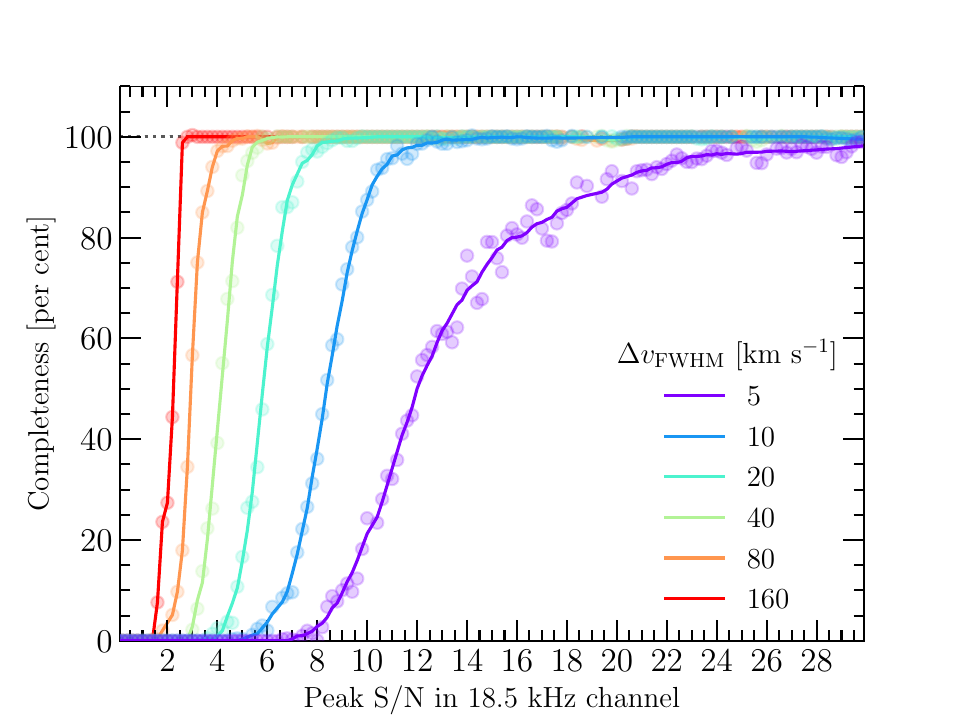}
    \caption{The completeness fractions plotted as a function of the peak signal-to-noise ratio for data from a good-quality observation (SB~34917). The colours represent a set of lines of different FWHMs, as indicated in the legend. The linefinder detection limit is set at ${\rm ln\,(B)}~> 20$ for the left-hand plot, and at ${\rm ln\,(B)}~> 30$ for the right-hand plot. Further details are provided in Section~8.6.}
    \label{fig:completeness}
\end{figure*}

We repeated this process over a wide range of combinations of S/N and line width. The completeness, i.e. the fraction of recovered lines, is plotted as a bivariate function for each peak S/N - line width combination (see e.g. Figure~\ref{fig:completeness}). The function is finally smoothed using a Savitzky-Golay filter \citep[][]{savitzky1964}. 

For these tests, we used spectra from two SBIDs, SB~34917 and SB~34581. The spectra of SB~34917 are known to be clean, without any significant RFI ducting features or ripples in the spectra. 
The completeness fractions with a detection limit of ln\,(B)~$ = 20$ show a relatively smaller scatter compared to those with ln\,(B)$=30$. Also, the curves are slightly shifted towards lower peak S/N values in the plot with ln\,(B)~$= 20$, implying that a larger fraction of weaker lines are detected.  

\begin{figure}[ht!]
    \centering
    \includegraphics[width=\textwidth]{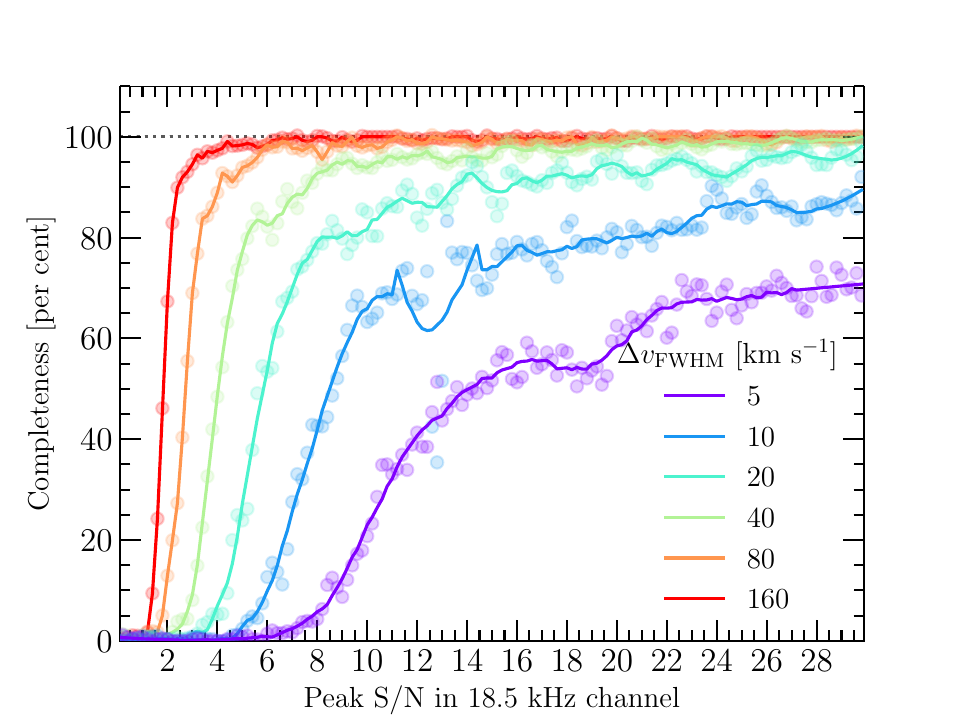}
    \caption{Similar to the left-hand plot of Figure \ref{fig:completeness}, but for an observation (SB~34681) where the data suffered from the spectral `ripples' discussed in Section \ref{sec:artefacts}.}
    \label{fig:completeness2}
\end{figure}

In Figure~\ref{fig:completeness2} we plot the completeness fractions for SB~34581, using a detection limit of ln\,(B)~$= 20$. The spectra in these observations are corrupted by ripples and ducting. The effects of these can be seen in the completeness plots; the detection fractions are not optimum for various FWHM and peak S/N combinations. 

We see from Figure~\ref{fig:completeness} that the completeness functions of SB~34917 are smooth, and reach $100\%$ detection fractions quickly for peak S/N $> 10$ and line widths $\rm > 10 \ km \ s^{-1}$. However, for SB~34581 in Figure \ref{fig:completeness2}, the functions are not smooth and the fractions do not reach $100\%$ for most combinations of the S / N peak and line width. This is because of the reduced efficacy of the automated line finder in identifying the induced Gaussian absorption feature among spectral ripples and ducting features.   

In our tests using the clean spectra (Figure~\ref{fig:completeness}, the right panel), we achieve 80 to 100\% completeness in recovering broader lines with widths above  80~km~s$^{-1}$ (red and orange lines), for a given peak S/N of 3. This corresponds to the significance of ln(B)~$= 30$ as detailed in \ref{sec:bayes}. Meanwhile, in order to reach a certain completeness level, narrower lines require higher peak S/N values. For instance, to achieve 50\%~completeness overall, peak S/N values of approximately 4, 6, 8, and 12 are needed for line widths of 40, 20, 10, and 5~km~s$^{-1}$, respectively (light green to purple lines). It is important to note that the channel resolution for each observation ranges from 8 to 19 km~s$^{-1}$.

\subsection{Comparison of repeat observations} 
\label{sec:repeat}

To test the robustness of the linefinder results, we compared data from FLASH Pilot Survey fields that were observed more than once. 
From the data in Table \ref{tab:finder} in \ref{sec:linefinder}, we identified 14 pairs of detected lines where the linefinder measurements could be compared. Ten of these were Pilot 2 re-ob\-ser\-va\-tions of lines 
detected in Pilot 1, and four were pairs of Pilot 2 observations.

\begin{figure*}[t]
\includegraphics[width=0.42\textwidth]{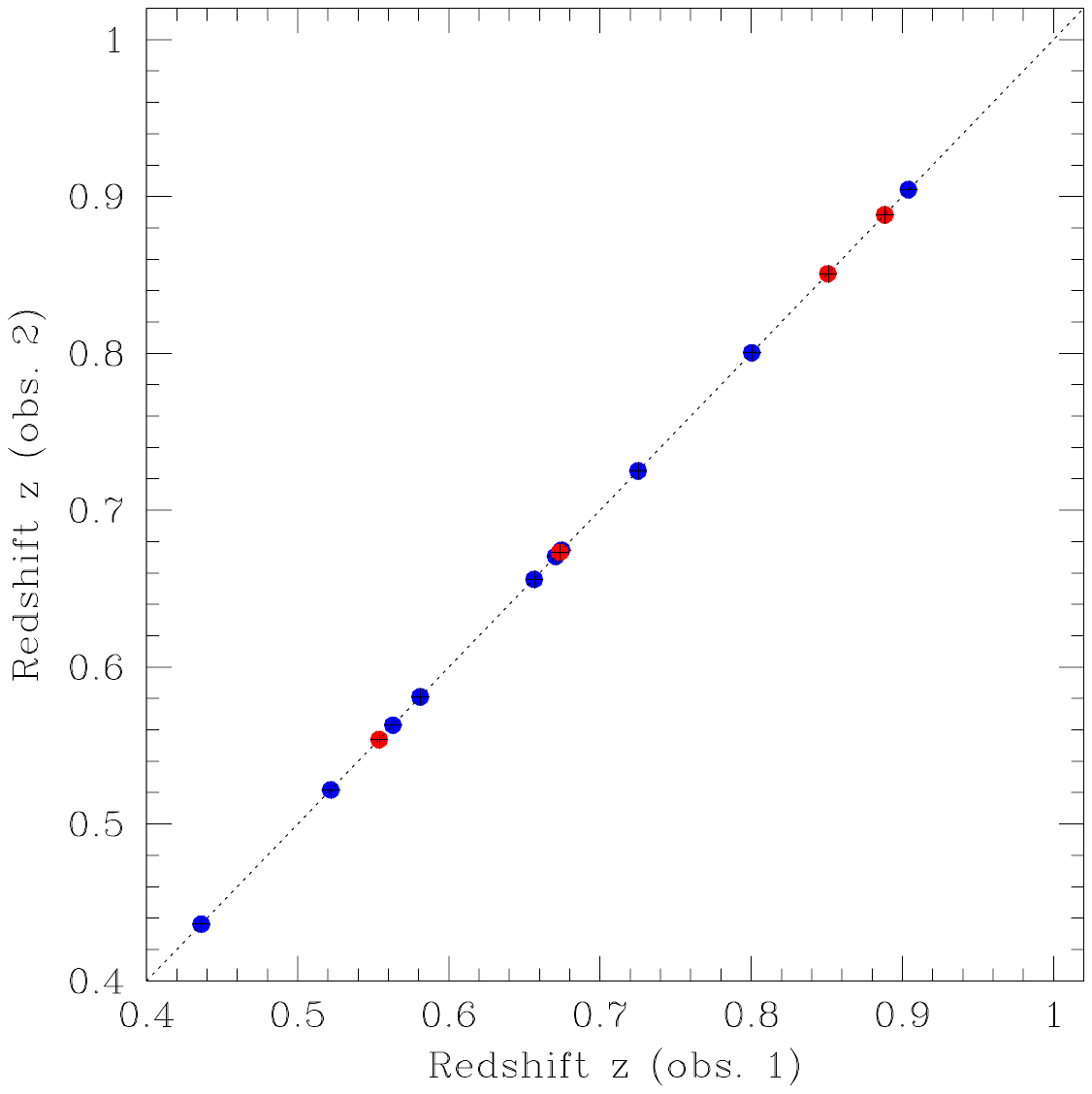}
\includegraphics[width=0.42\textwidth]{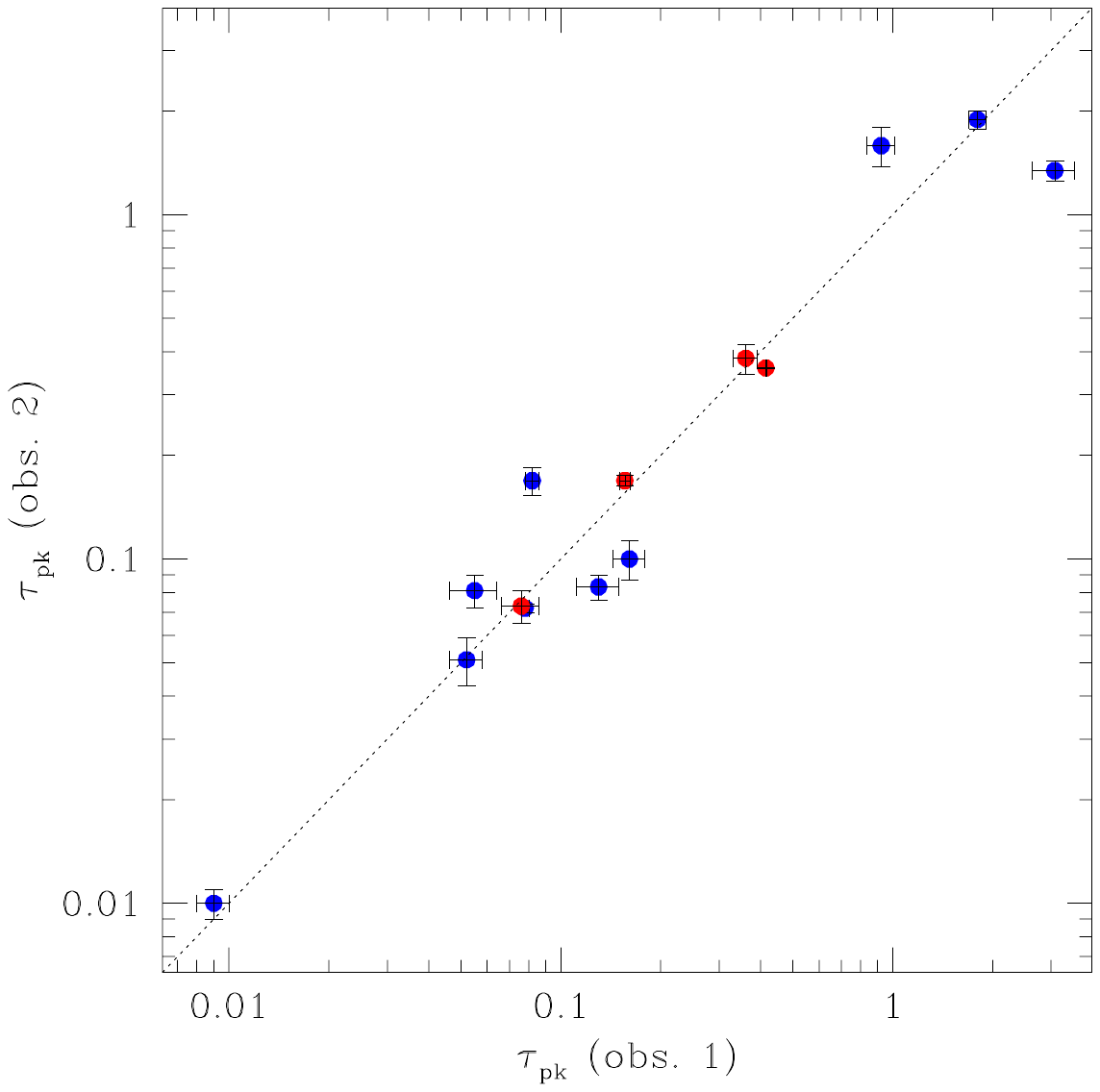}
\includegraphics[width=0.42\textwidth]{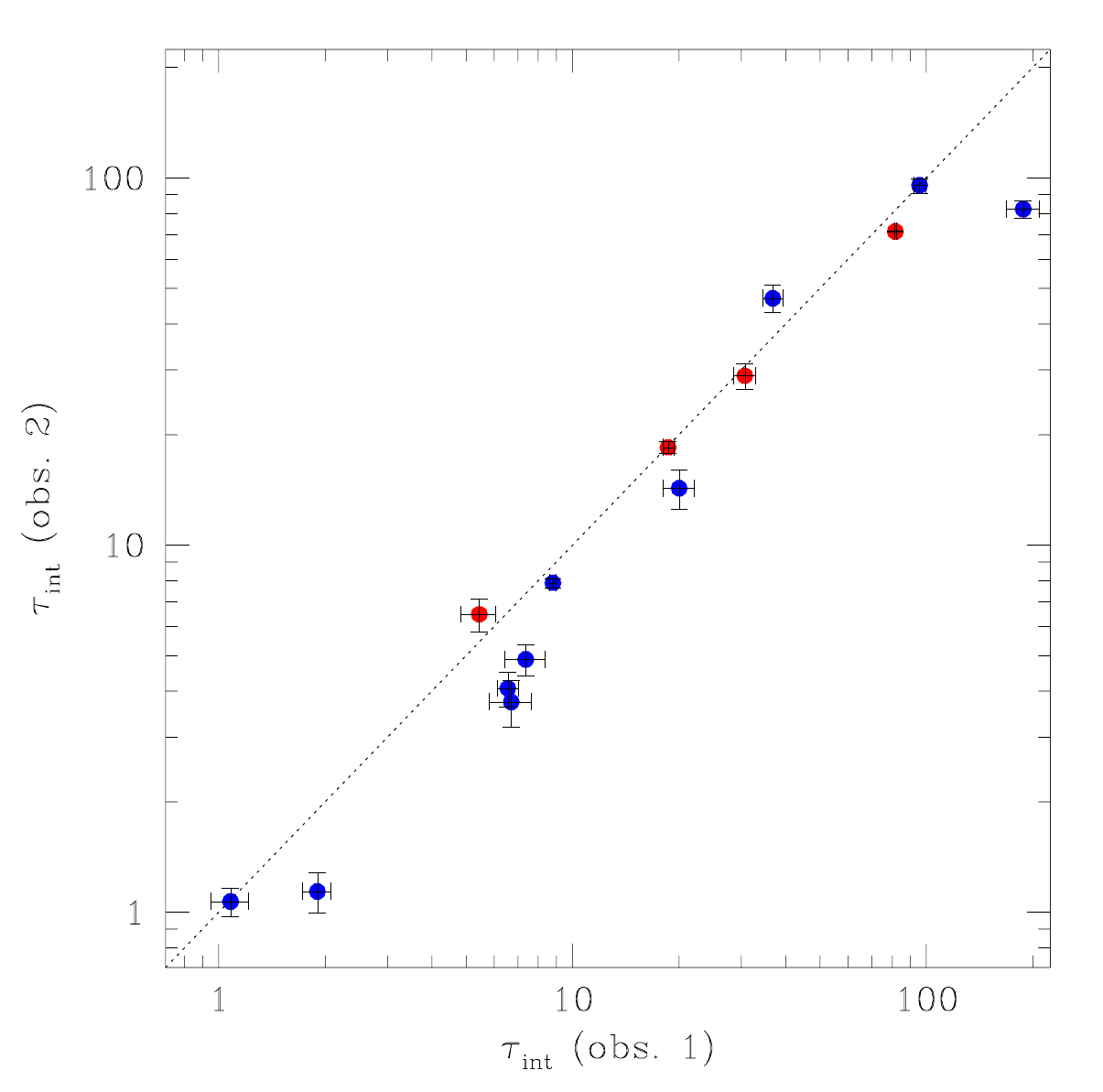}
\includegraphics[width=0.42\textwidth]{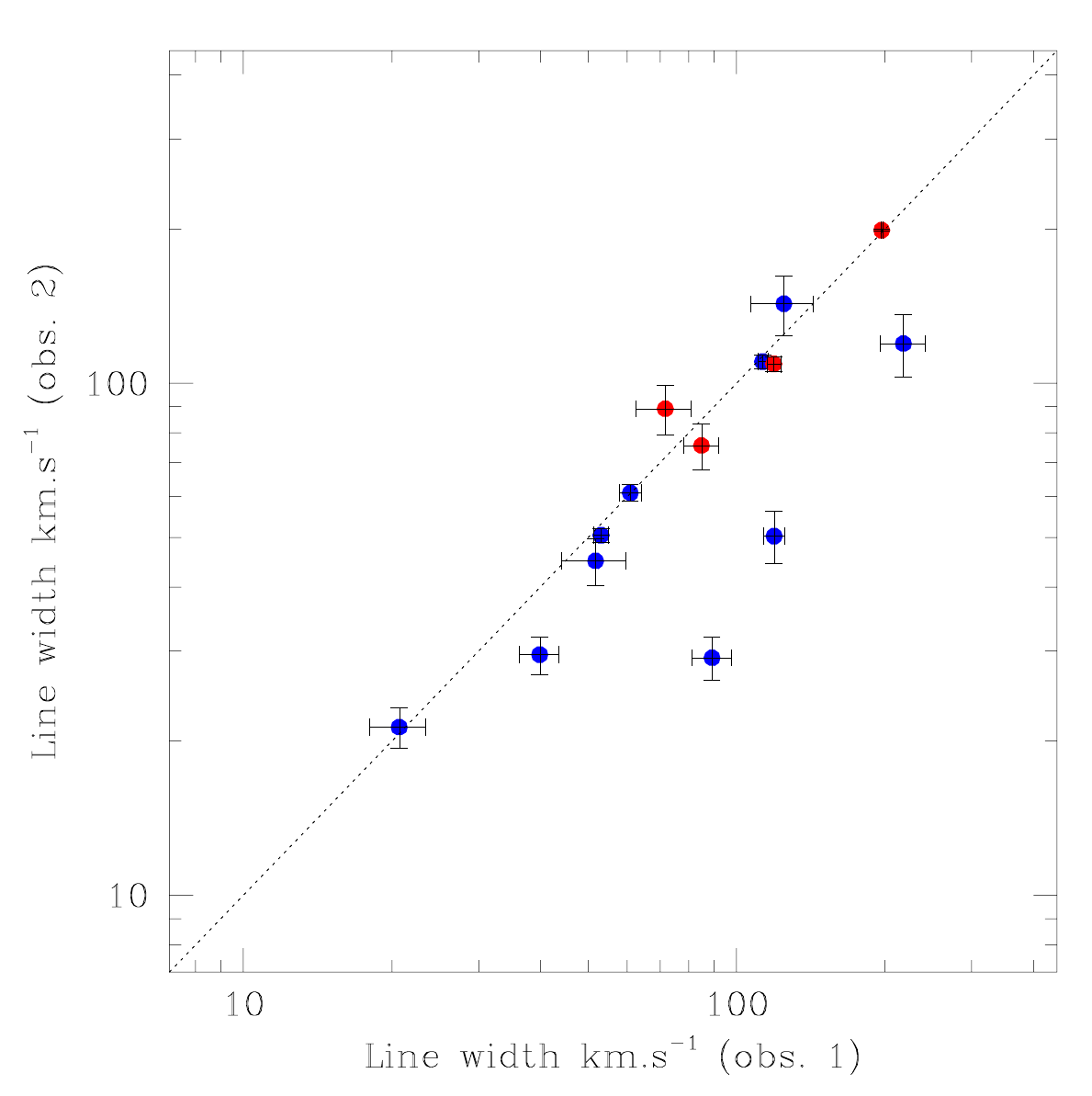}
\caption{Comparison of linefinder results for repeat observations. All values are for a single gaussian fit to the line profile, and the error bars show the uncertainties listed in the linefinder output. Blue points show a comparison between Pilot 1 and Pilot 2 data, and red points a comparison between two different Pilot 2 observations.}
\label{fig:repeats}
\end{figure*}

Figure \ref{fig:repeats}
shows the values of redshift $z$, peak optical depth $\tau_{\rm pk}$, integrated optical depth $\tau_{\rm int}$, and velocity width $\Delta{\rm V}$ measured for these pairs of lines. 
From this comparison, we find that redshift measurements are highly reproducible (differing by $<<0.1$\%). The peak and integrated optical depth measurements have typical uncertainties of around 25\% and the line width measurements (fitting a single component) have a typical uncertainty of around 20\%, though with a small number of outliers in each case.

\begin{figure*}[!h]
\begin{minipage}{0.48\linewidth}
\centering{\includegraphics[width=\linewidth]{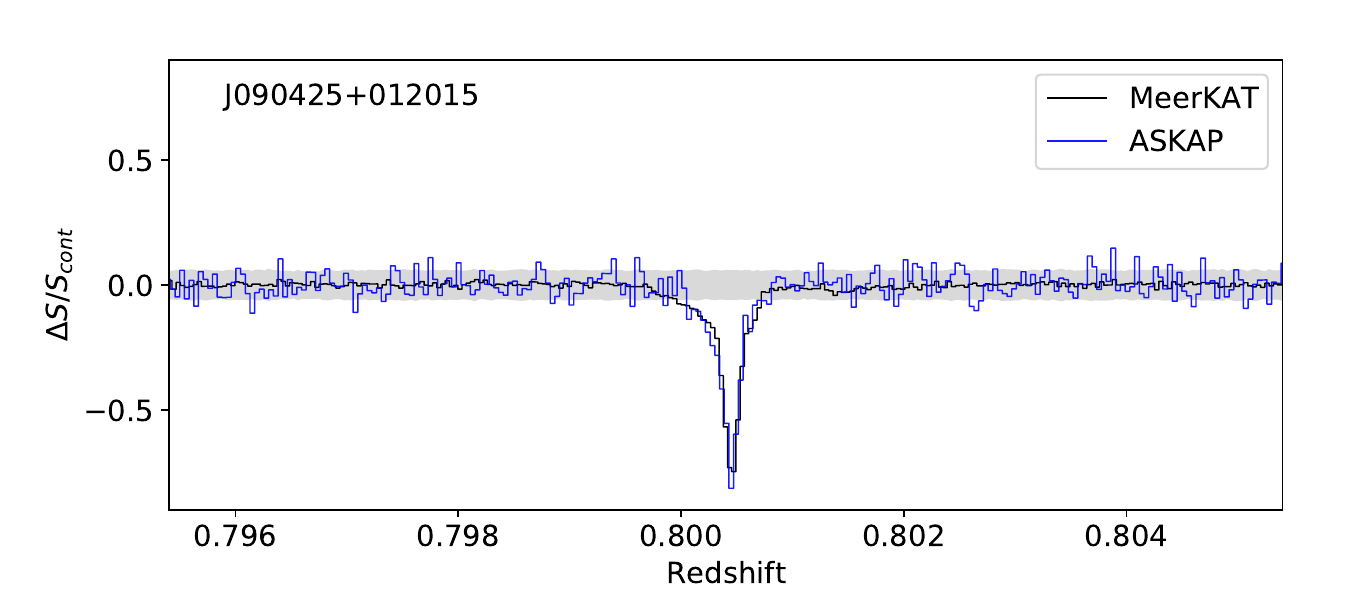}}
\end{minipage}
\begin{minipage}{0.48\linewidth}
\centering{\includegraphics[width=\linewidth]{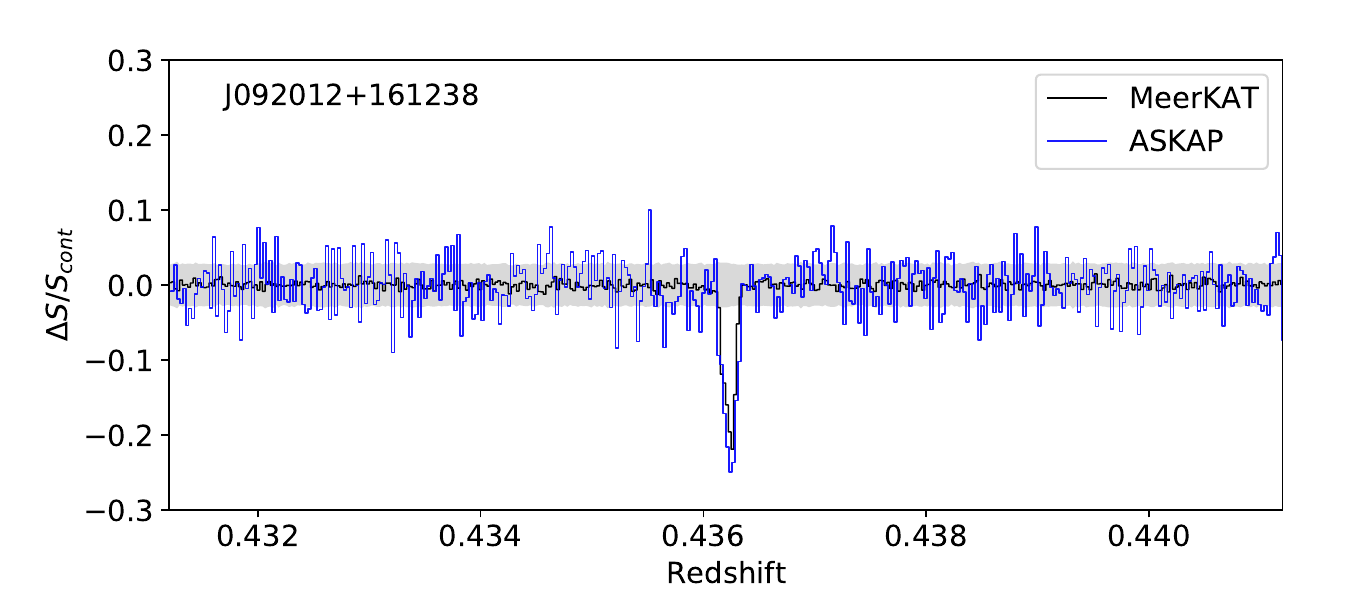}}
\end{minipage}
\caption{Spectra of FLASH detections observed with MeerKAT (black line) overlaid with the ASKAP spectrum (blue). The grey band denotes the 5-$\sigma$ noise level calculated from the MeerKAT spectra. 
\label{meerkatspectra}}
\end{figure*}

\subsection{Verification of FLASH data using MeerKAT}
\label{sec:meerkat}

A subset of the \hi absorption lines detected in the FLASH pilot surveys were observed with MeerKAT to better characterise the absorption lines with higher S/N. These observations were carried out under the program ID SCI-20210212-EM-01 and used the UHF band covering the frequency range of $580 - 1015$~MHz with the 32K mode correlator giving a similar spectral resolution of 16.6~kHz compared to ASKAP's 18.5~kHz. Each target was observed for a total integration time of 1~hr. A complete analysis of the MeerKAT observations will be the subject of future papers (Mahony et al. in prep., and Glowacki et al. in prep.), but we present initial results here for data verification purposes. 

Figure~\ref{meerkatspectra} shows the ASKAP-FLASH and MeerKAT spectra of two Pilot Survey detections (J090425+012015 at $z_{\rm HI}=0.80$ and J092012+161238 at $z_{\rm HI}=0.44$). With the exception of the higher S/N provided by the MeerKAT observations, the ASKAP and MeerKAT spectra match remarkably well and nicely demonstrate the spectral line capabilities and fidelity afforded by these two SKA precursor telescopes on radio-quiet sites. 

\begin{figure*}[!h]
\begin{center}
\includegraphics[width=0.49\columnwidth]{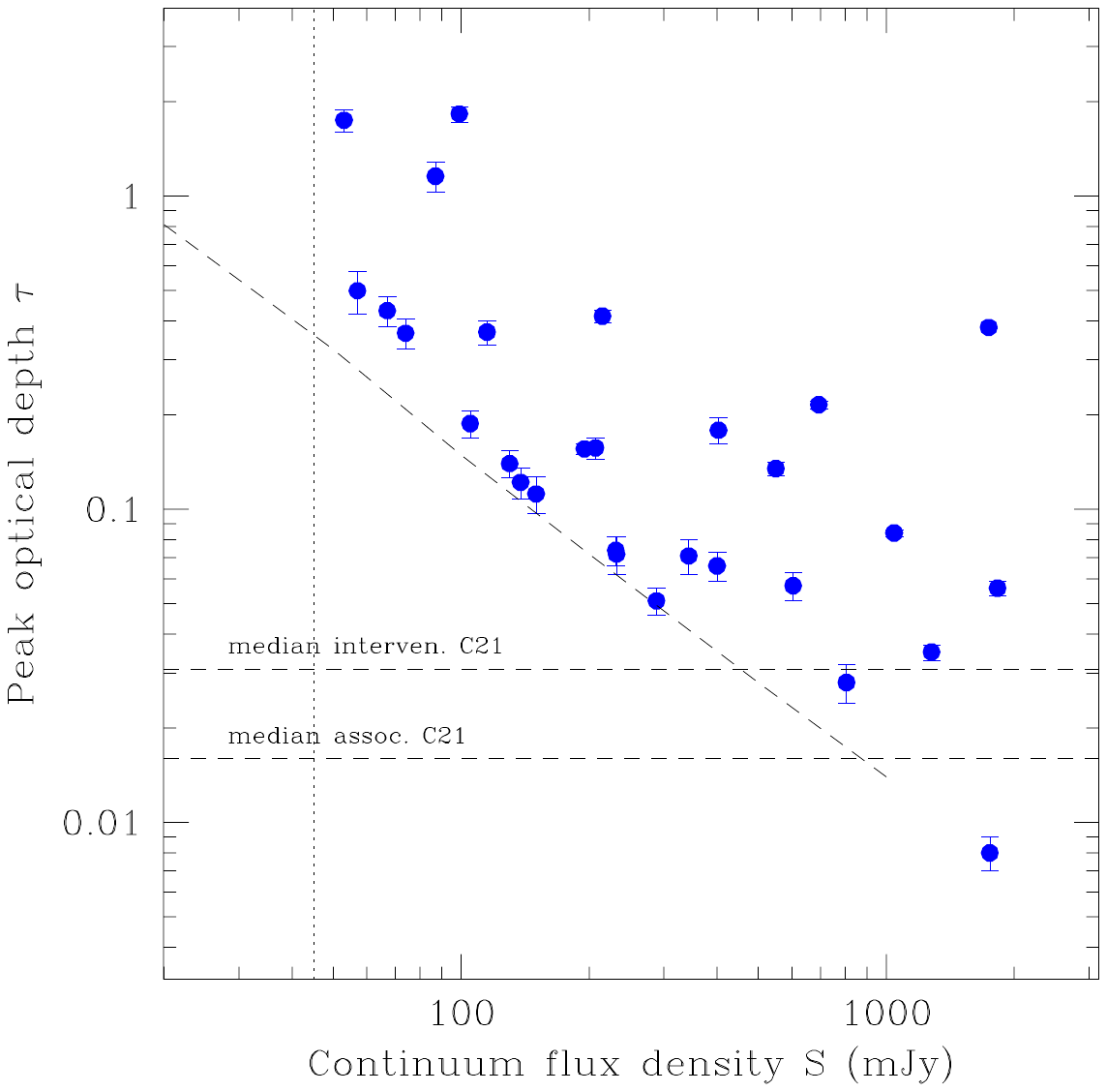}
\includegraphics[width=0.49\columnwidth]{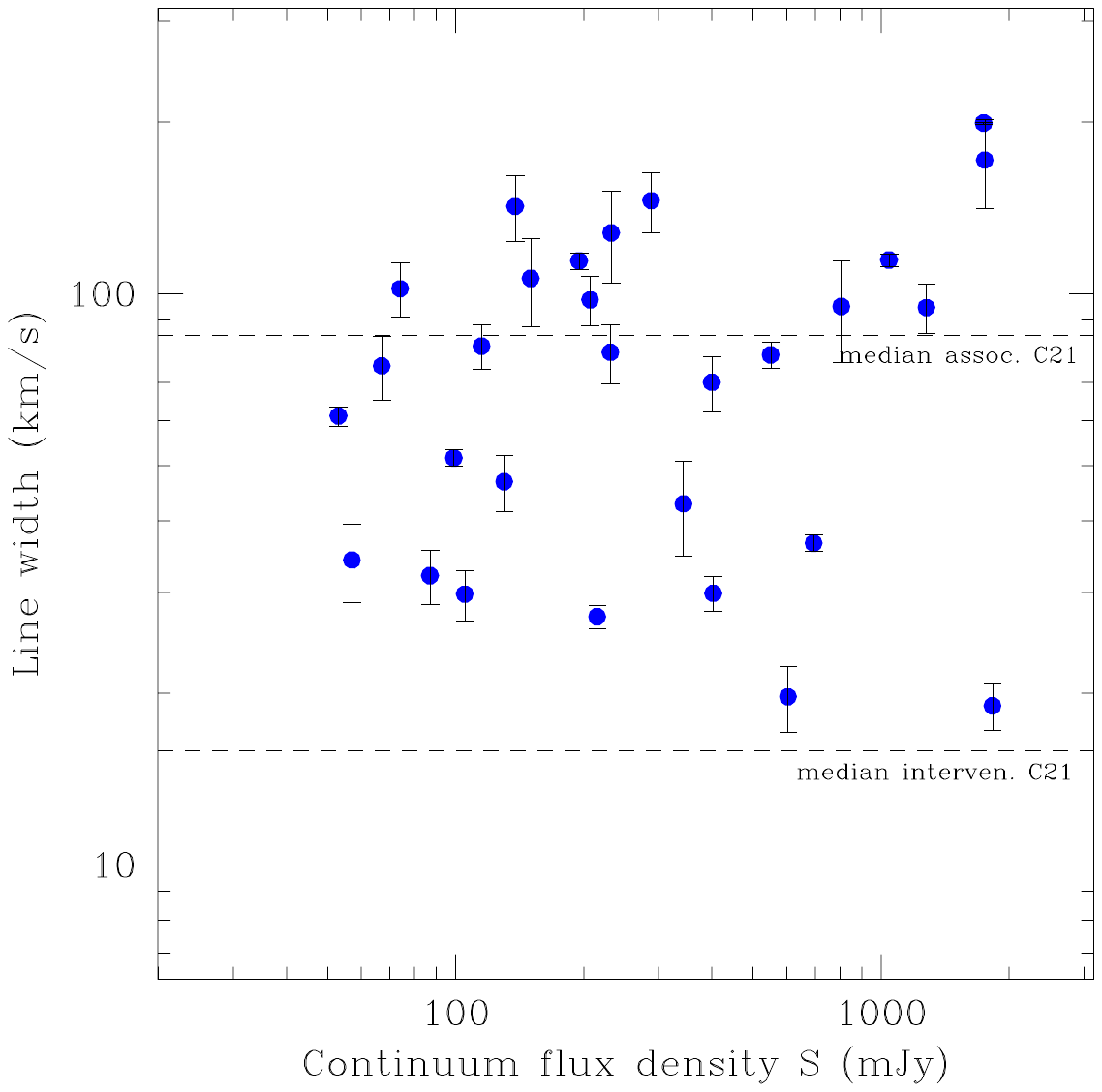}
\caption{Properties of the \hi lines detected against continuum sources of different flux densities. (left) The peak optical depth $\tau_{\rm pk}$, with a dashed line showing the expected sensitivity of the FLASH survey as calculated by \cite{allison12} and a vertical dotted line showing the 45\,mJy limit adopted for the \hi line search. (right) The line width in km s$^{-1}$, as listed in Table \ref{tab:finder2}. In both plots, horizontal dashed lines show the median values for the associated and intervening absorption lines in the literature compilation of \cite{curran2021}. 
}
\label{fig:props}
\end{center}
\end{figure*}

\subsection{General remarks on spectral-line data quality}
The quality of the spectral-line data taken at the radio-quiet ASKAP site can be extremely good (see Figure \ref{fig:spectrum}), with the FLASH band from 712--1000\,MHz generally free from terrestrial RFI. 

Overall though, the quality of the pilot survey data was variable and most of the processed data released in CASDA are classified as being of `uncertain' quality. This is perhaps to be expected from early observations with a new telescope coupled with a sophisticated data processing pipeline.  As discussed earlier, data from the first pilot survey suffered from correlator dropouts and needed post-processing because of an incorrect implementation of the beam-forming intervals, while the data from the second pilot survey was affected by several spectral-line artefacts. These problems have now been overcome, and should not affect future FLASH observations. 

\section{Discussion}
\label{sec:discussion}

\subsection{\hi absorption-line systems at intermediate redshift}
\label{sec:abs_discuss}
The 30 new \hi lines listed in Table \ref{tab:finder2} have almost doubled the number of known \hi absorption-line systems in the FLASH redshift range. Previously, \cite{curran2021} listed 44 known \hi absorbers at redshift $0.42<z<1$ across the whole sky, almost all detected in targeted observations of optically-selected quasars. Our untargeted absorption-line survey allows us to detect \hi lines against the full range of radio-source populations, and in the next section we discuss the radio continuum properties of the sources listed in Table \ref{tab:finder2}. 

\subsection{Radio continuum properties of \hi\-detected sources}
\label{sec:radioSED}

Spectral energy distribution (SED) modelling in the radio regime can provide insights into the nature of the AGN in a regime where its synchrotron emission typically dominates.  
All the continuum sources against which we detected \hi absorption were unresolved in the highest-resolution (12-15 arcsec beam) FLASH continuum images, and we can learn more about 
these compact sources by fitting the broadband radio continuum using flux density measurements from the literature. The fitting process assumes minimal broadband variability between epochs, but this appears reasonable given the low variability of most radio AGN at frequencies below a few GHz \citep[e.g.][]{bannister2011,mooley2016} 

Our SED construction and modelling was performed using \textsc{RadioSED}, a Bayesian inference package designed for use with broadband radio continuum data, and which also automatically compiles published flux densities from large area surveys \citep{kerrison2024a}\footnote{
For three sources, NVSS J002331+010114, NVSS J090425+012015, and NVSS J090331+010846, additional LOFAR data were acquired via private communication to better constrain the source SEDs in the megahertz regime. These data are drawn from the forthcoming LoTSS survey \citep{shimwell2022}.}.
This inference procedure can capture various degrees of spectral curvature allowing for the identification of peaked spectrum sources, thought to be radio AGN within a few thousand years of triggering \citep{ODea2021}.

Figure \ref{spectra1} shows the radio SEDs of the sources with \hi detections plotted alongside the \hi spectra. As discussed by \cite{kerrison2024b}, there is a clear excess of peaked-spectrum (PS) SEDs among the Pilot Survey detections, with 20/30 (67\%) showing a broadband peak. Previous studies of the radio population as a whole found a peaked fraction of 5-20\% \citep{ODea1998, callingham2017}, so it is remarkable to see so many PS sources in our untargeted search for \hi absorption. To quantify this further, if we assume that 20\% of all radio galaxies are PS sources \citep{ODea1998}, then the binomial probability of finding 21 or more PS sources in our \hi-absorbed sample is only $\text{Pr}(\textit{N}_{\text{PS}} \ge 21) = 3 \times 10^{-8}$. This number is even lower if we assume the 5\% population fraction from \cite{callingham2017}. More recent targeted searches have reported numerous \hi detections in PS sources at high redshift \citep[e.g.][and references therein]{aditya18a,glowacki19,aditya2024}.

In this untargeted survey, we find that the SED radio properties of the continuum sources against which we detect \hi absorption are not typical of the overall radio AGN population. To say more about this emerging trend requires larger statistical samples of absorbers, which the full FLASH survey will provide.

\subsection{Properties of the detected \hi\ lines}
 The \hi lines detected in the pilot survey span a wide redshift range from $z=0.43$
to $z=0.91$, and have velocity widths between 19 and 199 km\,s$^{-1}$. Figure \ref{fig:props} shows the distribution of optical depth and line width for the detected lines. 

The left-hand plot in Figure \ref{fig:props} shows that the limiting peak optical depth that can be detected increases at lower flux densities, as expected for an absorption-line search. The lower bound for the detected lines lies close to the diagonal line in this plot, which shows the expected FLASH sensitivity limit calculated by \cite{allison12}. In other words, we are reaching the optical-depth sensitivity expected when the FLASH survey was first proposed more than a decade ago. 

A second point to note from this plot is the high peak optical depth of most of our detected lines. 
Our sample contains three lines with a peak optical depth $\tau_{\rm pk}>1.0$, larger than any of the 136 lines in the \cite{curran2021} compilation of literature data. The median value of $\tau_{\rm pk}=0.148$ for our detected lines is significantly higher than the median $\tau_{\rm pk}=0.030$ for the literature sample compiled by \cite{curran2021}. 
This is consistent with the prediction by \cite{allison22} that FLASH would detect a population of lines with very high \hi optical depth. 

From the right-hand plot, we can see that the detections span a wide range in line width but there are very few lines narrower than about 25\,km\,s$^{-1}$. There is no significant difference in the median line width for sources across a wide range in continuum flux density. 

\subsection{The \hi\ detection rate in the pilot surveys} 

In the first pilot survey we detected 17 new lines\footnote{Excluding the line in NVSS\,J091256+030021, which was only detected in a 6-hr observation. } from 37 fields, a rate of 0.46$\pm0.11$ lines per field. In the second pilot survey the rate was lower, with 12 detections from the 38 fields not previously observed - a rate of 0.32$\pm0.09$ lines per field. 
The lower detection rate for the second pilot survey may be due to the large number of spectra with spectral ripples that reduce the completeness of the linefinder search (see Figure \ref{fig:completeness2}). 
We therefore adopt the Pilot 1 value of 0.5 lines per field as an indicative \hi detection rate for the pilot surveys. 

In Table 3 of their paper, \cite{allison22} estimate the total number of \hi detections expected from a FLASH survey covering 903 ASKAP fields. They also caution that their estimates are highly uncertain - particularly for the population of associated \hi lines. For an assumed \hi spin temperature T$_{\rm S}=300$\,K, \cite{allison21} estimated a detection rate of 3.1 lines per ASKAP field, with 70\% of these being associated lines and 30\% intervening lines. 

Our observed detection rate of $\sim0.5$ lines 
per ASKAP field is significantly lower than the \cite{allison22} estimate. There are at least two plausible reasons for this: 

\begin{enumerate}
    \item
{\bf Associated \hi absorption lines at redshift $0.4<z<1$ are less common than expected.} Recent work by \cite{su2022} and \cite{aditya2024} has shown that the detection rate of associated \hi absorption lines 
in radio AGN within the FLASH redshift range is only 2-3\%, significantly lower than the value of 10\% assumed by \cite{allison22} but consistent with the low detection rate (0 lines from 29 sources) found by \cite{murthy2022}
for radio galaxies at $0.7<z<1$. If we adjust the \cite{allison22} values to take into account the lower expected rate of associated lines, then the predicted \hi detection rate drops to $\sim1.5$ lines per FLASH field. 
\item 
{\bf Some narrow absorption lines may be missed in our survey.} 
The results shown in section \ref{sec:completeness}
imply that our sample will be less complete for narrow lines (FWHM $<$ 20\,km\,s$^{-1}$) than for broad lines across a wide range in peak S/N. How this affects the data shown in the right-hand plot of Figure \ref{fig:completeness} is harder to assess, since there is no simple relationship between peak S/N and continuum flux density 
(a weak line seen against a bright source and a
stronger line seen against a fainter source may have similar peak S/N values), but it is notable that we have so far detected vary few lines with velocity widths below 20-30\,km\,s$^{-1}$.
Such lines also lie close to the spectral resolution limit of the ASKAP data, and may be difficult to distinguish from the single-channel `glitches' seen in some pilot survey spectra. 
\end{enumerate}

\begin{figure}[h]
\begin{center}
\includegraphics[width=\linewidth]{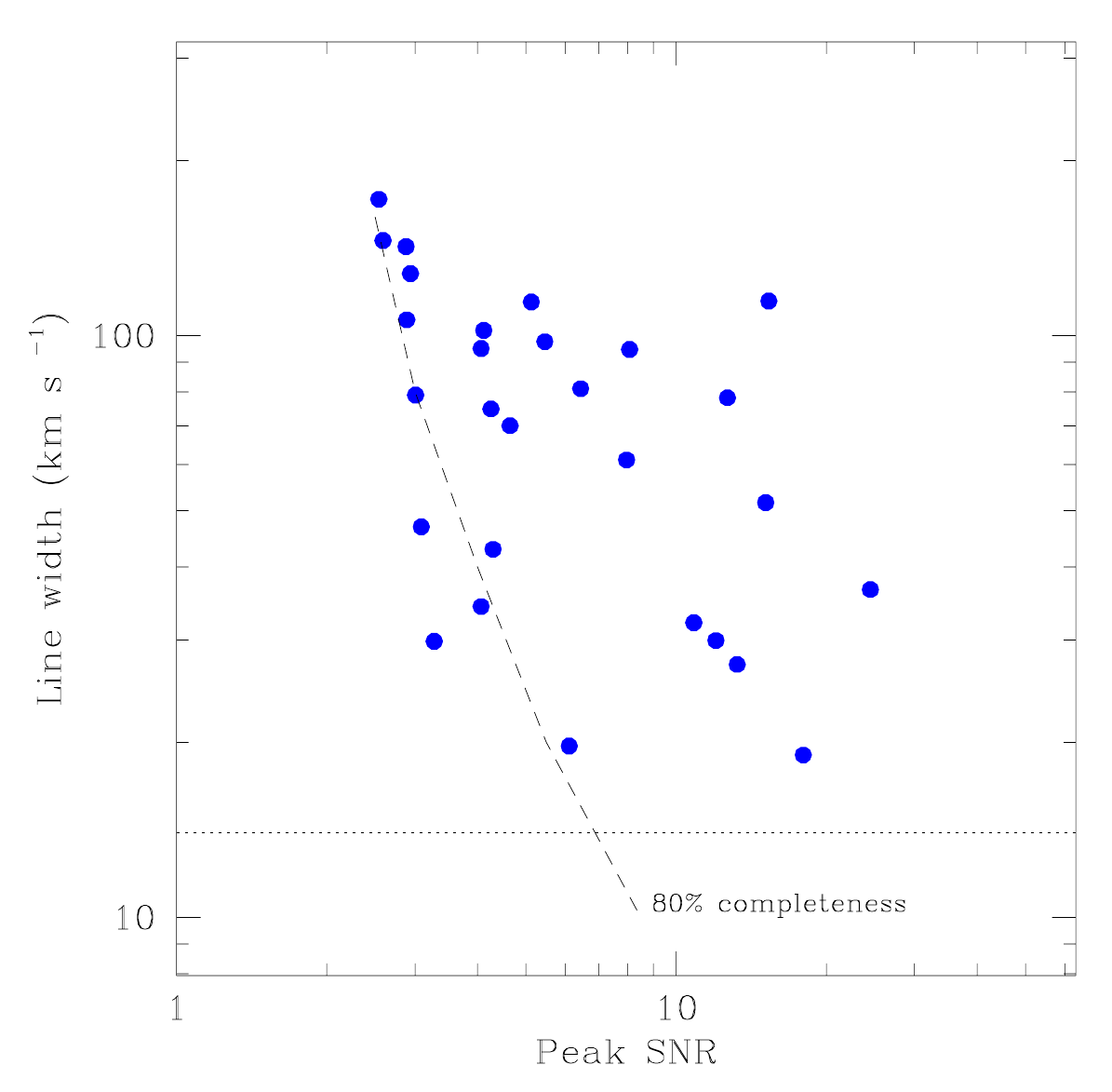}
\caption{Plot of velocity width against peak SNR for the detected lines. The dashed line shows the level at which completeness tests imply that at least 80\% of lines should be detected in a good-quality FLASH spectrum.}
\label{fig:compx}
\end{center}
\end{figure}

Figure \ref{fig:compx} shows the range of peak SNR and line width observed in the pilot survey detected lines, along with the 80\% completeness limit derived from the data shown in Figure \ref{fig:completeness}. While we do detect some lines down to the expected completeness limit, the variable quality of the Pilot Survey observations means that that not all fields will reach this completeness limit. 
A more detailed analysis of the observed detection rate will be carried out in a future paper using a larger, higher-quality dataset from the full FLASH survey. 

\subsection{Future work}
\label{sec:futurework}

Following the two phases of Pilot observations, we began the full five-year FLASH survey in late 2022. The full FLASH survey will cover 600 fields ASKAP fields (a total of 24,000 deg$^2$) rather than the 903 fields originally planned, with the allocated 1200~hours of observing time. The coverage will span declinations from $-90$ to $+18$ degrees at Galactic latitudes $|b|\geq8.5$~deg. Our pilot survey results suggest that the full FLASH survey will detect between 300 and 600 new absorption lines, with around 60\% of these being associated lines. 

The next paper in this series will identify the host galaxies of the \hi lines discovered in the pilot surveys, and will also discuss how best to determine whether an individual line is associated or intervening. 

\section{Summary}
\label{sec:summary}

The FLASH pilot surveys are pioneering studies that search for \hi absorption lines in and around distant galaxies without any optical preselection of targets. We have identified a sample of 30 new robustly-detected absorption lines, several of which have higher optical depths than any previously-known line. This work has led us to make new predictions about the detection rate for the full FLASH survey, highlighting its potential impact on our understanding of \hi absorption systems.

We successfully tested and demonstrated the automatic data pipeline for the survey and made public immediately after validation, with no proprietary period. We also used a powerful Bayesian line finder to identify and parameterise absorption lines in a consistent and uniform way. Observations of a subset of these lines with the MeerKAT telescope show a high degree of consistency, implying that both telescopes produce high-quality, reproducible spectral-line data. 

The three main aims of the FLASH Pilot Surveys have all been achieved. 

Firstly, we were able to test and demonstrate an 
end-to-end data pipeline for the FLASH survey and make line detections that are robust, reproducible, and have a statistical significance attached. We also identified several improvements that could be made in future, including masking out spectral-line artefacts before running the line finder, fitting multiple components to the absorption lines, and further investigating the completeness and detectability of faint and narrow spectral lines. 

Our second goal was to determine the \hi detection rate for associated absorption lines at $0.4<z<1$. This was done in two companion papers that were able to use optical redshifts as a prior for the line search \citep{su2022,aditya2024}. In the redshift range covered by FLASH, the detection rate for associated \hi lines is 2--3\% for the source population as a whole, rising to around 23\% for compact radio sources less than about 1\,arcsec in size. 

Finally, we have made new predictions of the detection rate for the full FLASH survey, showing that FLASH is likely to detect 300--600 new \hi absorption systems at intermediate redshifts ($0.4 < z < 1.0$). 
This represents an order of magnitude more detections than currently available in this redshift range, and should enable us to constrain the average spin temperature and/or \hi mass density at intermediate redshifts. 

\begin{acknowledgement}

This scientific work uses data obtained from Inyarrimanha Ilgari Bundara / the Murchison Radio-astronomy Observatory. We acknowledge the Wajarri Yamaji People as the Traditional Owners and native title holders of the Observatory site. CSIRO’s ASKAP radio telescope is part of the Australia Telescope National Facility (https://ror.org/05qajvd42). Operation of ASKAP is funded by the Australian Government with support from the National Collaborative Research Infrastructure Strategy. ASKAP uses the resources of the Pawsey Supercomputing Research Centre. Establishment of ASKAP, Inyarrimanha Ilgari Bundara, the CSIRO Murchison Radio-astronomy Observatory and the Pawsey Supercomputing Research Centre are initiatives of the Australian Government, with support from the Government of Western Australia and the Science and Industry Endowment Fund.

This research was supported by the Australian Research Council Centre of Excellence for All Sky Astrophysics in 3 Dimensions (ASTRO 3D) through project no.\,CE170100013. 

This work was supported by the Global-LAMP Program of the National Research Foundation of Korea (NRF) grant funded by the Ministry of Education (No. RS-2023-00301976). This work was supported by the National Research Foundation of Korea (NRF) grant funded by the Korea government (MSIT) (RS-2025-00516062).

RZ acknowledges support from the China Postdoctoral Science Foundation (Grant No. 2024M752979).

ACE acknowledges support from STFC grant ST/P00541/1. KG is supported by the Australian Research Council through the Discovery Early Career Researcher Award (DECRA) Fellowship (project number DE220100766) funded by the Australian Government. MG is supported by the Australian Government through the Australian Research Council’s Discovery Projects funding scheme (DP210102103), and through UK STFC Grant ST/Y001117/1. MG acknowledges support from the Inter-University Institute for Data Intensive Astronomy (IDIA). IDIA is a partnership of the University of Cape Town, the University of Pretoria and the University of the Western Cape. For the purpose of open access, the author has applied a Creative Commons Attribution (CC BY) licence to any Author Accepted Manuscript version arising from this submission.

The MeerKAT telescope is operated by the South African Radio Astronomy Observatory, which is a facility of the National Research Foundation, an agency of the Department of Science and Innovation.

We thank the anonymous referee for the valuable comments and suggestions that allowed us to improve our manuscript. We acknowledge technical support from the Australian SKA Regional Centre (AusSRC). We express our gratitude to Dr. Jonghwan Rhee and Prof. Nissim Kanekar for their insightful suggestions.

\end{acknowledgement}

\bibliography{ref}

\clearpage
\appendix
\label{sec:appendix}

\setcounter{table}{0}
\renewcommand{\thetable}{A\arabic{table}}
\setcounter{figure}{0}
\renewcommand{\thefigure}{A\arabic{figure}}

\section{A note on Bayesian ln\,(B) values and statistical significance}\label{note:lnB}
\label{sec:bayes}

In this paper we have used a Bayesian evidence parameter ln\,(B) to identify absorption lines in ASKAP spectra. Here, we briefly discuss the relationship between ln\,(B) and the statistical n\,$\sigma$ uncertainty often used by the wider astronomy community to assess the significance of a result. 

The calibration from frequentist p-values, or  `sigma significance levels' to the Bayes factor has been discussed in detail by \cite{sellke2001} and \cite{trotta2008}. \cite{trotta2008} also provides an empirical scale for evaluating the strength of evidence when comparing two models, with ln\,(B) = 5 corresponding to `strong evidence' (a 3.6\,$\sigma$ result, with p $=0.0003$) and ln\,(B) = 11 to a 5$\sigma$ value. \cite{benneke2013} apply Bayesian methods to a specific problem in exoplanet atmospheres and provide an expanded version of the \cite{trotta2008} Bayes to frequentist `translation table'.

In the context of our FLASH absorption-line finding problem, the Bayes value is the odds ratio of obtaining the observed data (assuming the absorption-line model is correct) versus just gaussian noise. The p-value is the probability of obtaining data features that are at least as extreme as those observed assuming just Gaussian noise. In both cases no consideration is made for non-Gaussian noise or artefacts - which we know are present in the ASKAP Pilot Survey data (see Figure~\ref{fig:artefacts}).

We therefore stress that all statistics quoted in this paper (whether B, p-values and/or sigmas) are calculated with respect to the assumed Gaussian noise and not any other non-Gaussian noise features. 

To examine the empirical relationship between ln\,(B) and the signal-to-noise ratio (SNR) in our current set of line detections, we calculated an approximate peak SNR for each FLASH detection using the single spectrum in which the line was first detected. We then calculated the size of the `dip’ in the continuum at the line peak from the continuum flux density and the observed peak optical depth $\tau_{\rm pk}$ and divided this by an assumed noise of 5.5\,mJy per beam in a single channel. 

\begin{figure}
\includegraphics[width=0.94\columnwidth]{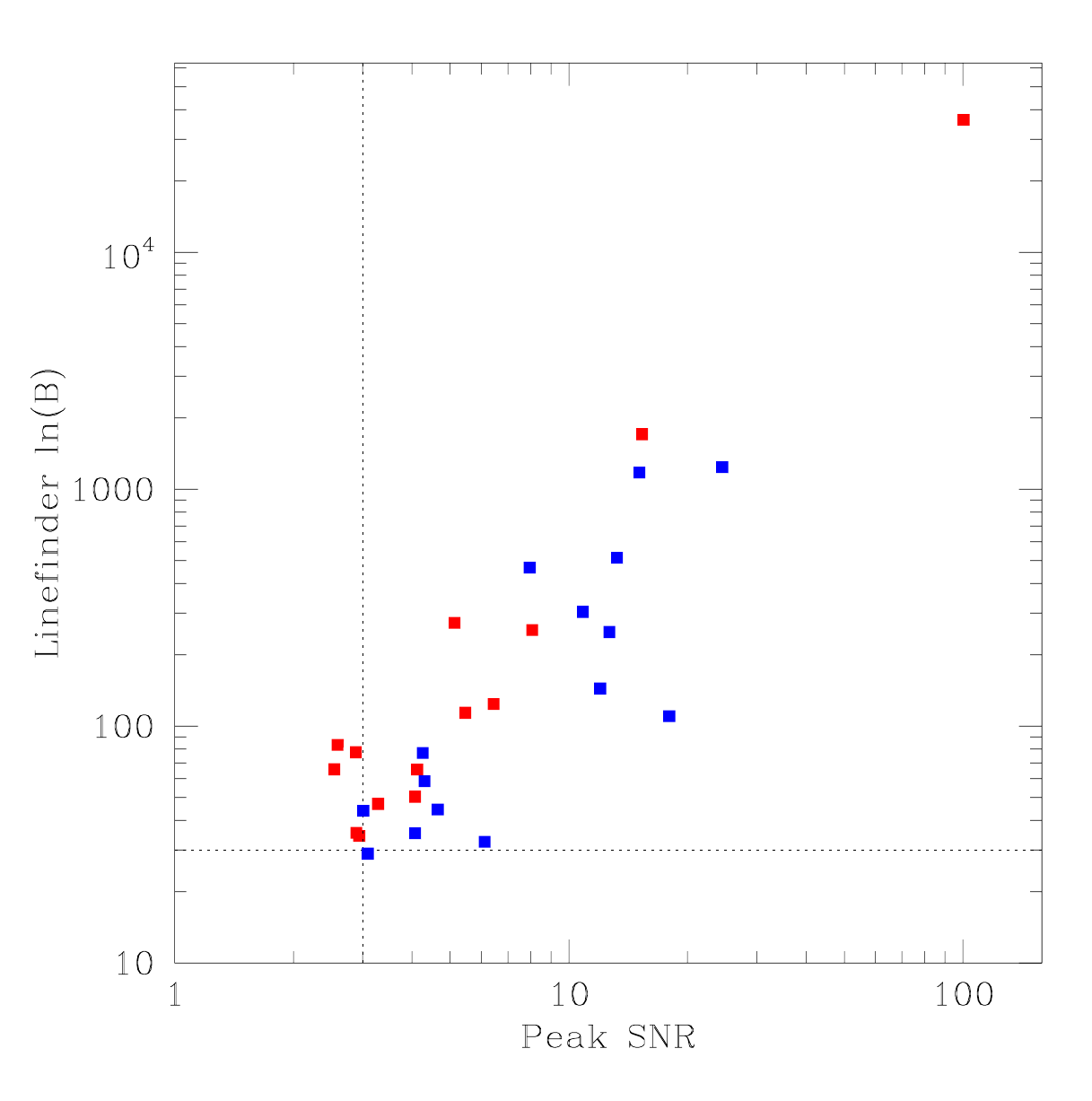}
\caption{Relationship between the linefinder ln\,(B) value and estimated peak SNR for the \hi absorption lines detected in the Pilot Survey. 
Blue points represent lines with velocity width $<80$\,km\,s$^{-1}$, and red points show broader lines with velocity width $\geq80$\,km\,s$^{-1}$. The horizontal line at ln\,(B) = 30 shows the cutoff value used in this paper to select line detections and the vertical dotted line corresponds to a 3$\sigma$ detection of the line peak. } 
\label{fig:A1}
\end{figure}

The results are shown in Figure\,\ref{fig:A1}, where dotted lines show a peak SNR of 3\,$\sigma$ and our adopted cutoff for detections of ln\,(B) = 30.
We need to keep in mind that the SNR in this plot is a peak value, while ln\,(B) is calculated for the whole line rather than a single channel. Nevertheless, there is a reasonable correlation that suggests we are detecting lines with peak SNR greater than about 2.5 to 3\,$\sigma$.
 
\begin{figure}
\includegraphics[width=0.94\columnwidth]{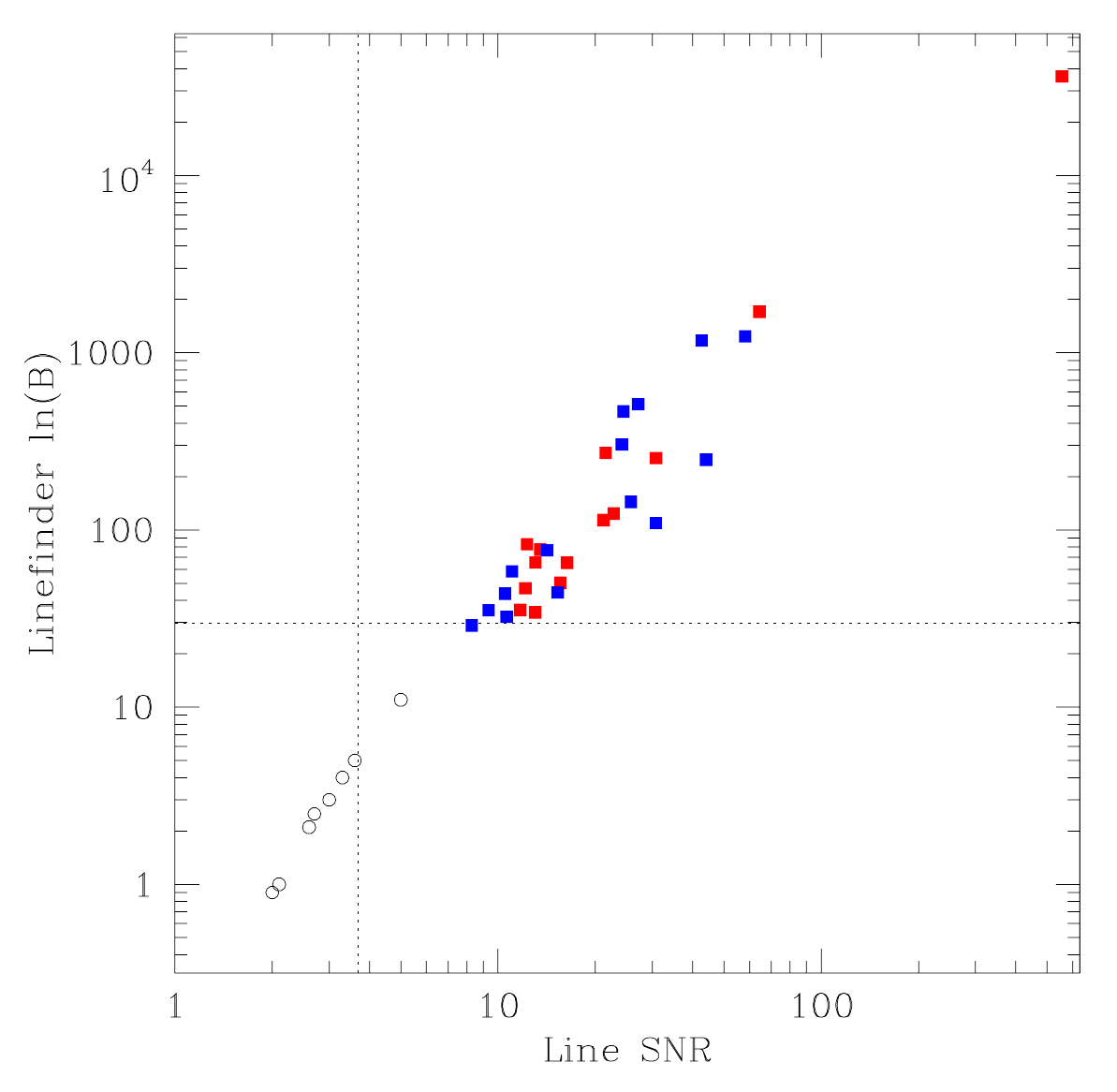}
\caption{As in Figure \ref{fig:A1}, except that the SNR values plotted are now integrated over the whole line, assuming a gaussian profile.  In this plot, black open circles at lower ln\,(B) show values tabulated by \cite{benneke2013} and the vertical dotted line corresponds to the 3.6\,$\sigma$ value considered by \cite{trotta2008} to be `strong' evidence. 
}
\label{fig:det1}
\end{figure}

Figure\,\ref{fig:det1} shows SNR values integrated over the whole line, which takes into account the range of line wdiths present in our data.  Values from Table~2 of \cite{benneke2013} are plotted as open circles, and our data extend the \cite{benneke2013} values out to much higher values of ln\,(B). This plot implies that an absorption line with ln\,(B)$\geq30$ has a line SNR $\geq7-8$ and is therefore a highly significant detection. It also implies that (in the absence of artefacts) we should be able to detect lines down to values of ln\,(B) $\sim10$. Such detections may become  possible in future if we can apply machine learning algorithms to distinguish artefacts from astronomical signals in a reliable way. 

\section{Notes on individual objects}
\label{sec:notes}

\noindent
{\bf PKS\,0011-023} (J001425-020556): \cite{dunlop1989} list an optical redshift of $z=2.08$ for the host galaxy, and give the reference for this as `Spinrad (private communication)'. Although no further redshift information is available in the literature, PKS\,0011-023  is generally assumed to be a high-redshift radio galaxy \citep[e.g.][]{pentericci2000}. If the redshift of 2.08 is correct, then our FLASH 21\,cm \hi detection at $z=0.678$ arises in an intervening system.\\

\noindent
{\bf NVSS\,J002331+010114} (J002331+010114): The unpublished LOFAR data point at 144\,MHz in the SED plot in Figure \ref{spectra1} was supplied by J. Callingham (private communication) and implies that this is a peaked-spectrum source.\\

\noindent
{\bf MRC\,0023-482} (J002604-475618): This object is classified as a peaked-spectrum radio source by \cite{callingham2017}. The \hi absorption line at $z=0.6746$ shown in Figure~\ref{spectra1} is clearly visible in both Pilot Survey spectra of this source, but was only identified by the linefinder in the second (SBID 37448) spectrum.\\

\noindent
{\bf NVSS J015516-251423} (J015516-251423): The peak \hi\ optical depth in this source ($\tau=1.83$) is the highest seen in our Pilot Survey sample.\\

\noindent
{\bf PKS\,0253-259} (J025544-254739): An absorption line was independently detected in two adjoining FLASH fields, 308 (SB 15212) and 309 (SB 13269). In both cases the source lies quite close to the edge of the ASKAP field, but the detection can be considered reliable. As can be seen from Figure \ref{spectra1}, the \hi line appears to have two narrow velocity components separated by about 80\,km\,s$^{-1}$.\\

\noindent
{\bf PKS\,0405-280} (J040757-275705): This object was identified as a peaked-spectrum radio source by both \cite{snellen2002} and \cite{callingham2017}. \\

\noindent
{\bf NVSS\,J051806-245502} (J051805-245502): This compact ICRF source was observed with the VLBA by \cite{petrov2008} and shows resolved structure on scales of a few milliarcseconds.\\

\noindent
{\bf MRC\,0531-237} (J053354-234430): 
This source has a deep and broad \hi absorption line at $z=0.851$, which is discussed in more detail by \cite{aditya2024}. \cite{mccarthy1996} list an optical redshift of $0.851\pm0.050$, from which \cite{aditya2024} identify this as an associated absorption-line system. It was identified as a peaked-spectrum radio source by \cite{callingham2017}. \\

\noindent
{\bf NVSS\,J090331+010846} (J090331+010847): 
This object lies in the GAMA G09 field \citep{driver11,liske2015}. The line detected in the pilot survey is discussed in detail by \cite{su2022}, who identified it as an associated \hi absorption system at $z=0.52$.\\

\noindent
{\bf NVSS\,J090425+012015} (J090425+012015): This absorber lies in the GAMA G09 field, with an \hi redshift of $z_{\rm HI}=0.8004$. No optical redshift is available. 
The line is detected in two adjoining fields in the second Pilot Survey, 546 (SBID 34559) and 547 (SBID 34539), but the radio source is very close to the edge of field 547, and the line detection has lower significance in this field.\\

\noindent
{\bf NVSS\,J091256+030021} (J091256+03002): This 33\,mJy source lies below the 45\,mJy flux density limit used to extract spectra for our linefinder search in the 2-hour Pilot Survey fields. A lower flux density limit of 30\,mJy was used for the linefinder search in the 6-hour GAMA fields, and this line was identified at $z_{\rm HI}=0.859$.  
The continuum source has been observed with the VLBA by \cite{deller2014}, who measured a source size of $6.4\times3.8$\,mas at 1.4 GHz.\\

\noindent
{\bf PKS\,0917+18} (J092011+175324): This object was identified as a peaked-spectrum source by \cite{callingham2017}. 
Two photometric redshift estimates have been published, $z\sim0.68$ \citep{yan2012} and $z\sim0.42$ \citep{ma2019}. Our detection of 21 cm absorption at $z_{\rm HI}=0.904$ is at higher redshift than either of these values, suggesting that the photometric redshifts may be underestimated because of a significant non-thermal contribution in optical and infrared light. \\

\noindent
{\bf NVSS\,J094650-202044} (J094650-202045): This source lies within the 6\,deg$^2$ region of sky covered by the CENSORS survey \citep{best2003}, which has deep K-band imaging and optical spectroscopy available for many sources. \cite{brookes2008} derive a redshift of $z=0.91$ for the optical counterpart of NVSS J094650-202044, based on the detection of a single emission line of [OII].\\

\noindent
{\bf NVSS\,J113622+004850} (J113622+004850): 
This object lies in the GAMA G12 field \citep{driver11,liske2015}. It is discussed in detail in a recent paper by \cite{su2022}, who identified it as an associated \hi absorption system at $z=0.56$. \\

\noindent
{\bf NVSS\,J150506+022927} (J150506+022928): 
\cite{callingham2017} identified this object as a peaked-spectrum radio source. \cite{hewett2010} list an optical redshift of $z=0.8055\pm0.0004$ based on an SDSS spectrum. \\

\noindent
{\bf NVSS\,J170135-294918} (J170135-294917): This source is at low Galactic latitude ($b=-7.0$\,deg). \\

\noindent
{\bf PKS\,2007-245} (J201045-242545): This object was identified as a peaked-spectrum radio source by \cite{callingham2017}. \\

\noindent
{\bf NVSS\,J215924-241752} (J215924-241752): This source is in the MRC-1Jy sample \citep{mccarthy1996}, and was identified by \cite{baker1999} as a quasar at $z=0.862$. 
The object is discussed in more detail by \cite{aditya2024}, who identified it as an associated \hi\ absorption system and noted that the \hi absorption line (at $z=0.868$) is redshifted relative to the optical lines. \\

\noindent
{\bf PKS\,2311$-$477} (J231351-472911): This object was identified as a peaked-spectrum radio source by \cite{callingham2017}.\\

\noindent
{\bf SUMSS\,J233432-585646} (J233432-585646):  This object was identified as a peaked-spectrum radio source by \cite{callingham2017}.\\

\noindent
{\bf Note}: The following three objects, detected during the pilot observations, are already well-studied sources and therefore not included in this paper. These serve as good test cases for our observations: \\

\noindent 
{\bf PKS 1610-77} (J161749-77171): An intervening \hi\ absorption line was detected at $z=0.450$ in ASKAP commissioning data by \cite{sadler20}. This object and its environment have been studied in detail by \cite{weng2022}.\\

\noindent 
{\bf PKS 1740-517} (J174425-514442): \cite{allison15} detected a strong associated \hi\ absorption line at $z=0.438$ in ASKAP commissioning data. The field containing this source was observed in the second Pilot Survey as a test, but the poor quality of the data meant that the linefinder results were not reliable. \\ 

\noindent 
{\bf PKS 1830-21} (J183339-210339): This bright (11~Jy) radio source is a well-studied system in which a $z=2.51$ quasar is gravitationally lensed by a foreground galaxy at $z=0.885$ \citep{subrahmanyan1990,wiklind1996,lidman1999}. The radio spectrum shows broad intervening \hi and OH absorption in the foreground galaxy \citep{chengalur1999,allison2017}. The line provides a good test of the 5~MHz beam-forming intervals used for the Pilot~2 observations, and so PKS\,1830-211 was used as a `quality-gate' test field at the start of the second FLASH pilot survey.\\

\section{Additional detection using a redshift prior}
\label{sec:extra}

Here, we show one example of an \hi absorption line that was detected in the Pilot Survey by using a known optical redshift (and previous \hi detection) as a prior for the line search. Although this line has a ln\,(B) value below our detection cutoff (ln\,(B) = 30), we accept as a genuine detection because of the additional information from the redshift prior. It is likely that other lines of this kind are present in the Pilot Survey data.\\

\begin{figure}[ht]
\centering
\includegraphics[width=1.1\columnwidth]{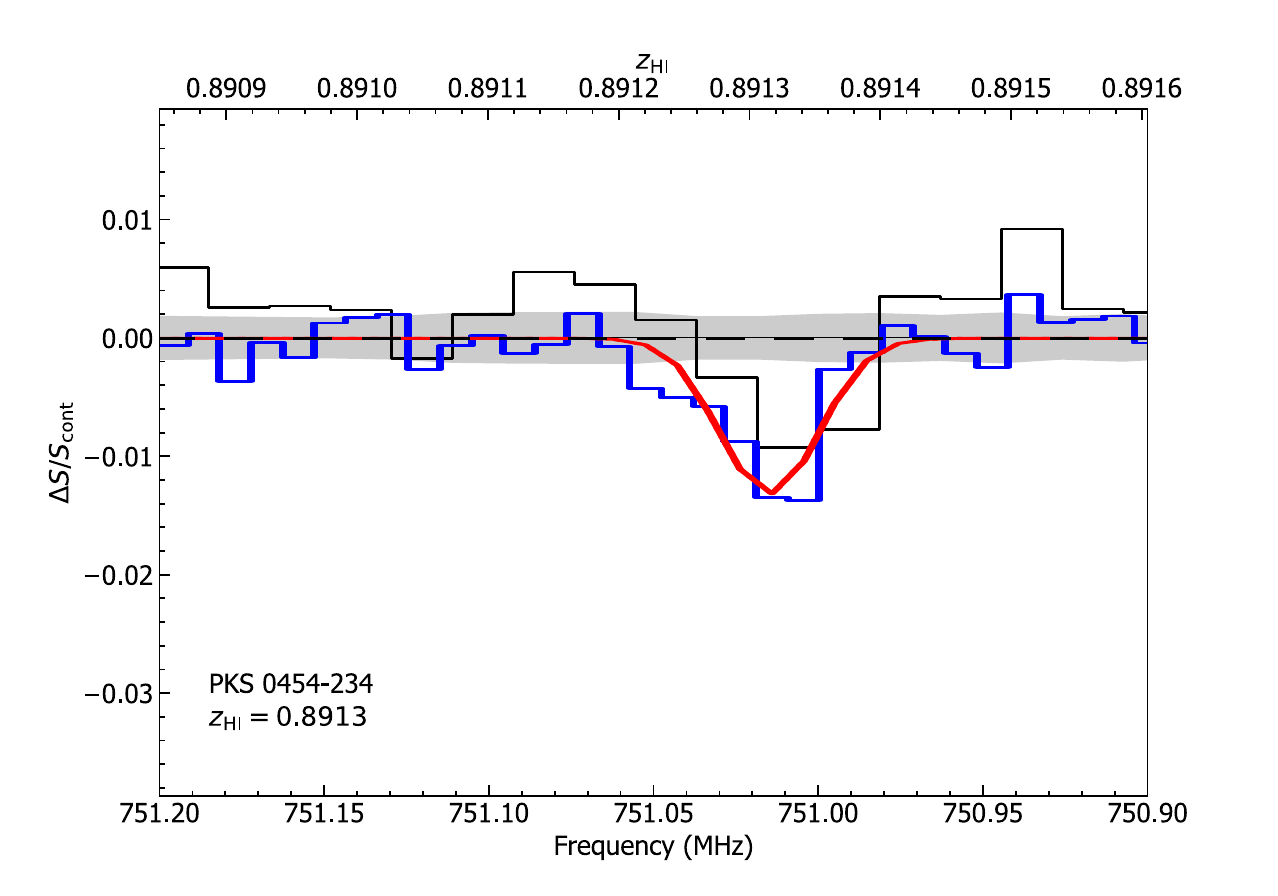}
\caption{Line profile of PKS\,0454-23. Blue and red lines are from \citet{gupta2012}. Black is from our observation (SBID 34547\_1a). The grey shaded region is the 1$\sigma$ level of the rms spectral-line noise.}
\label{fig:0454}
\end{figure}

\noindent
{\bf PKS\,0454-23} (J045703-232453): This object is a well-studied quasar at redshift $z=1.003$ \citep{stickel1989} and with a continuum flux density of 3.18\,Jy in the ASKAP band. \cite{gupta2012} detected an intervening \hi absorption line at $z=0.891$ in a targeted observation using the redshift of a known \mgii absorption system. 
The line seen by \cite{gupta2012} is weak and narrow (peak optical depth 0.013, FWHM $14\pm2$\,km\,s$^{-1}$). 
It was not found in our original linefinder search, but when we use the \cite{gupta2012} redshift as a prior then the line is marginally detected (ln\,(B) = 8.53) in our ASKAP spectrum (see Figure \ref{fig:0454}). We measure a peak optical depth of $0.009\pm0.002$ and velocity width of $13.9\pm2.7$\,km.s$^{-1}$, in reasonable agreement with the \cite{gupta2012} values. \\

\section{Individual spectral-line plots} 
\label{sec:plots}

In Figure~\ref{spectra1}, we plot the line profiles from the co-added spectra for each of the 30~detections listed in Table \ref{tab:finder2} along with the radio continuum SED fits described in Section~\ref{sec:radioSED} and based on the methodology of \cite{kerrison2024a}. Each caption includes a brief note on individual objects.

\begin{figure*}[!h]
\captionsetup[subfigure]{labelformat=empty}
\centering

\begin{subfigure}[b]{\textwidth}
\includegraphics[width=0.53\linewidth]{figures/coadded/coadded_spectrum_201.png}
\includegraphics[width=0.47\linewidth]
{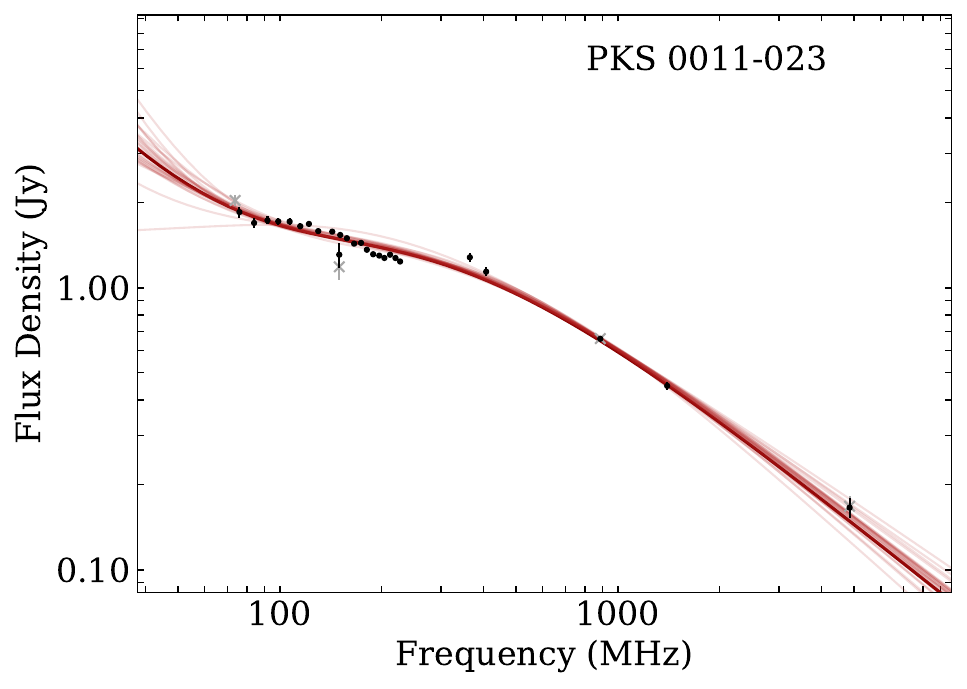}
\caption{1.\ PKS\,0011-023}
\end{subfigure}

\begin{subfigure}[b]{\textwidth}
\includegraphics[width=0.53\linewidth]{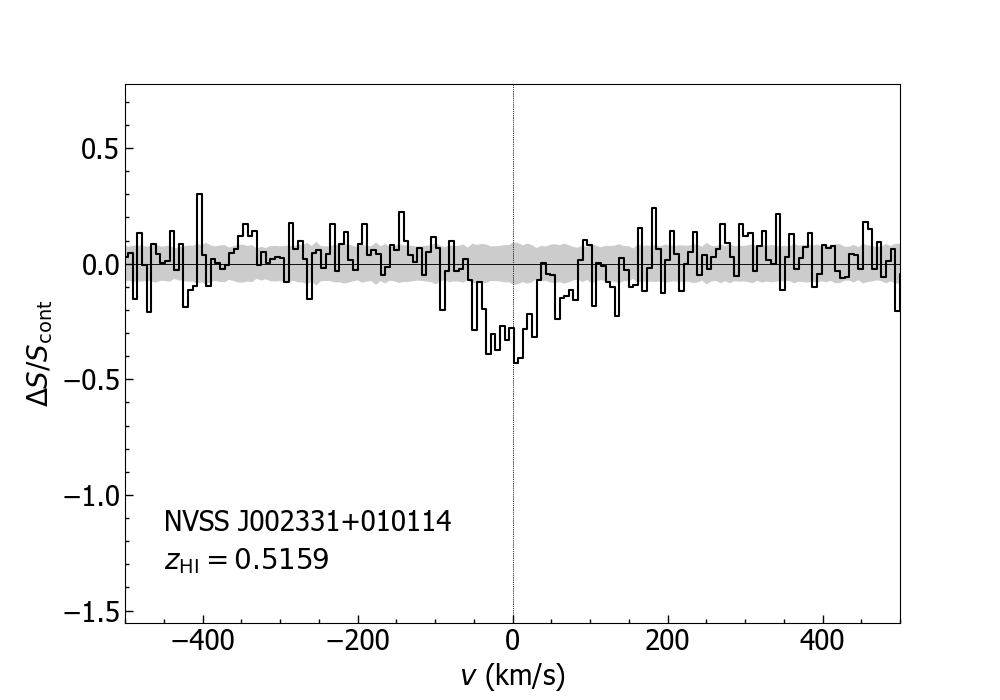}
\includegraphics[width=0.47\linewidth]{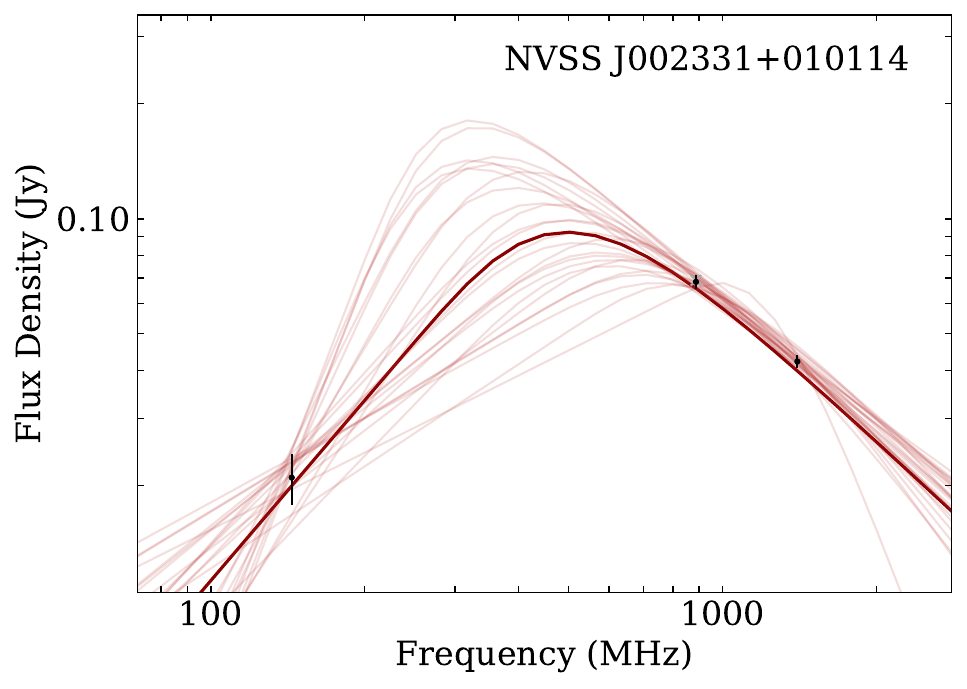}
\caption{2.\ NVSS\,J002331+010114}
\end{subfigure}

\begin{subfigure}[b]{\textwidth}
\includegraphics[width=0.53\linewidth]{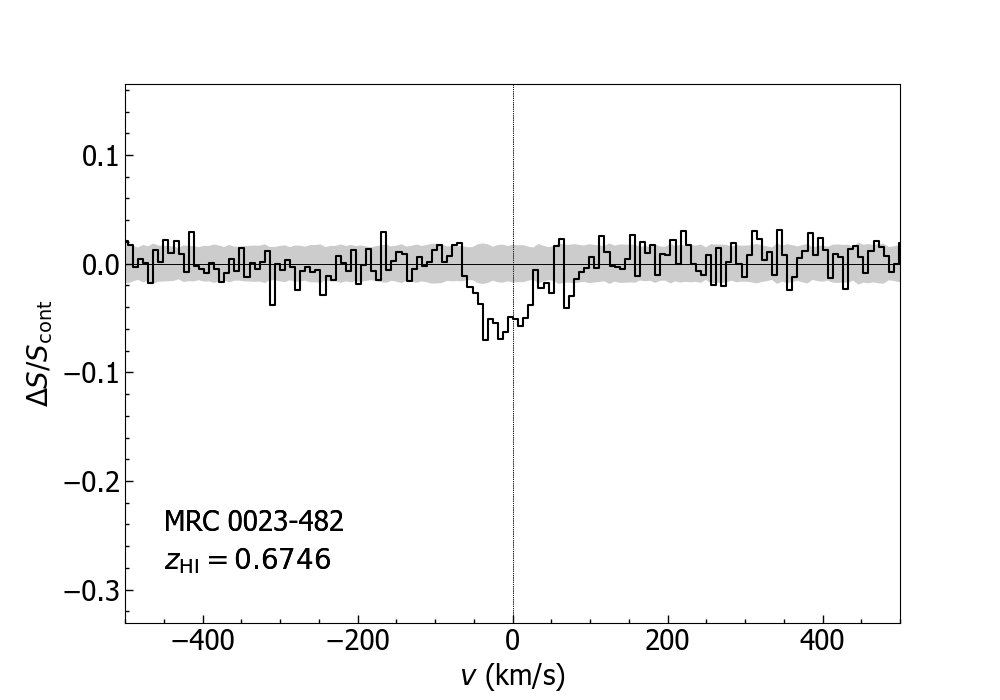}
\includegraphics[width=0.47\linewidth]
{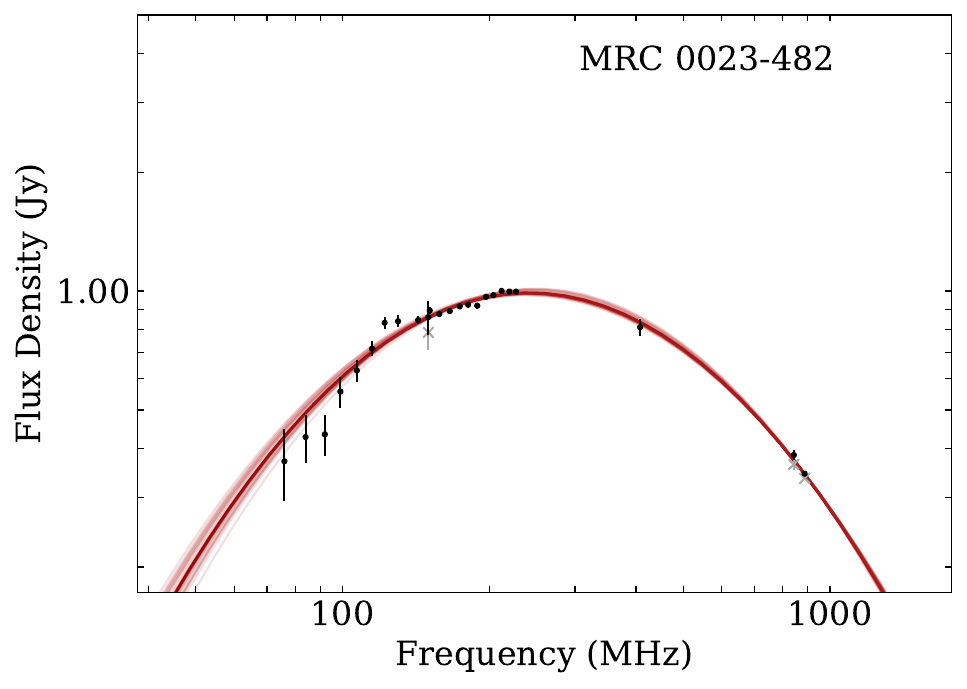}
\caption{3.\ MRC\,0023-482}
\end{subfigure}
\caption{Spectral-line profiles and radio SED fits for: 1.\ PKS\,0011-023 (top), 2.\ NVSS\,J002331+010114 (middle), and 3.\ MRC\,0023-482 (bottom). The SED for NVSS\,J002331+010114 uses LOFAR data at 144\,MHz from the LoTSS survey (private communication with J. Callingham)}
\label{spectra1}
\end{figure*}

\setcounter{figure}{3}
\begin{figure*}[!h]
\captionsetup[subfigure]{labelformat=empty}
\centering

\begin{subfigure}[b]{\textwidth}
\includegraphics[width=0.53\linewidth]{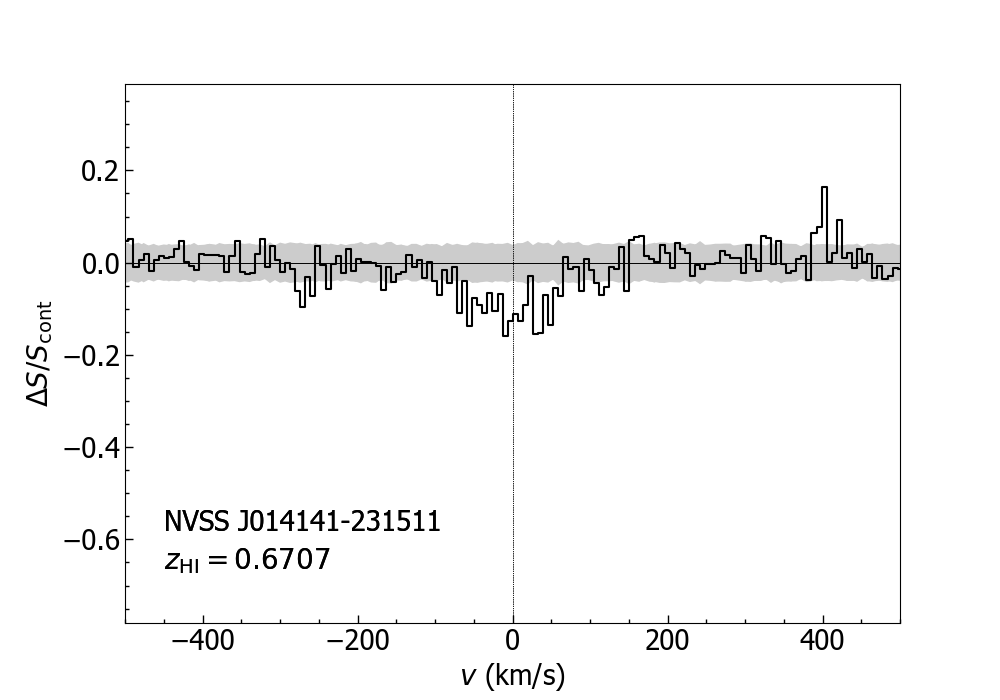}
\includegraphics[width=0.47\linewidth]{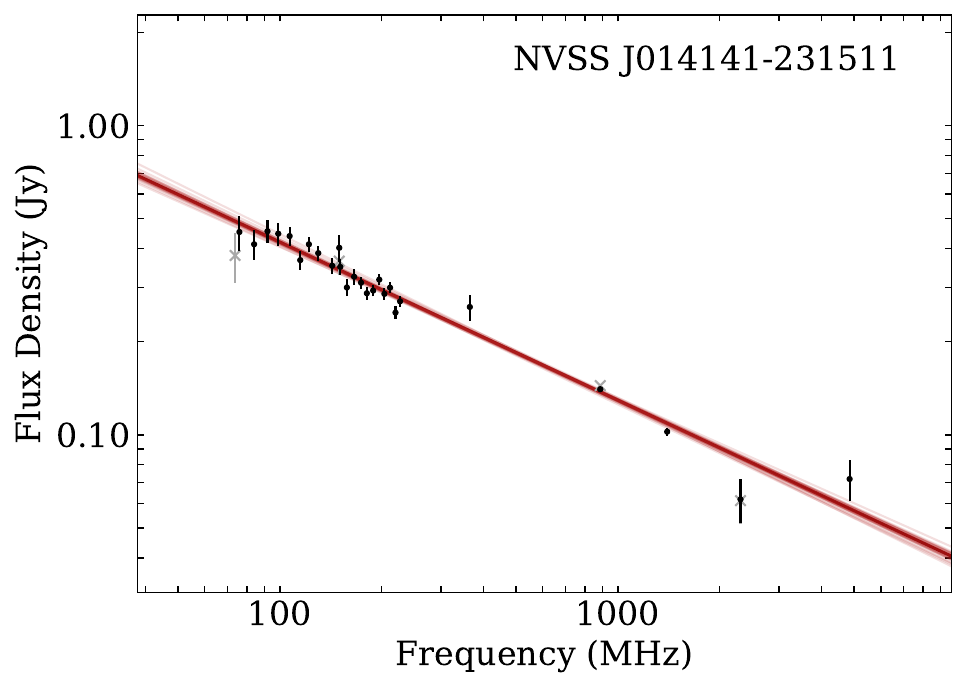}
\caption{4.\ NVSS\,J014141-231511}
\end{subfigure}

\begin{subfigure}[b]{\textwidth}
\includegraphics[width=0.53\linewidth]{figures/coadded/coadded_spectrum_103.png}
\includegraphics[width=0.47\linewidth]{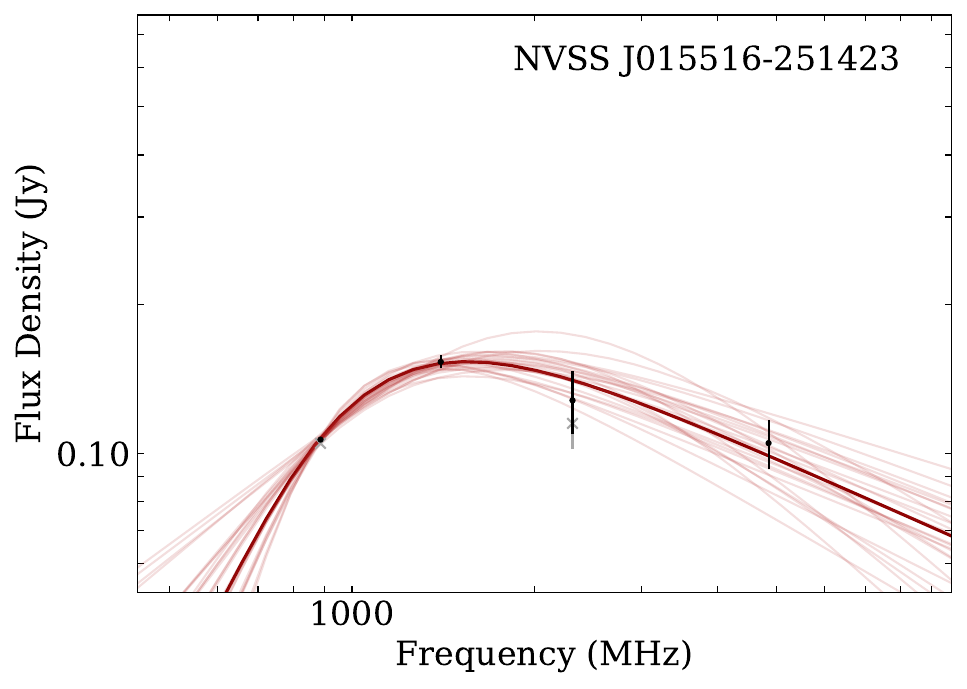}
\caption{5.\ NVSS J015516-251423}
\end{subfigure}

\begin{subfigure}[b]{\textwidth}
\includegraphics[width=0.53\linewidth]{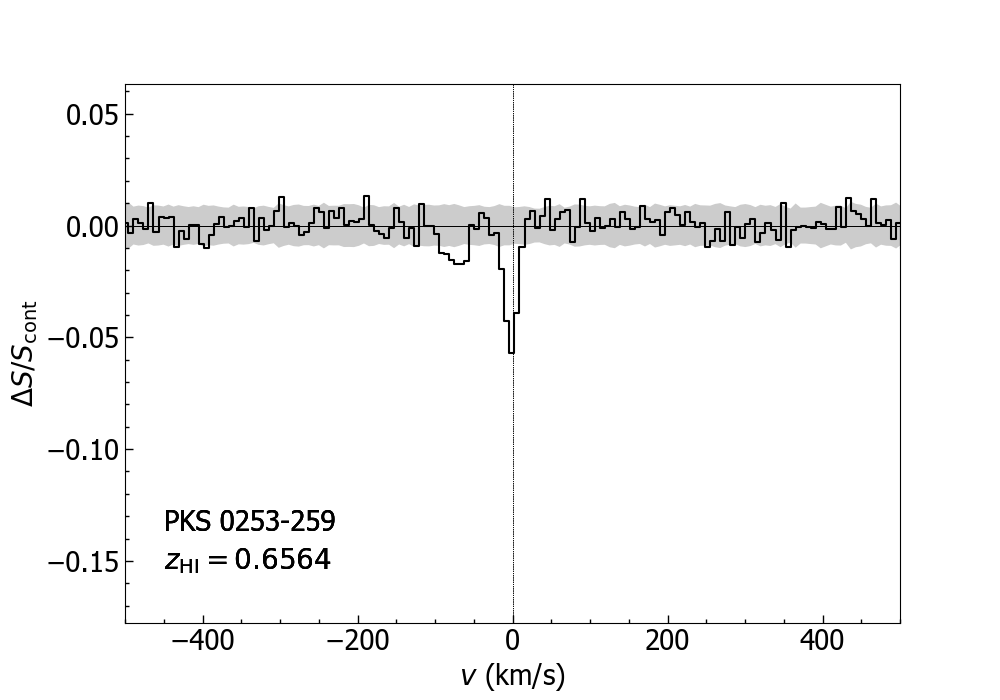}
\includegraphics[width=0.47\linewidth]
{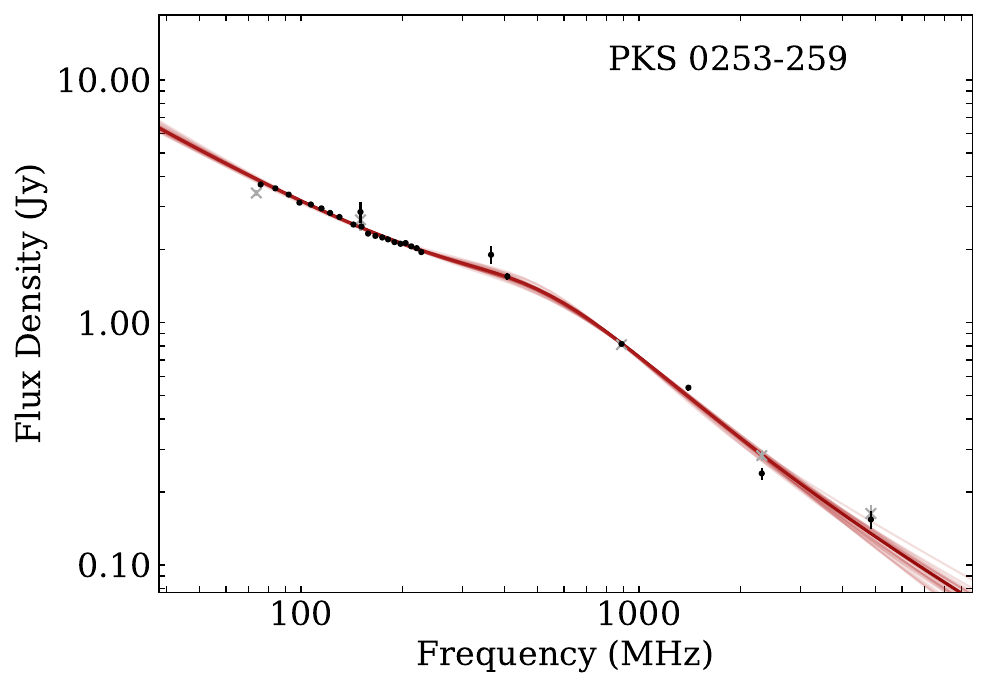}
\caption{6.\ PKS\,0253-259}
\end{subfigure}
\caption{(continued) Spectral-line profiles and radio SED fits for: 4.\ NVSS\,J014141-231511 (top), 5.\ NVSS J015516-251423 (middle), and 6.\ PKS\,0253-259 (bottom). }
\end{figure*}

\setcounter{figure}{3}
\begin{figure*}[!h]
\captionsetup[subfigure]{labelformat=empty}
\centering

\begin{subfigure}[b]{\textwidth}
\includegraphics[width=0.53\linewidth]{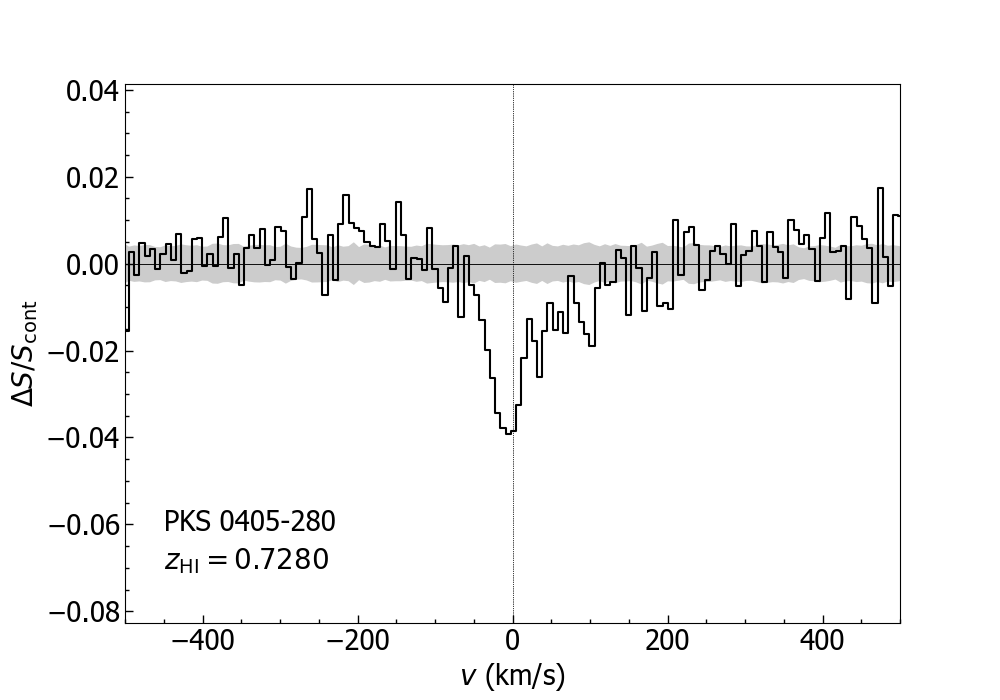}
\includegraphics[width=0.47\linewidth]
{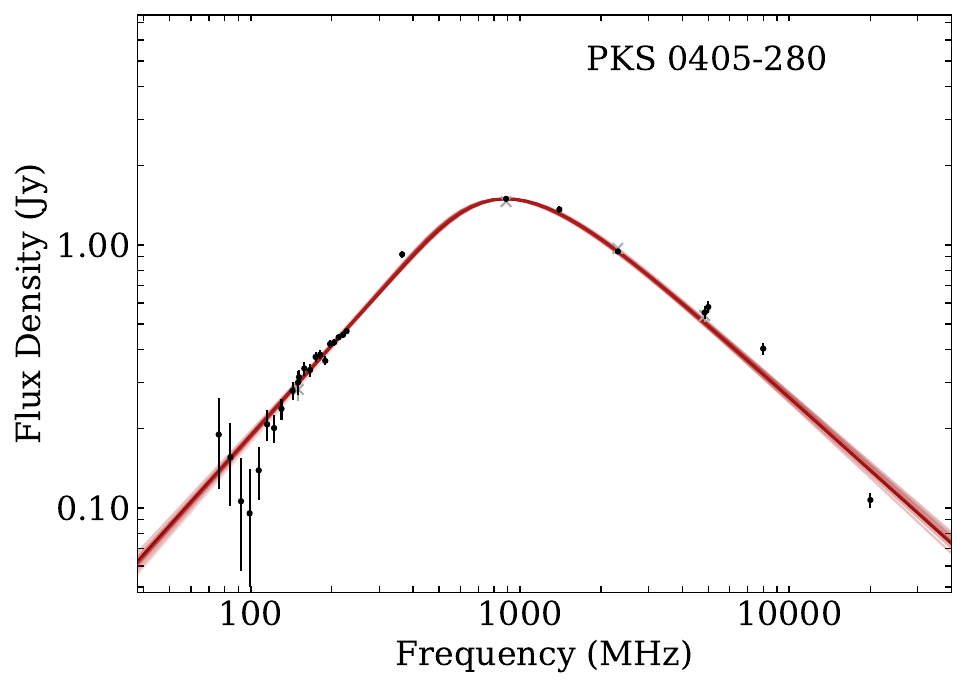}
\caption{7.\ PKS\,0405-280}
\end{subfigure}

\begin{subfigure}[b]{\textwidth}
\includegraphics[width=0.53\linewidth]{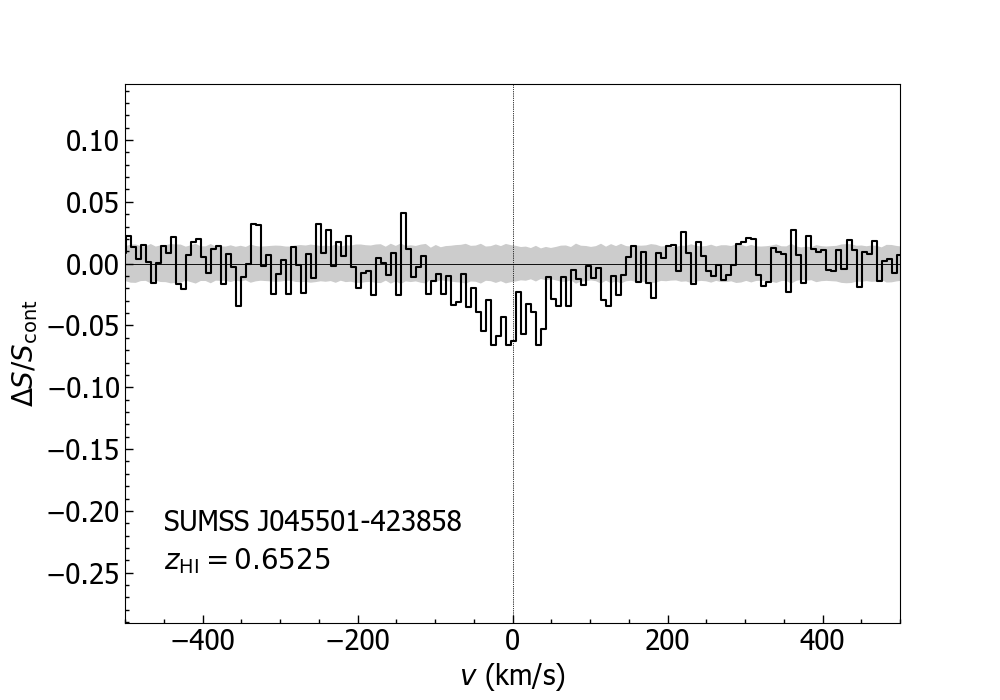}
\includegraphics[width=0.47\linewidth]
{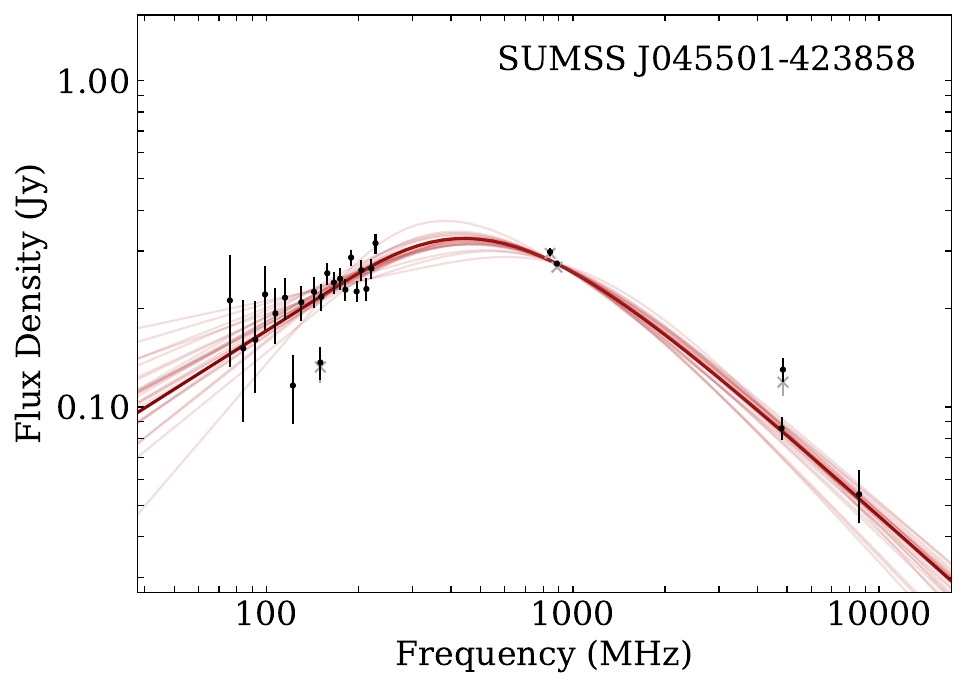}
\caption{8.\ SUMSS\,J045501-423858}
\end{subfigure}

\begin{subfigure}[b]{\textwidth}
\includegraphics[width=0.53\linewidth]{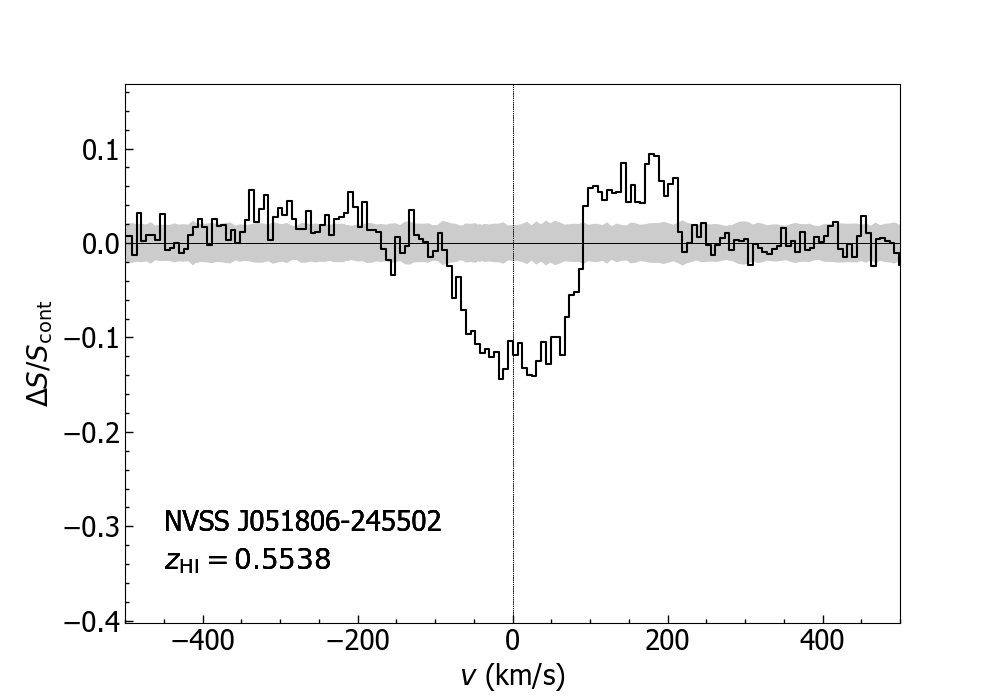}
\includegraphics[width=0.47\linewidth]{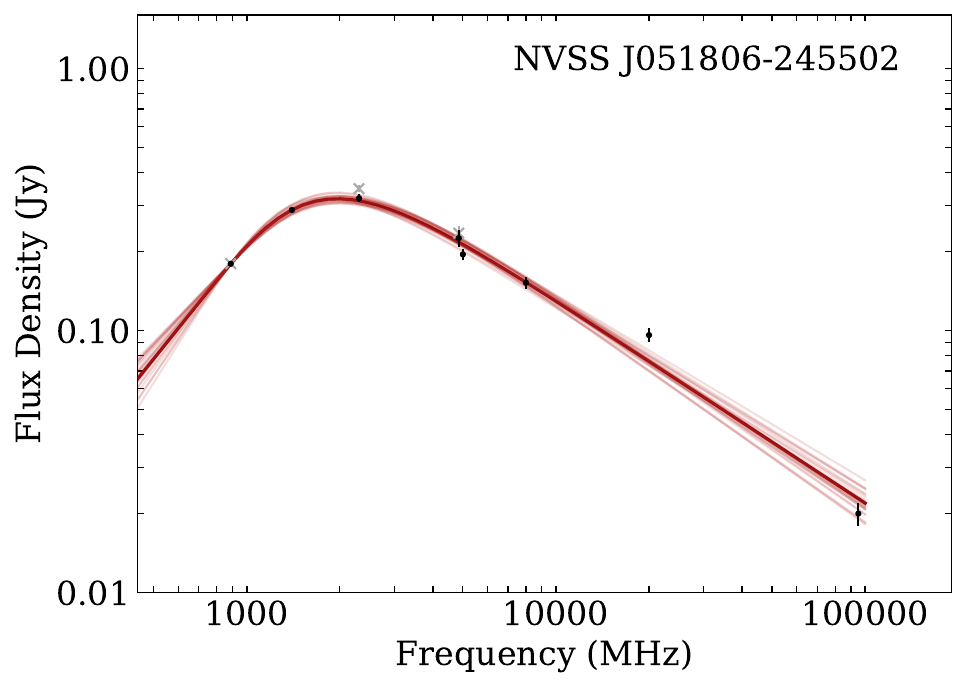}
\caption{9.\ NVSS~J051806-245502}
\end{subfigure}
\caption{(continued) Spectral-line profiles and radio SED fits for: 7.\ PKS\,0405-280 (top), 8.\ SUMSS\,J045501-423858 (middle), and 9.\ NVSS~J051806-245502 (bottom).}
\end{figure*}

\setcounter{figure}{3}
\begin{figure*}[!h]
\captionsetup[subfigure]{labelformat=empty}
\centering

\begin{subfigure}[b]{\textwidth}
\includegraphics[width=0.53\linewidth]{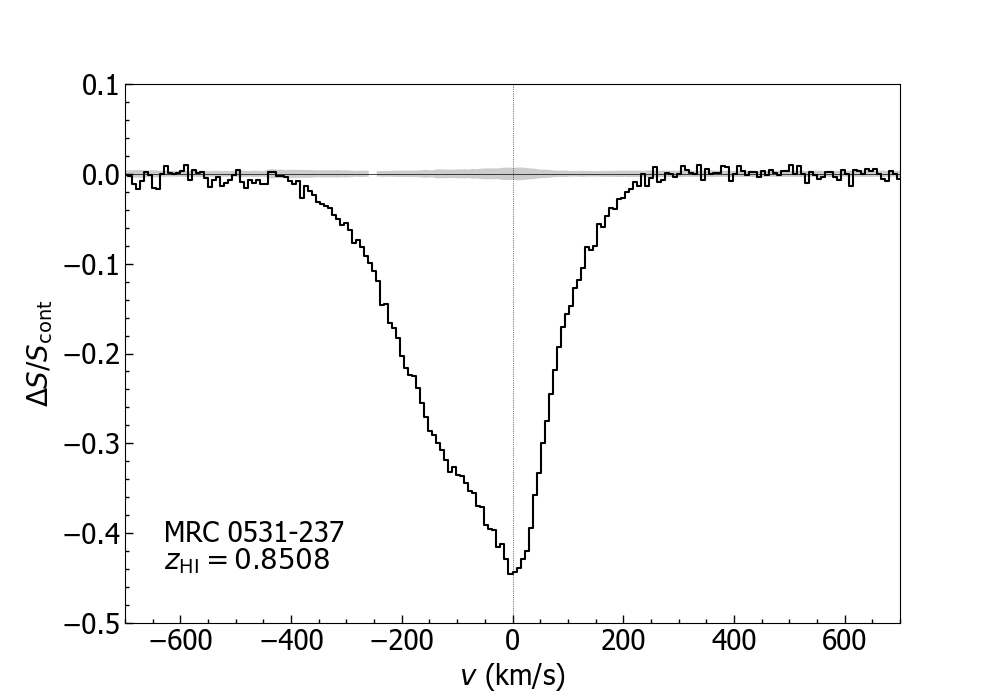}
\includegraphics[width=0.47\linewidth]
{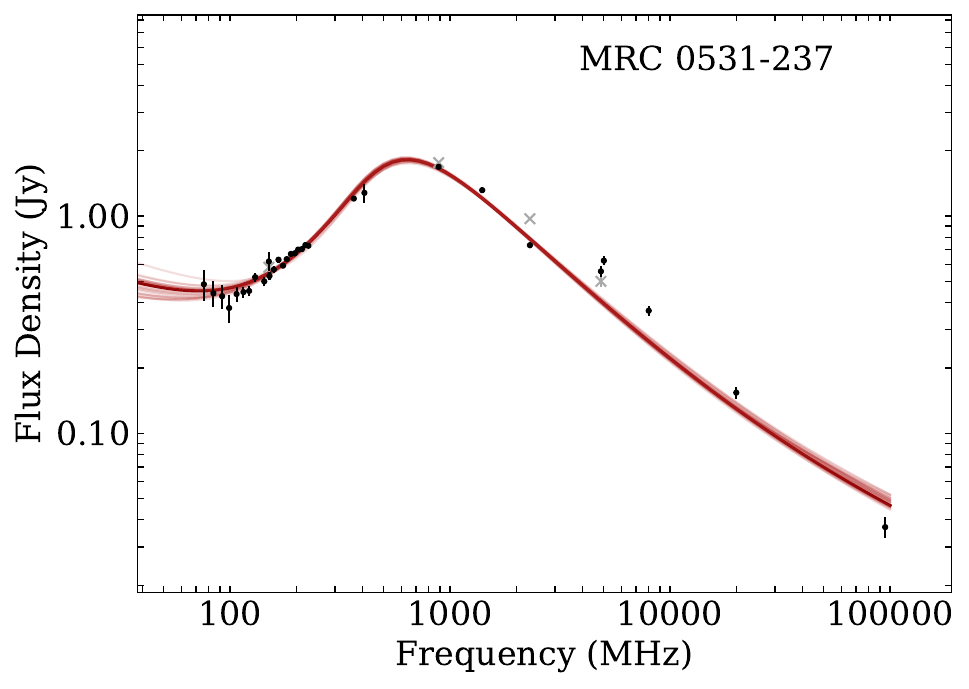}
\caption{10.\ MRC~0531-237}
\end{subfigure}

\begin{subfigure}[b]{\textwidth}
\includegraphics[width=0.53\linewidth]{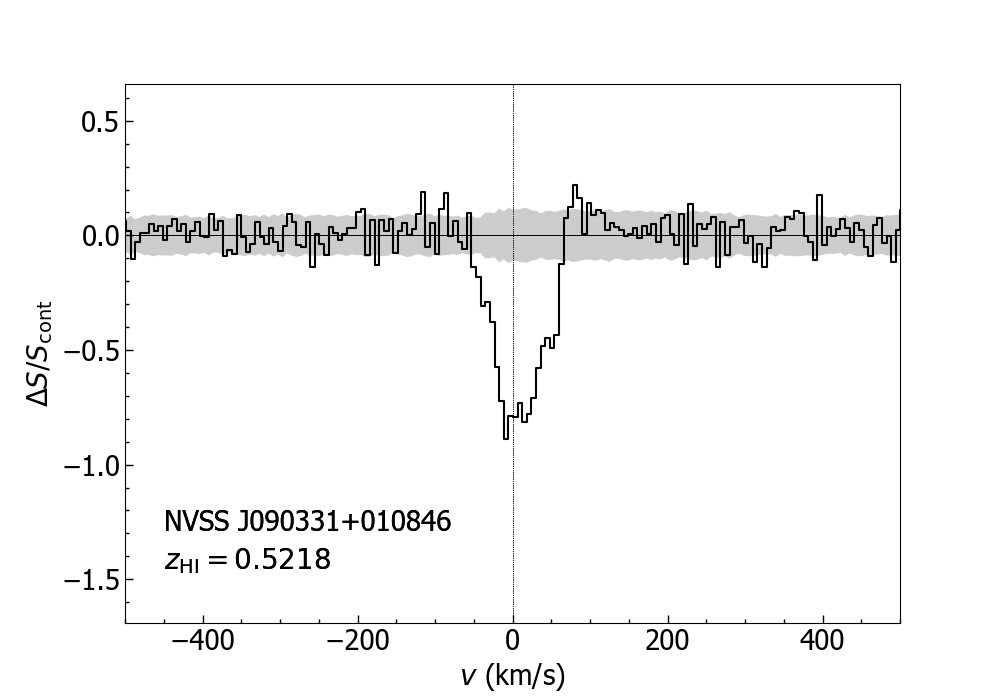}
\includegraphics[width=0.47\linewidth]{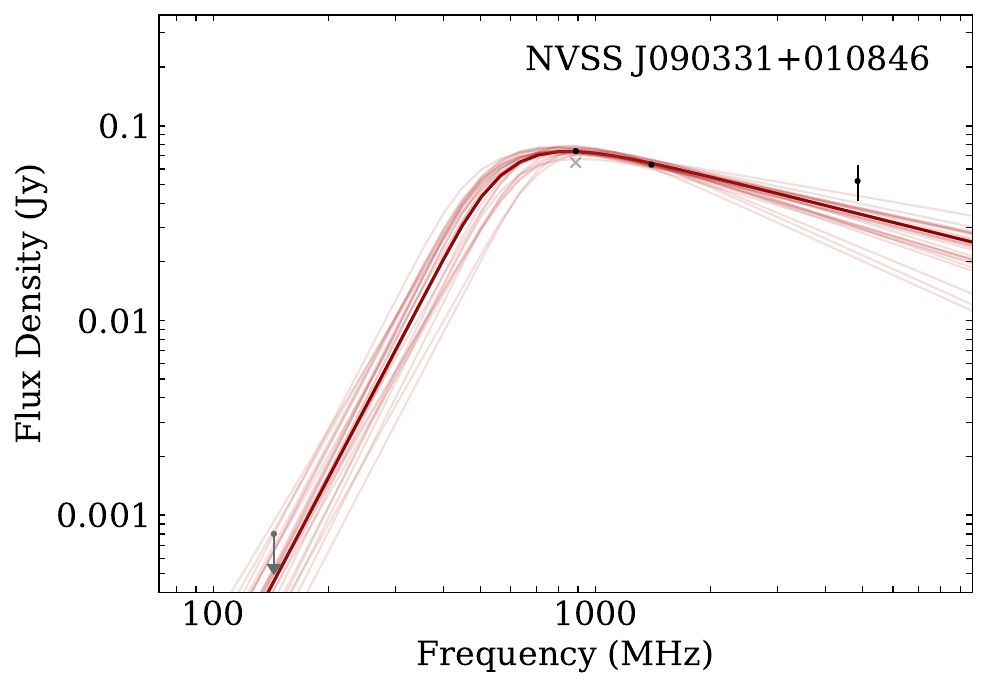}
\caption{11.\ NVSS\,J090331+010846}
\end{subfigure}

\begin{subfigure}[b]{\textwidth}
\includegraphics[width=0.53\linewidth]{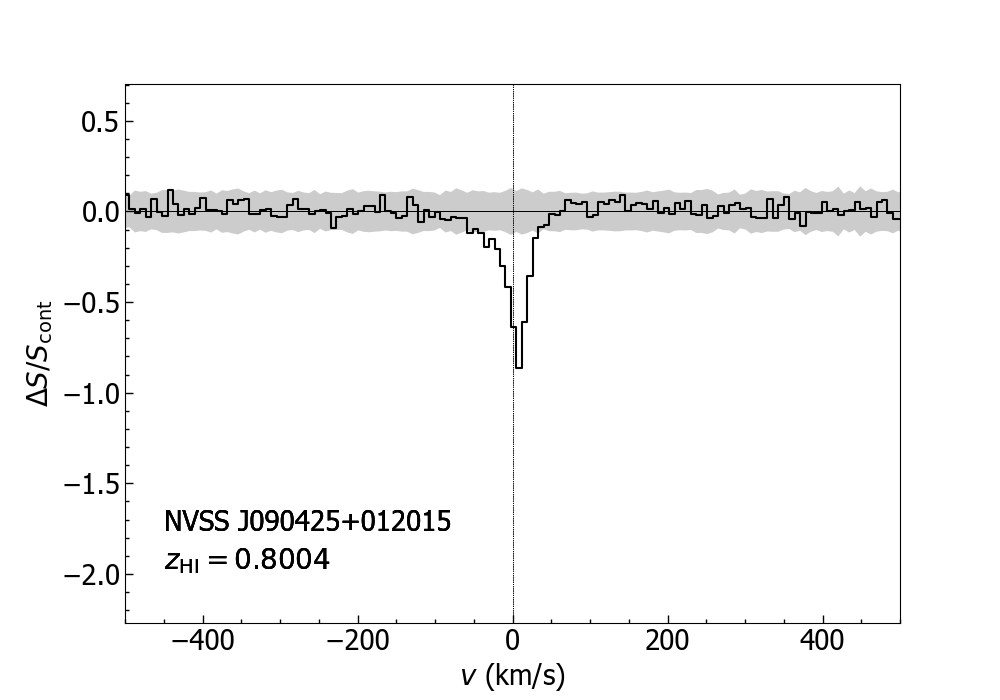}
\includegraphics[width=0.47\linewidth]{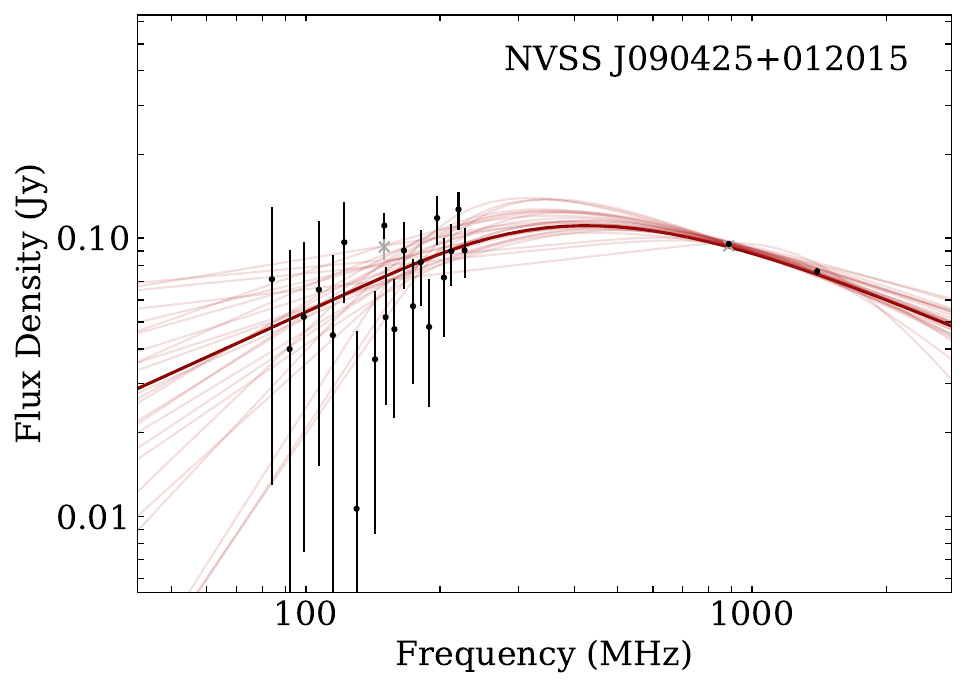}
\caption{12.\ NVSS\,J090425+012015}
\end{subfigure}
\caption{(continued) Spectral-line profiles and radio SED fits for: 10.\ MRC~0531-237 (top), 11.\ NVSS\,J090331+010846 (middle), and 12.\ NVSS\,J090425+012015 (bottom). Note that in the case of NVSS\,J090331+010846, the SED makes use of an upper limit from LOFAR at 144\,MHz (grey arrow) supplied by J. Callingham \& T. Shimwell (private communication). The SED for NVSS\,J090425+012015 uses LOFAR data at 144\,MHz from the LoTSS survey supplied by J. Callingham (private communication).}
\end{figure*}

\setcounter{figure}{3}
\begin{figure*}[!h]
\captionsetup[subfigure]{labelformat=empty}
\centering

\begin{subfigure}[b]{\textwidth}
\includegraphics[width=0.53\linewidth]{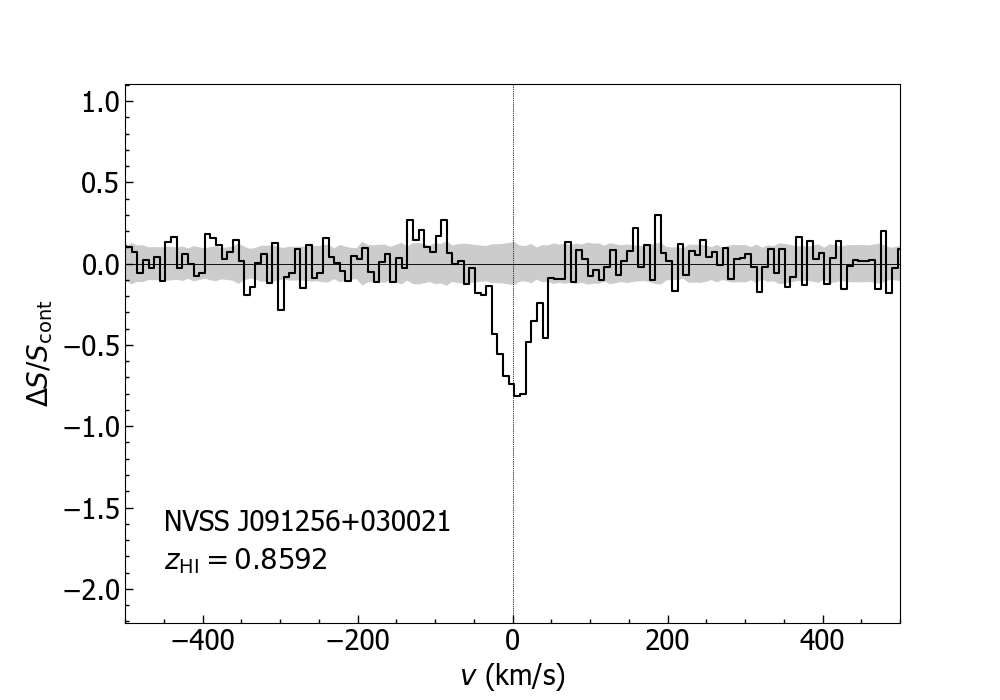}
\includegraphics[width=0.47\linewidth]{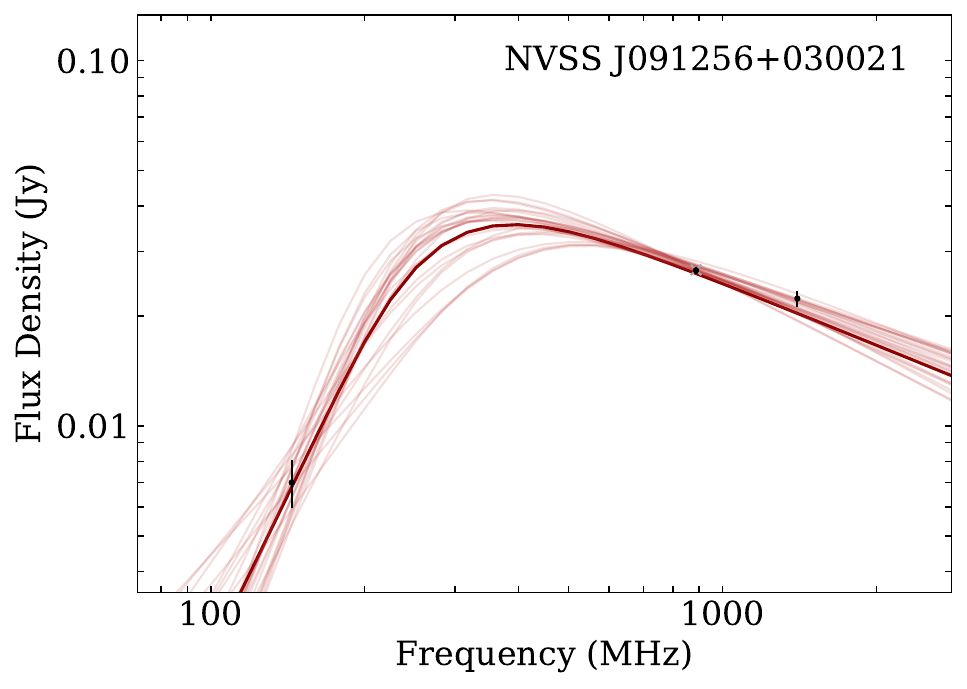}
\caption{13.\ NVSS\,J091256+030021}
\end{subfigure}

\begin{subfigure}[b]{\textwidth}
\includegraphics[width=0.53\linewidth]{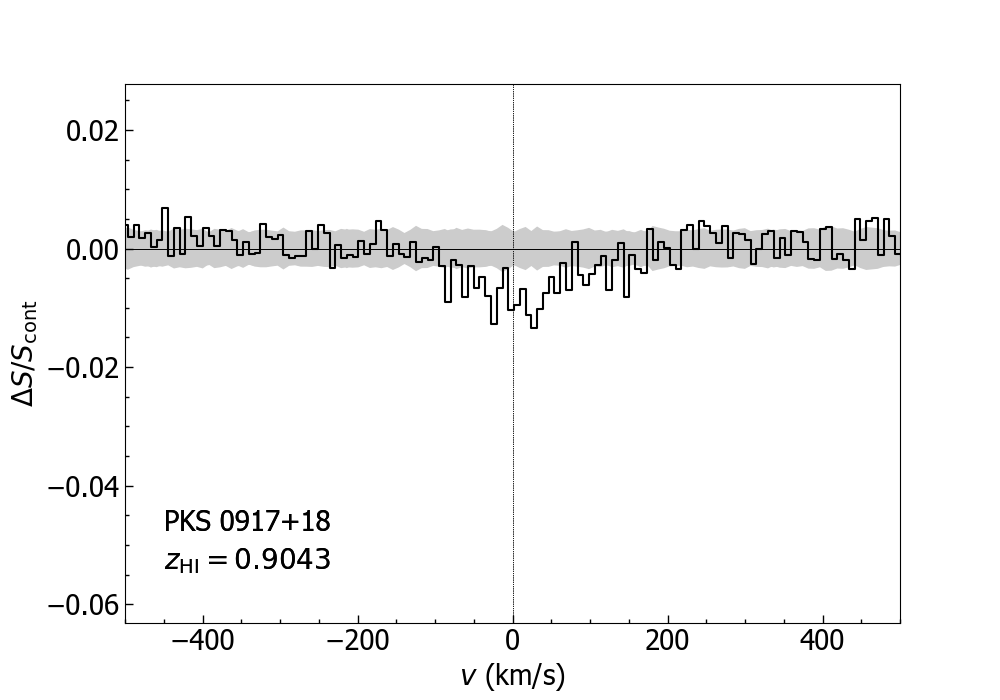}
\includegraphics[width=0.47\linewidth]{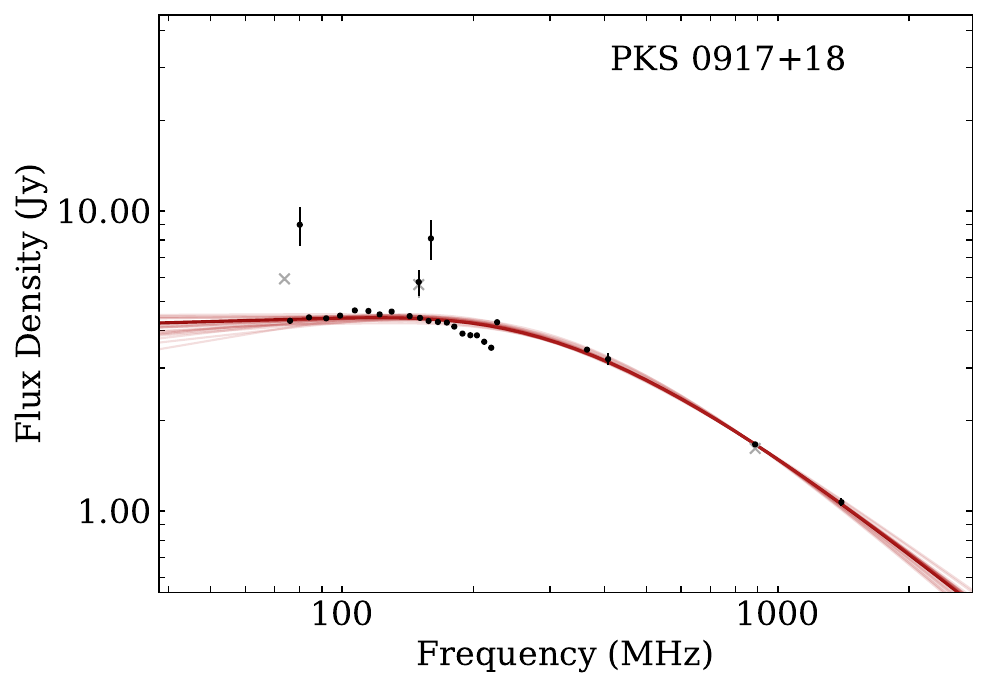}
\caption{14.\ PKS\,0917+18}
\end{subfigure}

\begin{subfigure}[b]{\textwidth}
\includegraphics[width=0.53\linewidth]{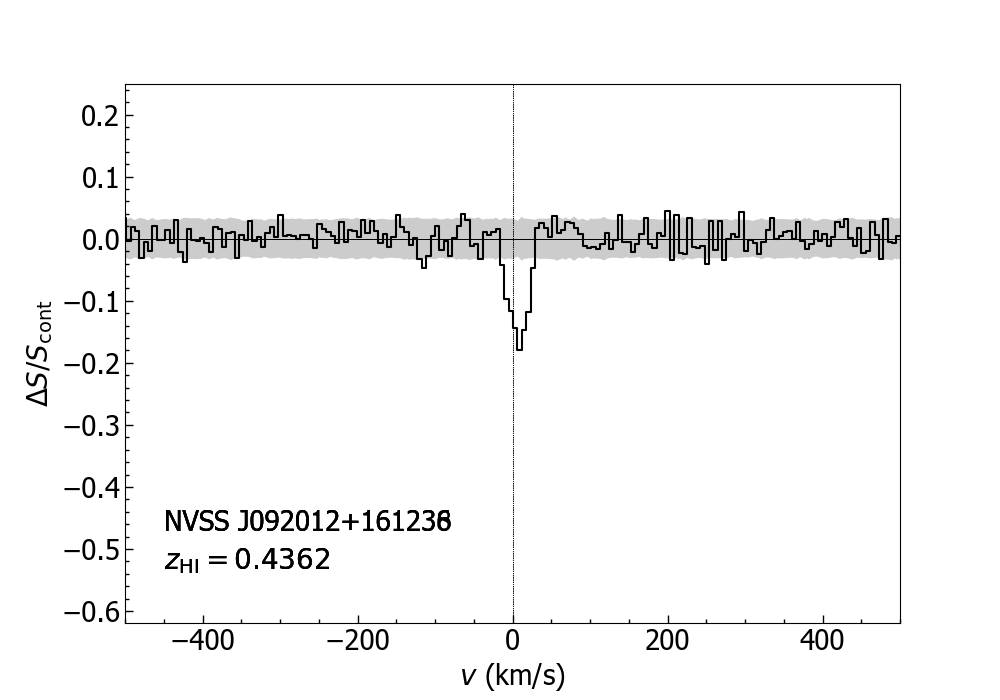}
\includegraphics[width=0.47\linewidth]{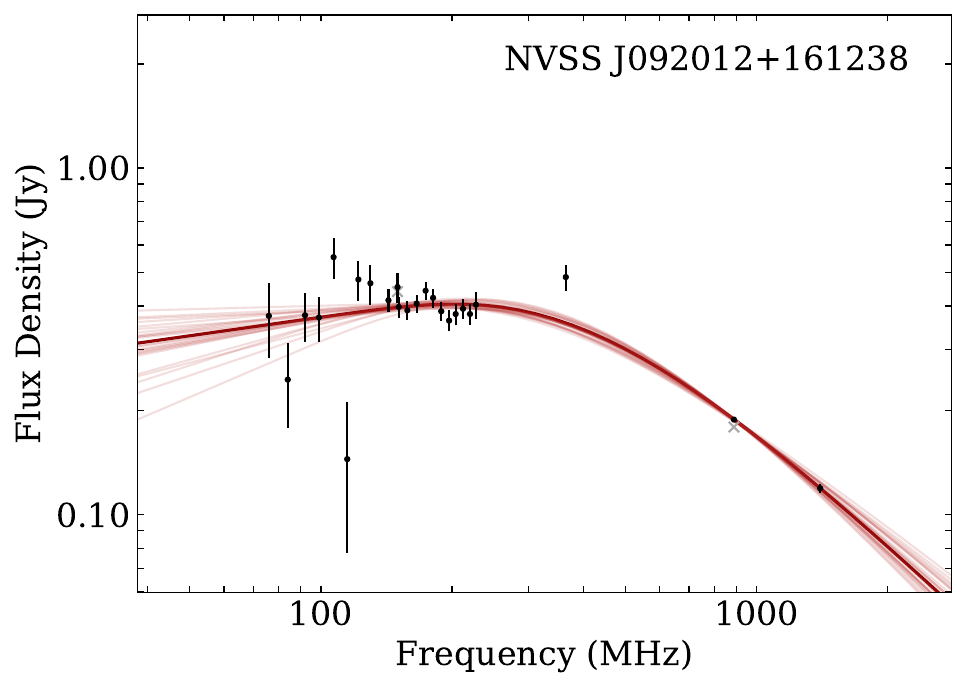}
\caption{15.\ NVSS\,J092012+161238}
\end{subfigure}
\caption{(continued) Spectral-line profiles and radio SED fits for: 13.\ NVSS\,J091256+030021 (top), 14.\ PKS\,0917+18 (middle), and 15.\ NVSS\,J092012+161238 (bottom).}
\end{figure*}

\setcounter{figure}{3}
\begin{figure*}[!h]
\captionsetup[subfigure]{labelformat=empty}
\centering

\begin{subfigure}[b]{\textwidth}
\includegraphics[width=0.53\linewidth]{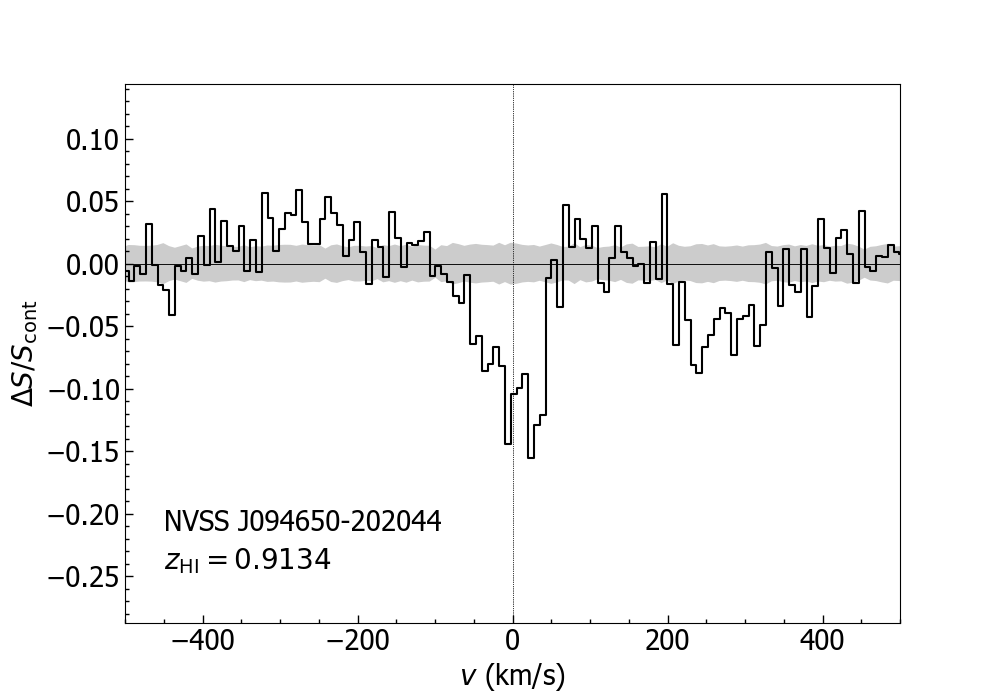}
\includegraphics[width=0.47\linewidth]
{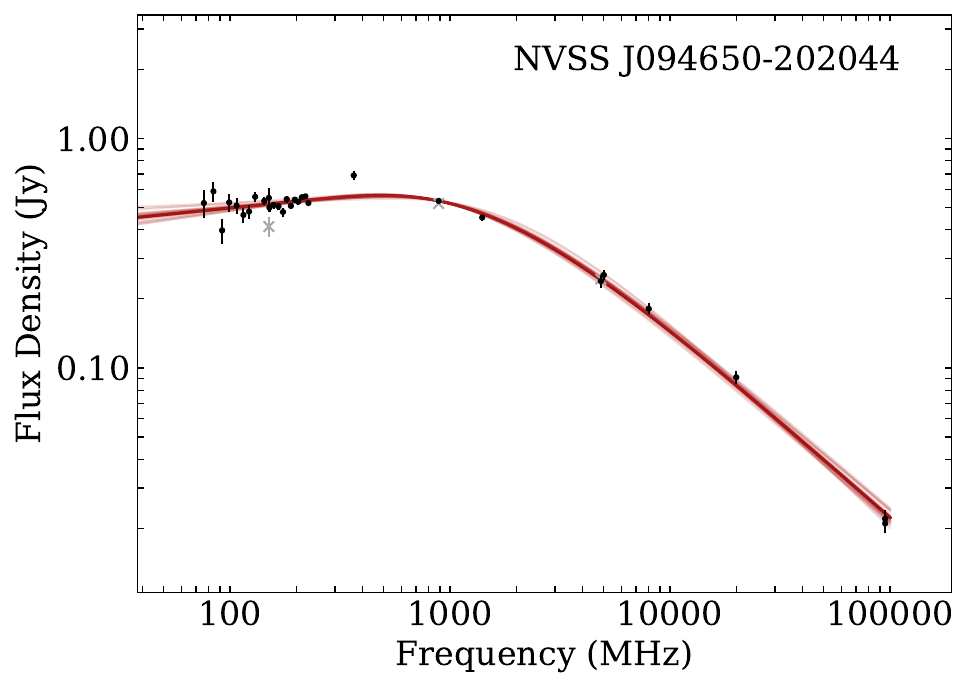}
\caption{16.\ NVSS~J094650-202044}
\end{subfigure}

\begin{subfigure}[b]{\textwidth}
\includegraphics[width=0.53\linewidth]{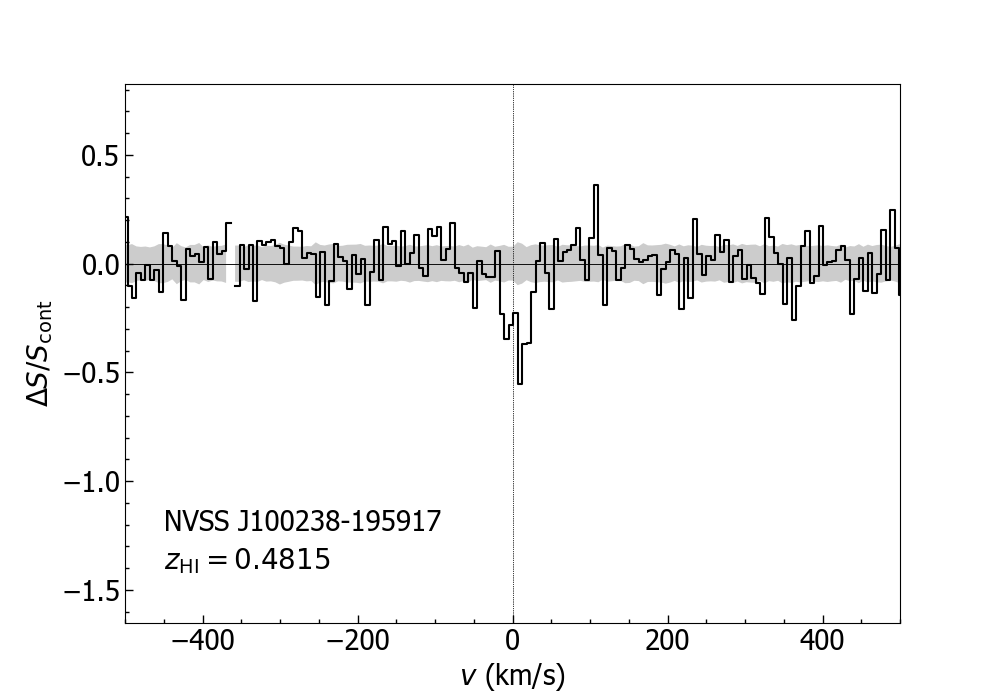}
\includegraphics[width=0.47\linewidth]{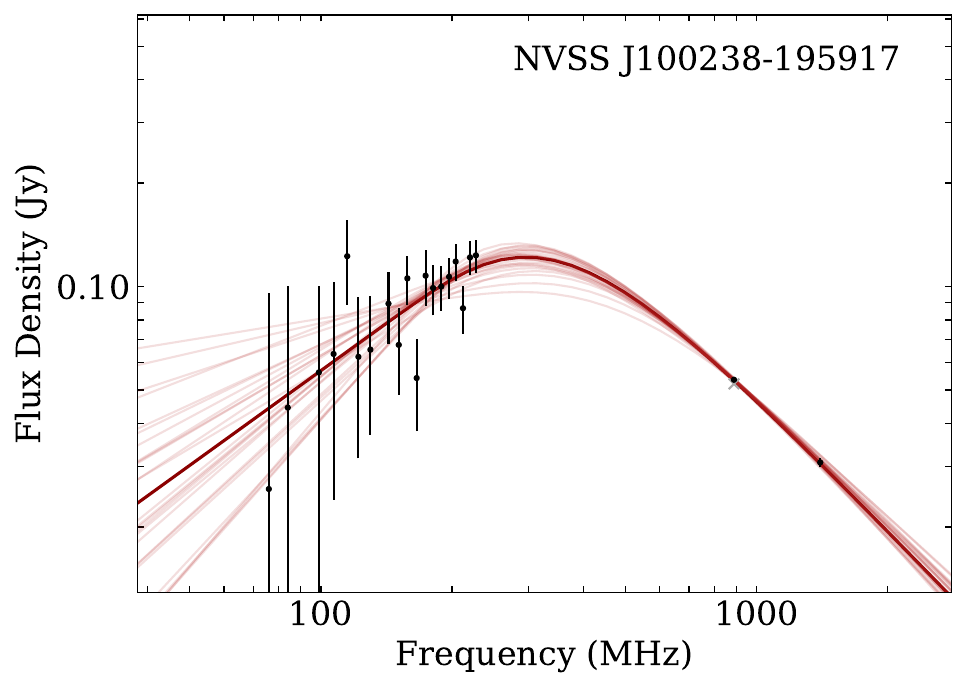}
\caption{17.\ NVSS\,J100238-195917}
\end{subfigure}

\begin{subfigure}[b]{\textwidth}
\includegraphics[width=0.53\linewidth]{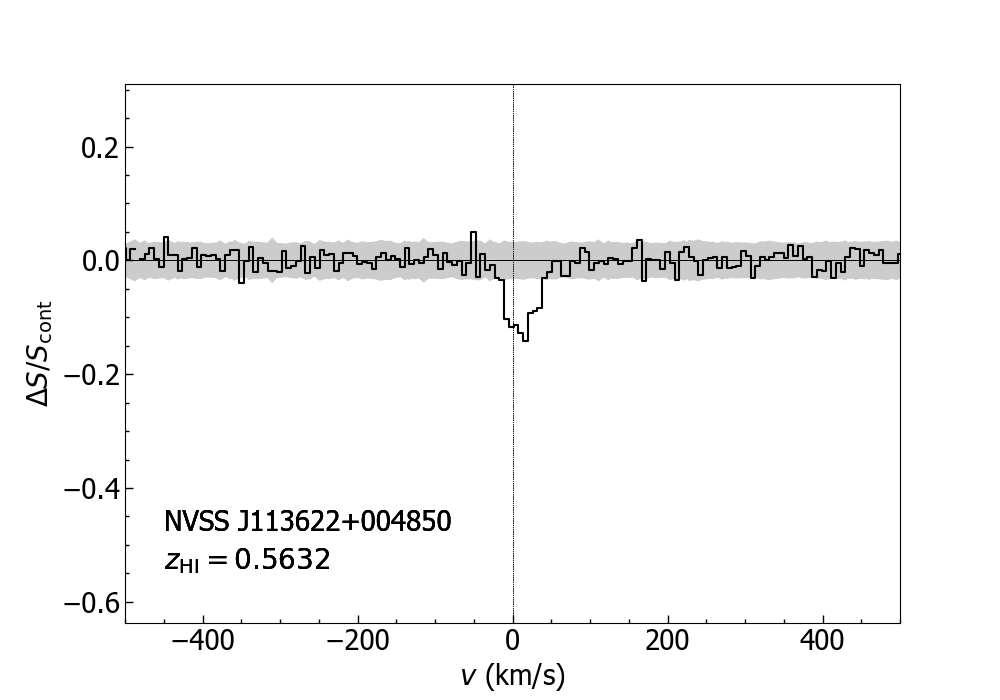}
\includegraphics[width=0.47\linewidth]{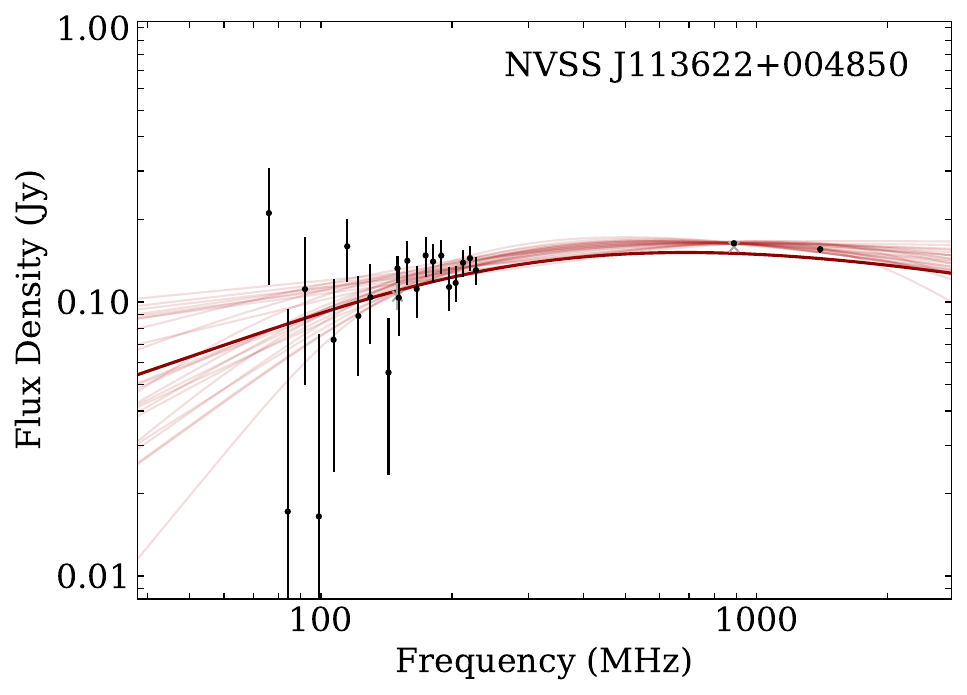}
\caption{18.\ NVSS\,J113622+004850}
\end{subfigure}
\caption{(continued) Spectral-line profiles and radio SED fits for: 16.\ NVSS~J094650-202044 (top), 17.\ NVSS\,J100238-195917 (middle), and 18.\ NVSS\,J113622+004850 (bottom).}
\end{figure*}

\setcounter{figure}{3}
\begin{figure*}[!h]
\captionsetup[subfigure]{labelformat=empty}
\centering

\begin{subfigure}[b]{\textwidth}
\includegraphics[width=0.53\linewidth]{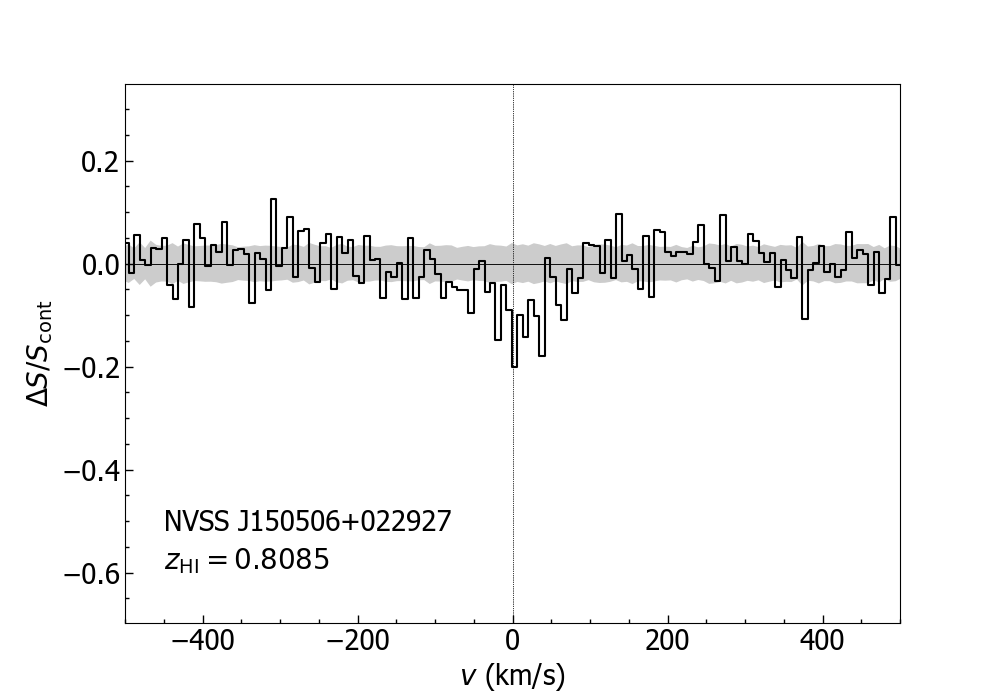}
\includegraphics[width=0.47\linewidth]{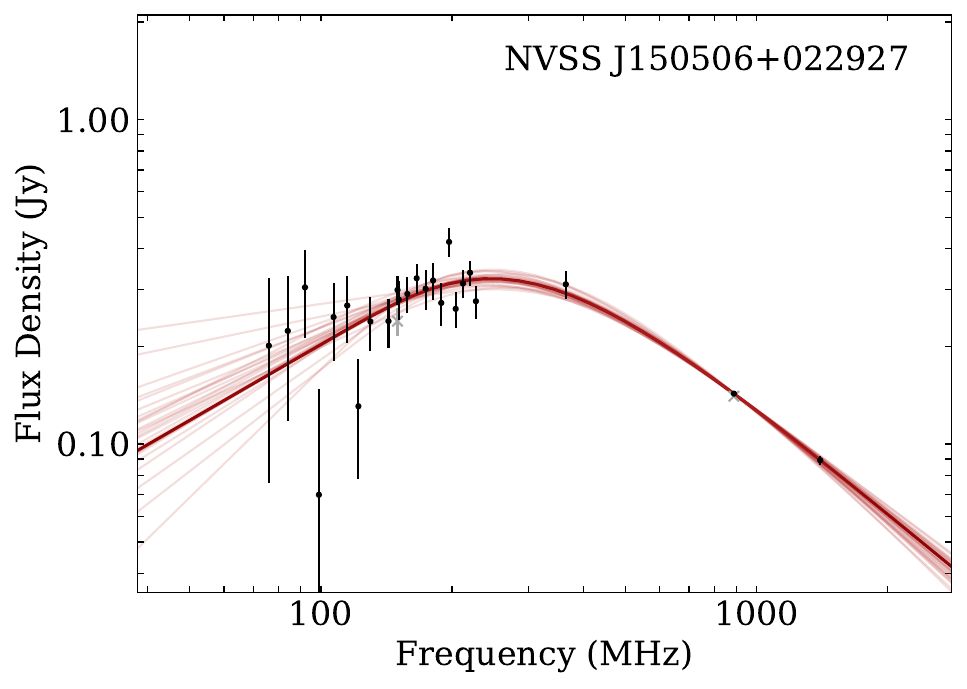}
\caption{19.\ NVSS\,J150506+022927}
\end{subfigure}

\begin{subfigure}[b]{\textwidth}
\includegraphics[width=0.53\linewidth]{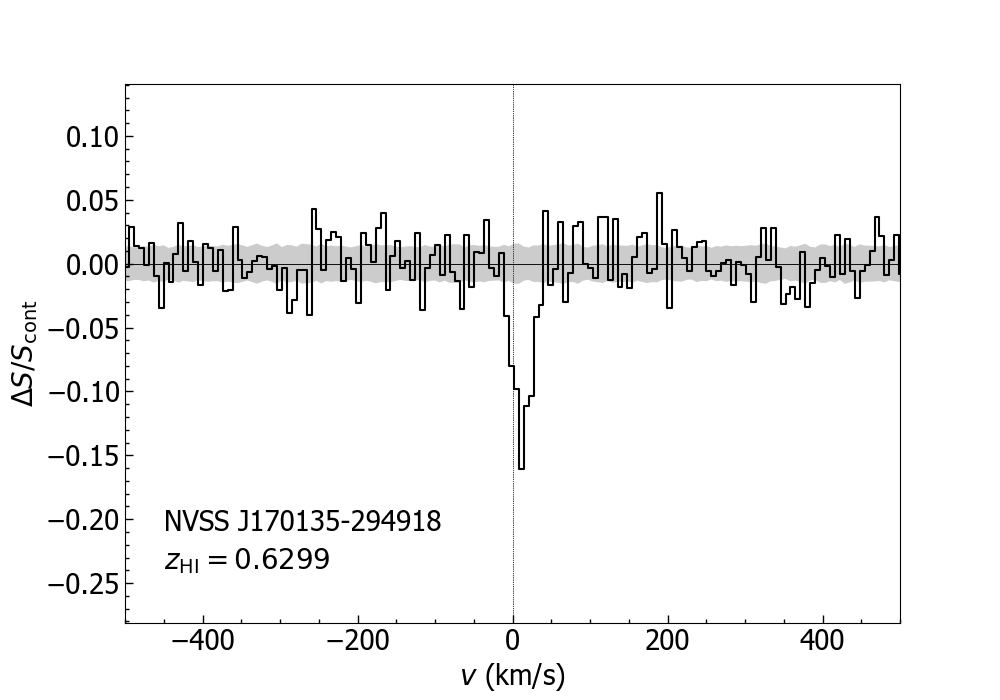}
\includegraphics[width=0.47\linewidth]{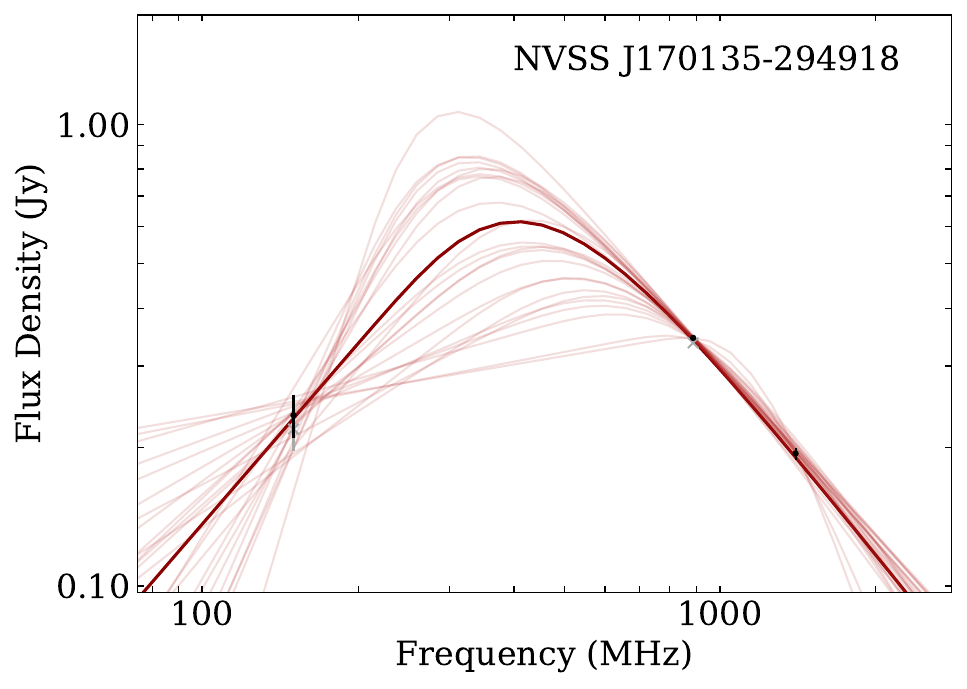}
\caption{20.\ NVSS\,J170135-294918}
\end{subfigure}

\begin{subfigure}[b]{\textwidth}
\includegraphics[width=0.53\linewidth]{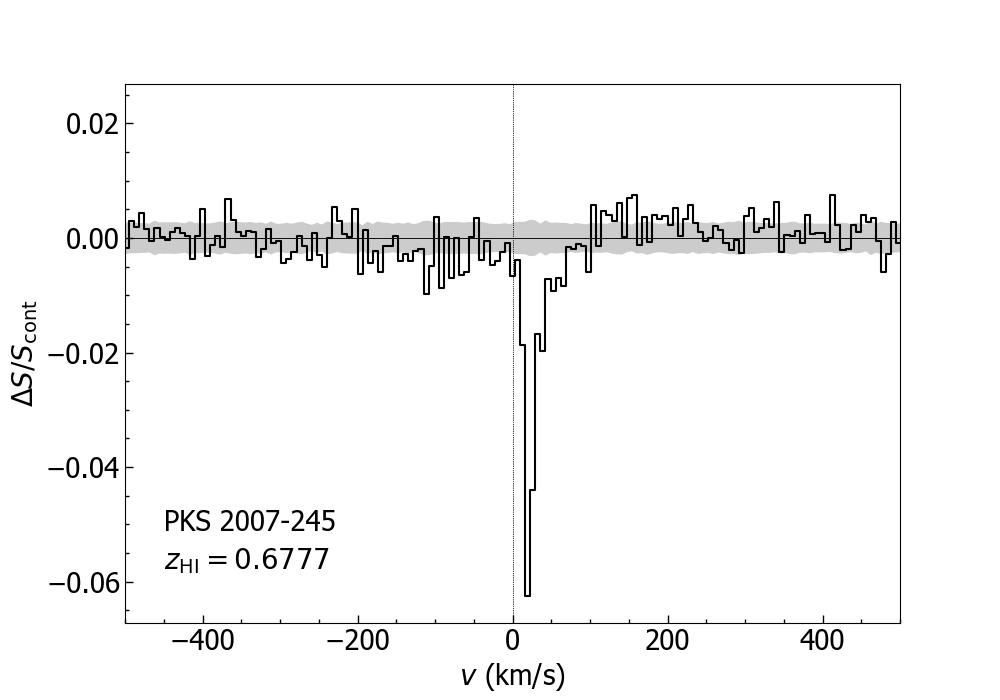}
\includegraphics[width=0.47\linewidth]
{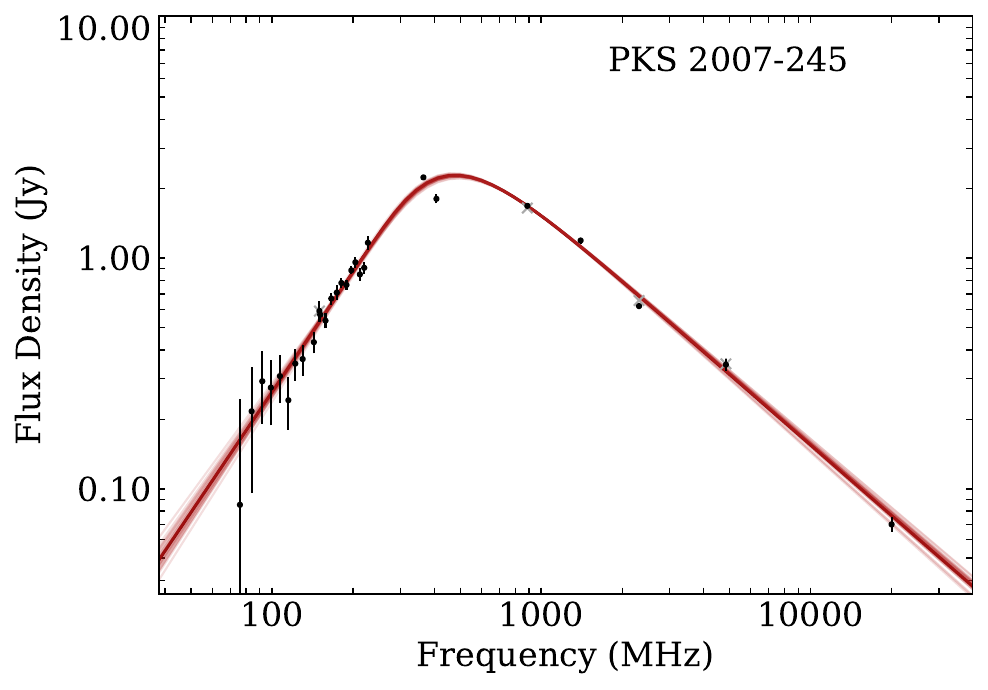}
\caption{21.\ PKS\,2007-245}
\end{subfigure}
\caption{(continued) Spectral-line profiles and radio SED fits for: 19.\ NVSS\,J150506+022927 (top), 20.\ NVSS\,J170135-294918 (middle), and 21.\ PKS\,2007-245 (bottom).}
\end{figure*}

\setcounter{figure}{3}
\begin{figure*}[!h]
\captionsetup[subfigure]{labelformat=empty}
\centering

\begin{subfigure}[b]{\textwidth}
\includegraphics[width=0.53\linewidth]{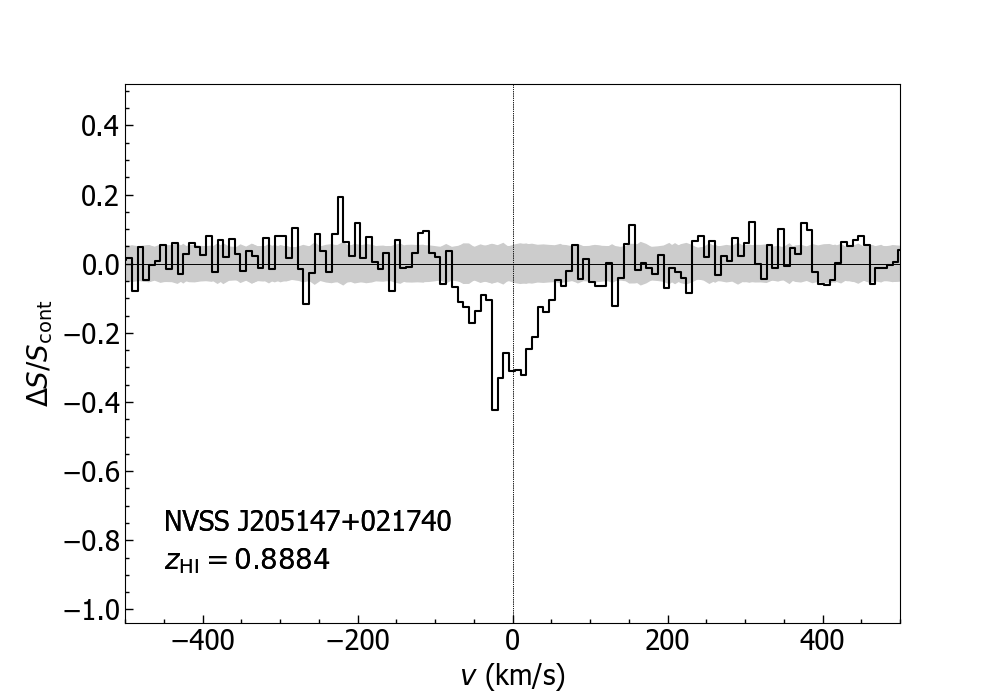}
\includegraphics[width=0.47\linewidth]{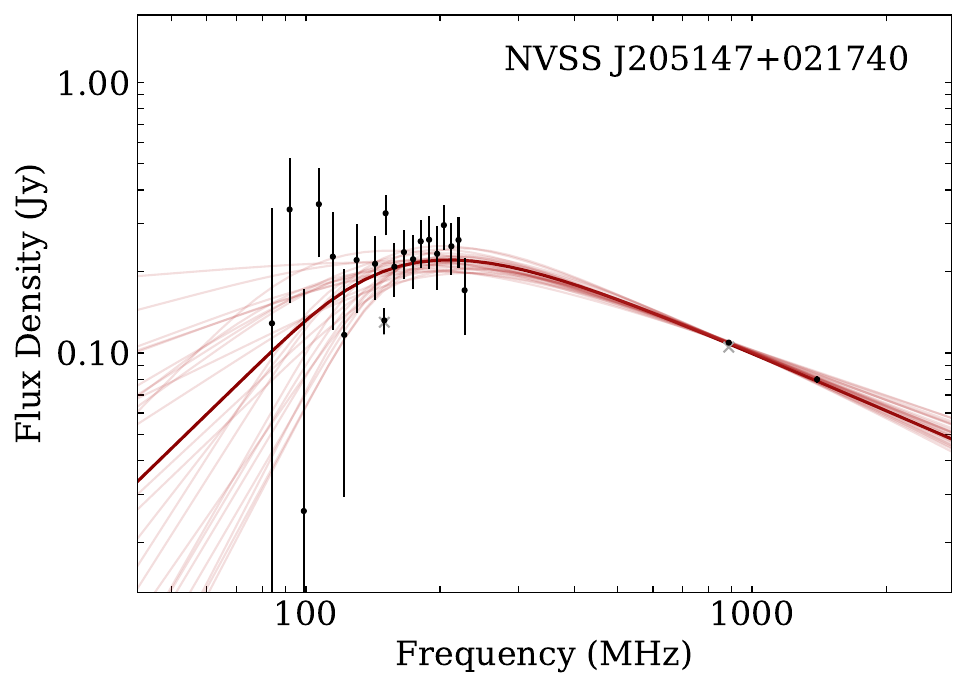}
\caption{22.\ NVSS\,205147+021740}
\end{subfigure}

\begin{subfigure}[b]{\textwidth}
\includegraphics[width=0.53\linewidth]{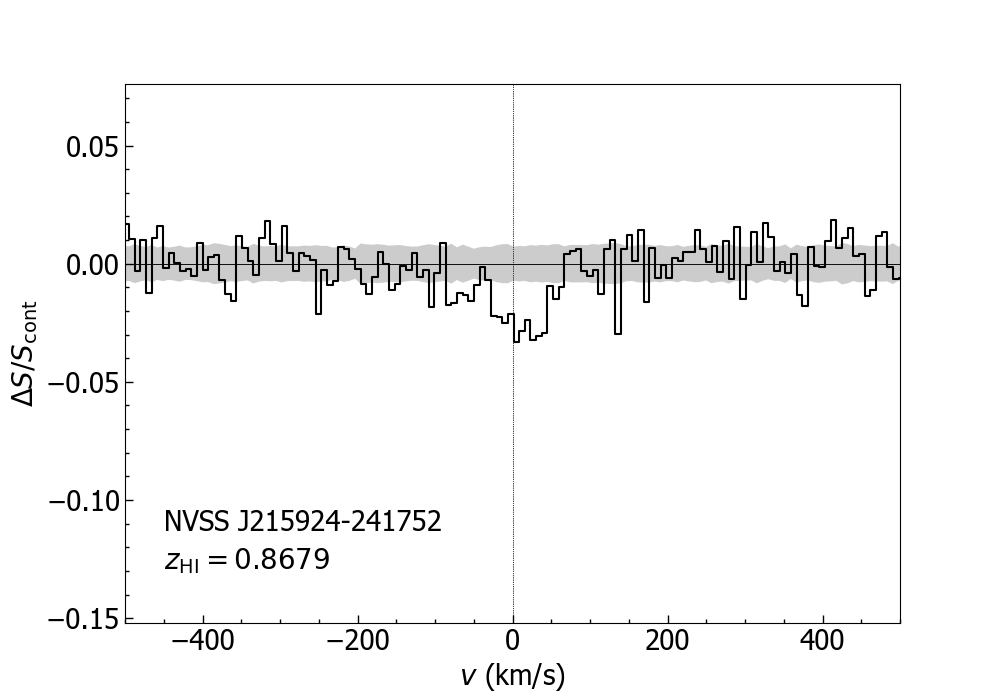}
\includegraphics[width=0.47\linewidth]{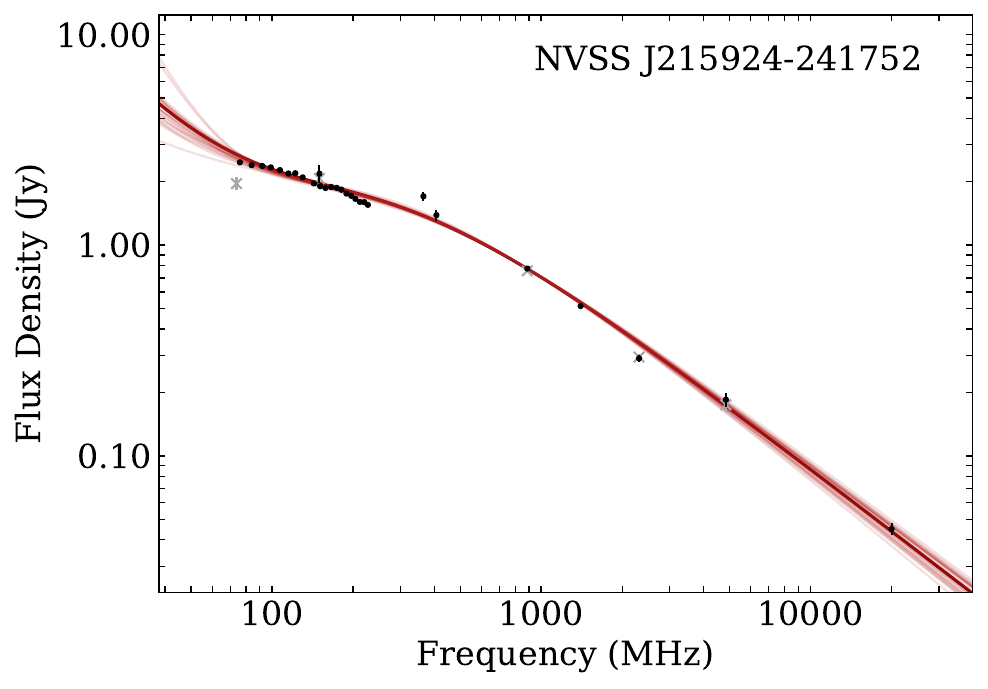}
\caption{23.\ NVSS\,J215924-241752}
\end{subfigure}

\begin{subfigure}[b]{\textwidth}
\includegraphics[width=0.53\linewidth]{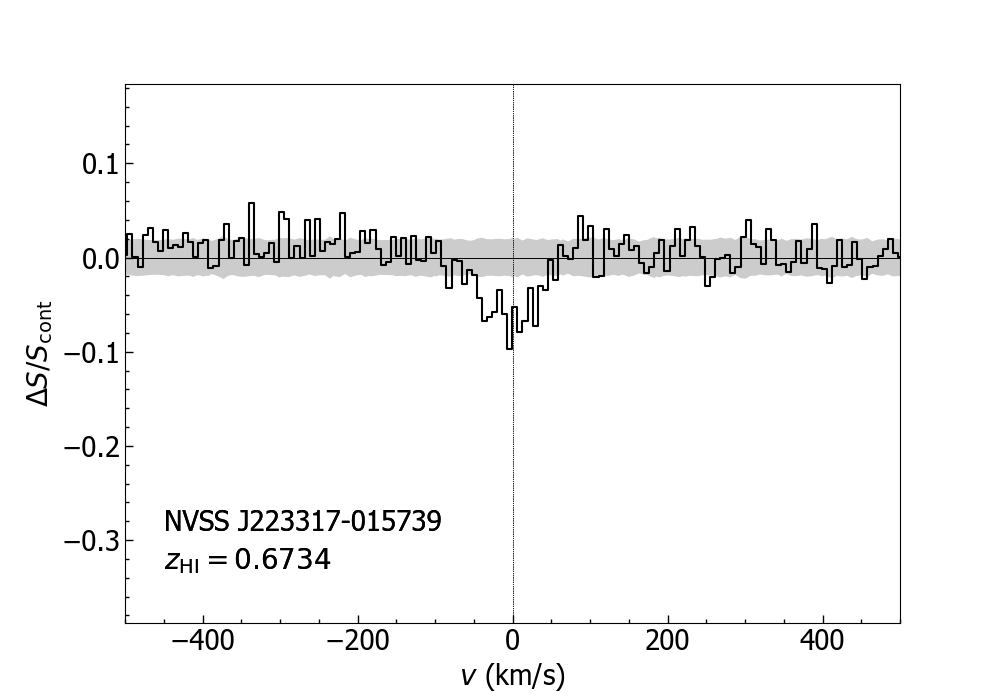}
\includegraphics[width=0.47\linewidth]{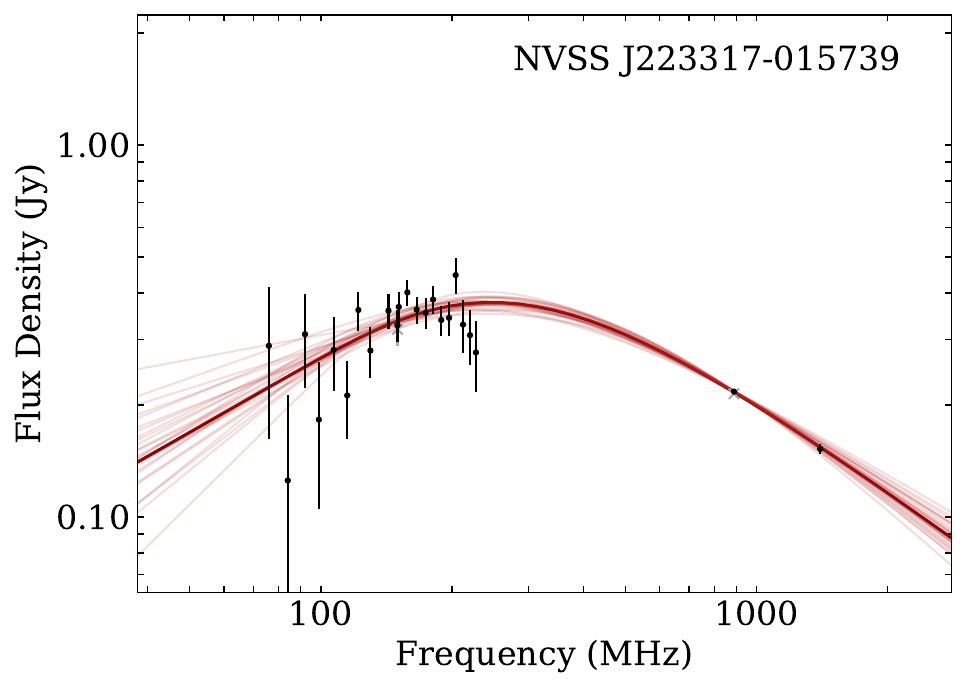}
\caption{24.\ NVSS\,J223317-015739}
\end{subfigure}
\caption{(continued) Spectral-line profiles and radio SED fits for: 22.\ NVSS\,205147+021740 (top), 23.\ NVSS\,J215924-241752 (middle), and 24.\ NVSS\,J223317-015739 (bottom).}
\end{figure*}

\setcounter{figure}{3}
\begin{figure*}[!h]
\captionsetup[subfigure]{labelformat=empty}
\centering

\begin{subfigure}[b]{\textwidth}
\includegraphics[width=0.53\linewidth]{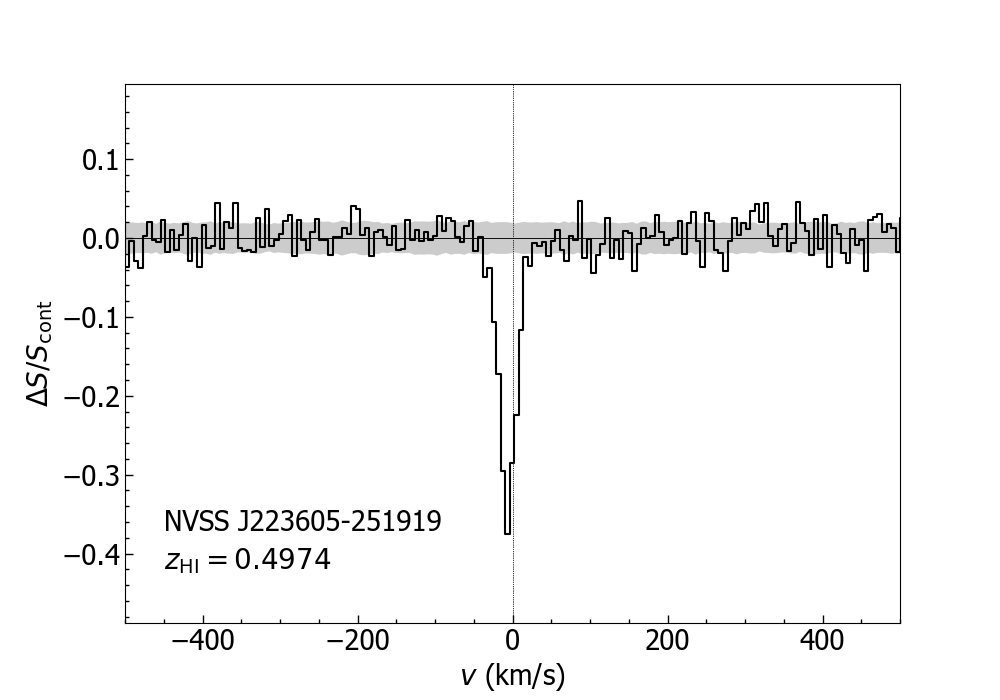}
\includegraphics[width=0.47\linewidth]{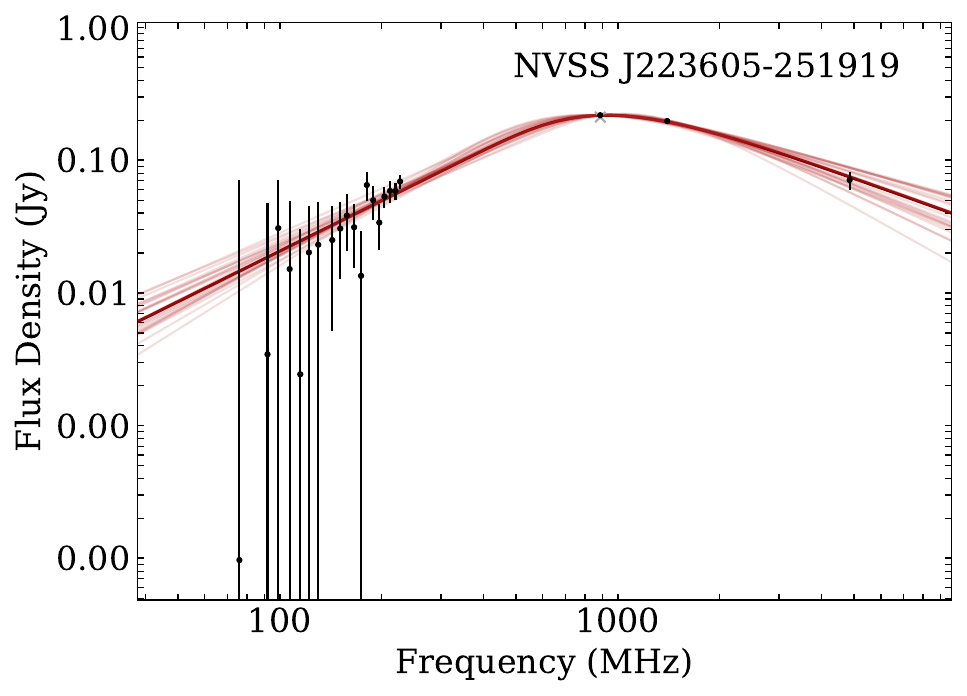}
\caption{25.\ J223605-251919}
\end{subfigure}

\begin{subfigure}[b]{\textwidth}
\includegraphics[width=0.53\linewidth]{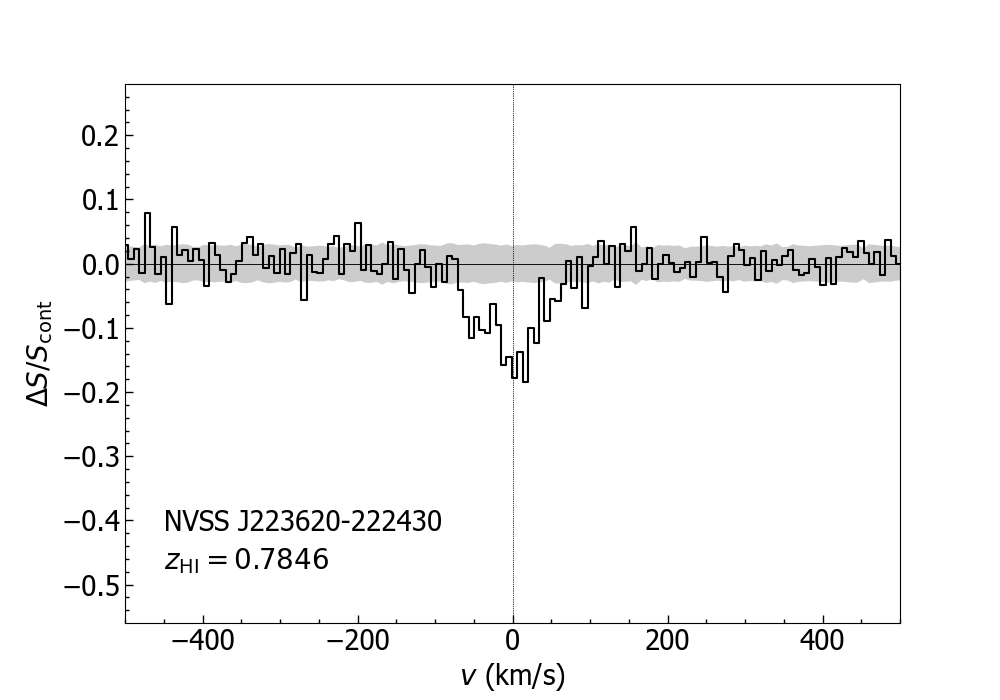}
\includegraphics[width=0.47\linewidth]{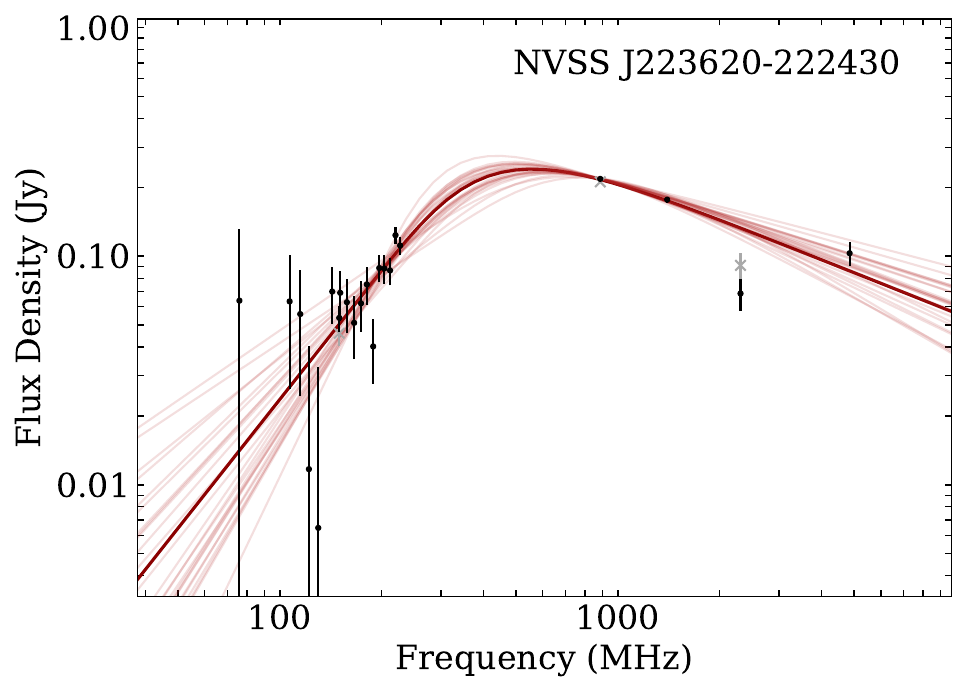}
\caption{26.\ NVSS\,J223620-222430}
\end{subfigure}

\begin{subfigure}[b]{\textwidth}
\includegraphics[width=0.53\linewidth]{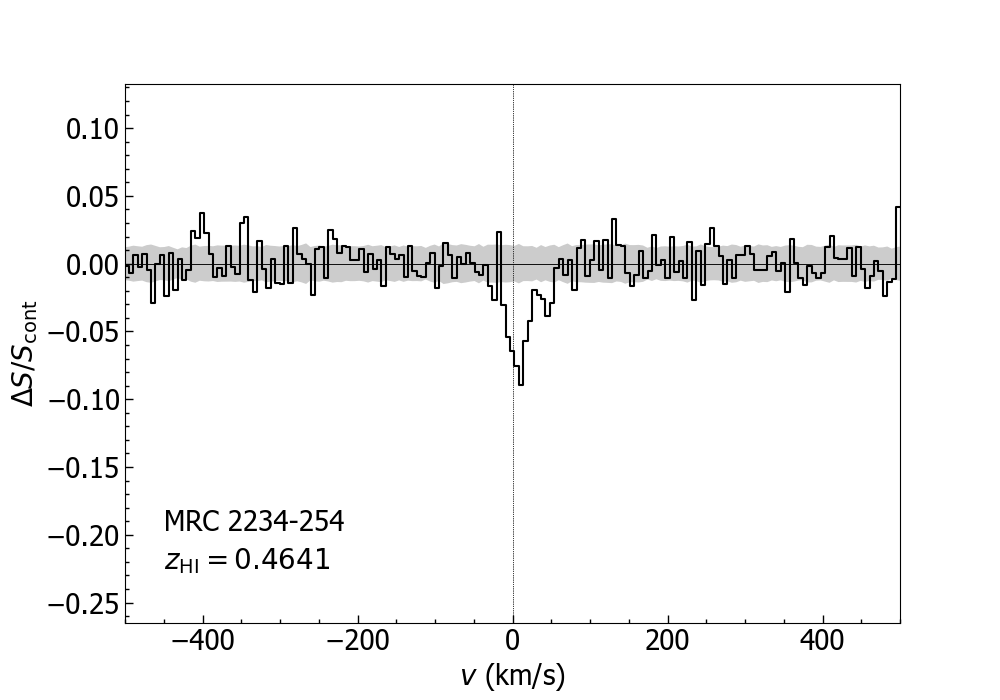}
\includegraphics[width=0.47\linewidth]
{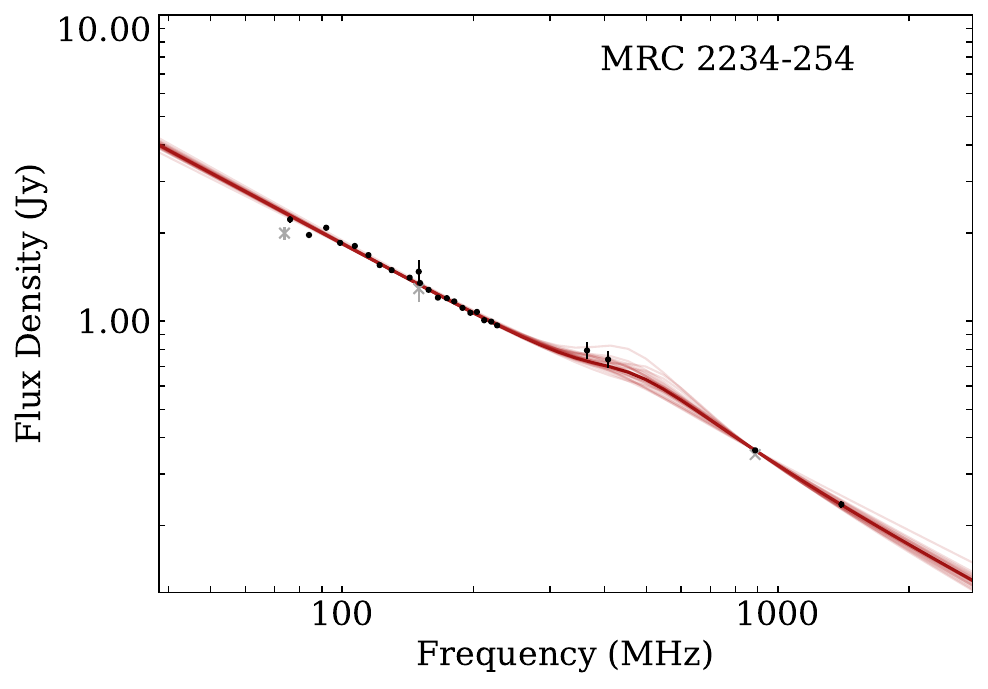}
\caption{27.\ MRC\,2234-254}
\end{subfigure}
\caption{(continued) Spectral-line profiles and radio SED fits for: 25.\ J223605-251919 (top), 26.\ NVSS\,J223620-222430 (middle), and 27.\ MRC\,2234-254 (bottom).}
\end{figure*}

\setcounter{figure}{3}
\begin{figure*}[!h]
\captionsetup[subfigure]{labelformat=empty}
\centering

\begin{subfigure}[b]{\textwidth}
\includegraphics[width=0.53\linewidth]{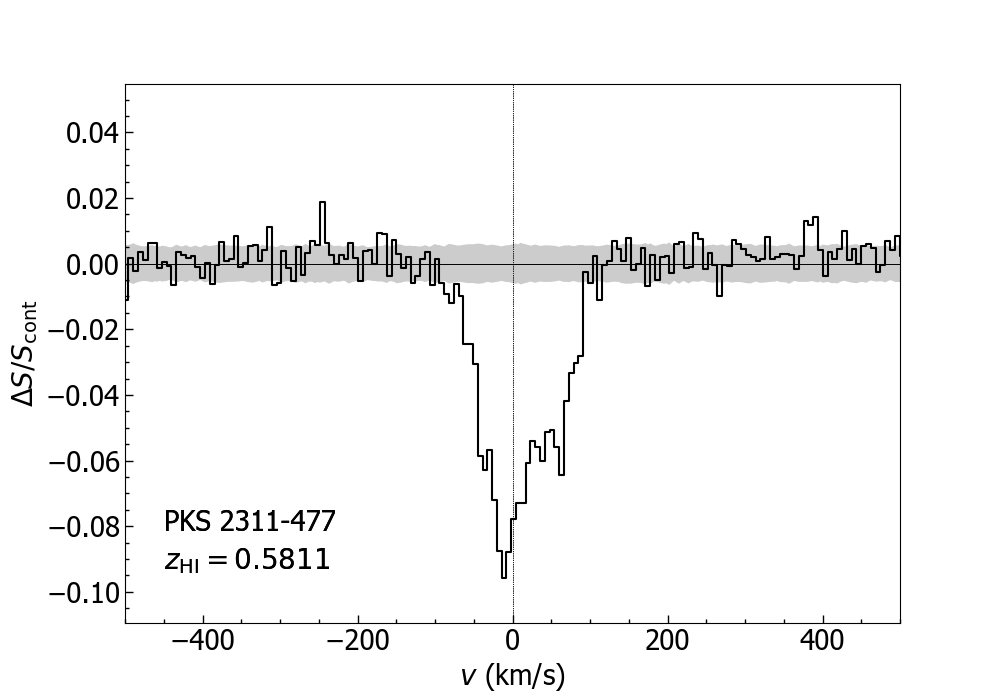}
\includegraphics[width=0.47\linewidth]
{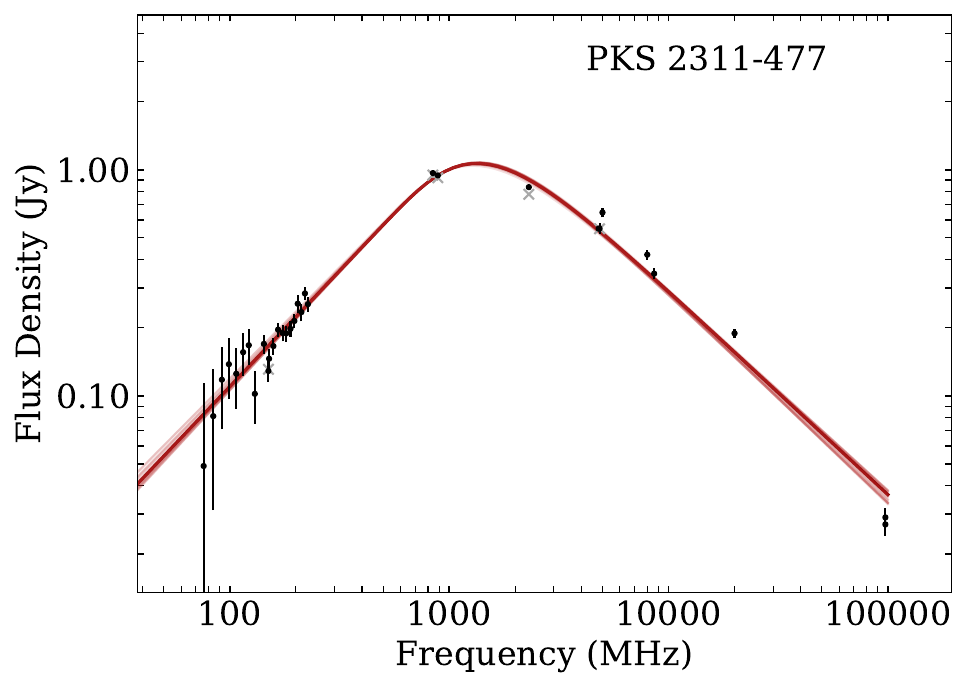}
\caption{28.\ PKS\,2311-477}
\end{subfigure}

\begin{subfigure}[b]{\textwidth}
\includegraphics[width=0.53\linewidth]{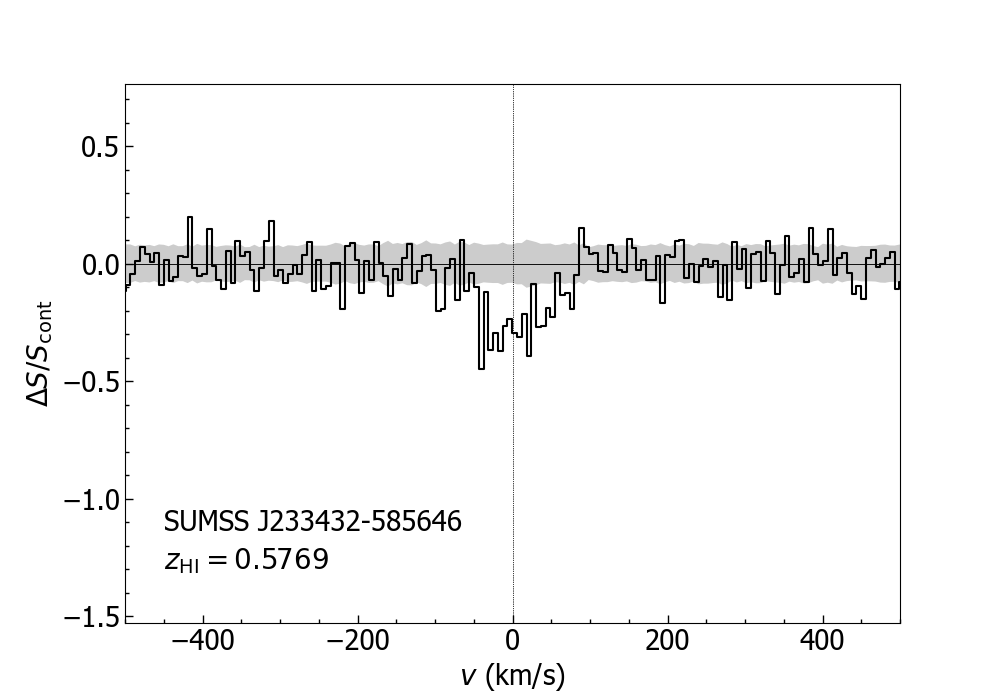}
\includegraphics[width=0.47\linewidth]{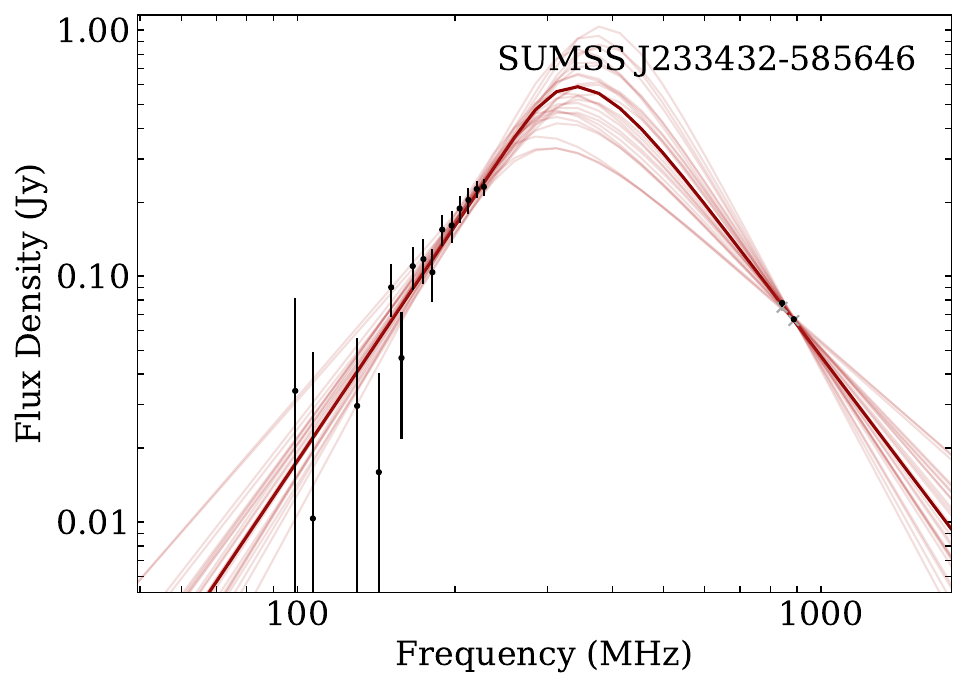}
\caption{29.\ SUMSS\,J233432-585646}
\end{subfigure}

\begin{subfigure}[b]{\textwidth}
\includegraphics[width=0.53\linewidth]{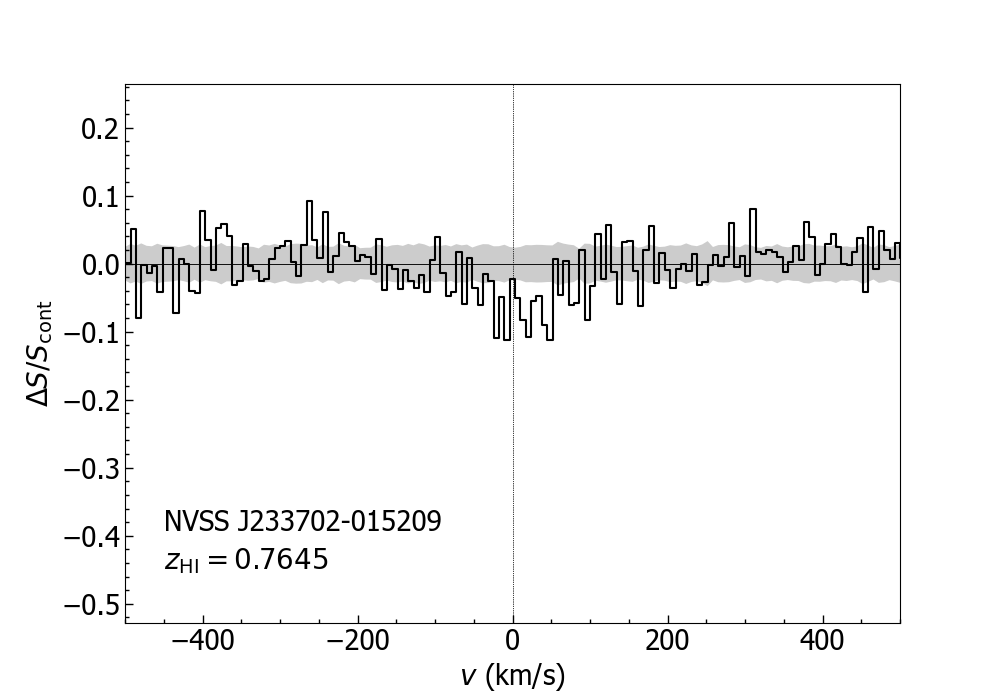}
\includegraphics[width=0.47\linewidth]{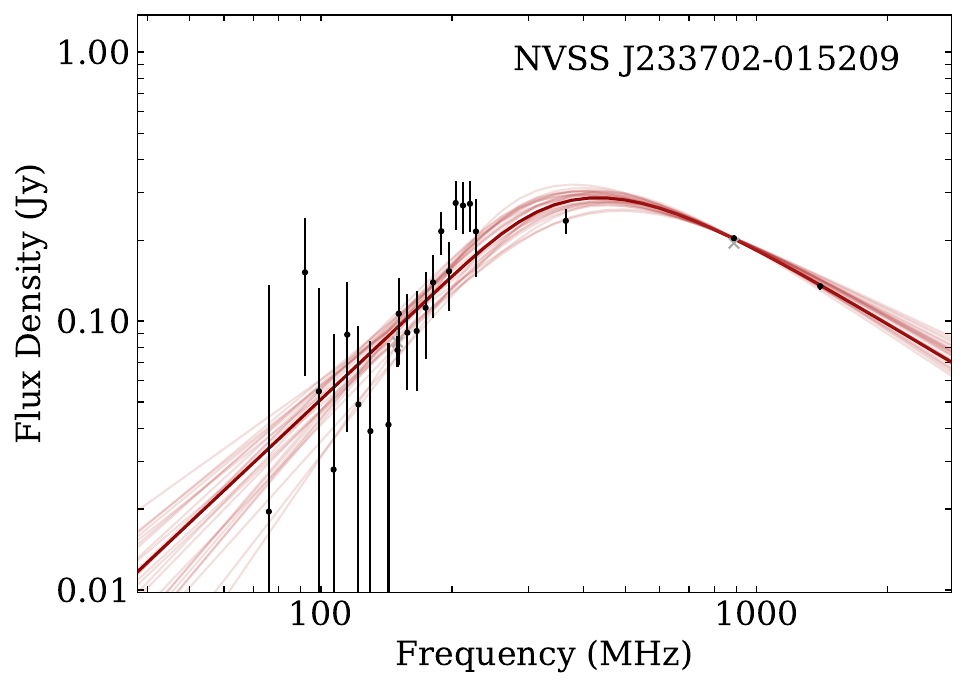}
\caption{30.\ NVSS\,J233702-015209}
\end{subfigure}
\caption{(continued) Spectral-line profiles and radio SED fits for: 28.\ PKS\,2311-477 (top), 29.\ SUMSS\,J233432-585646 (middle), and 30.\ NVSS\,J233702-015209 (bottom).}
\end{figure*}

\clearpage
\section{Individual detection and linefinder results}
\label{sec:linefinder}

\clearpage
\onecolumn 
\begin{table}
\centering
\tabcolsep 2pt
\caption{FLASH Pilot Survey detections and FLASHfinder results; (1) ID; (2) Selavy ID; (3) Component name; (4) Field name; (5) Right Ascension (deg); (6) Declination (deg); (7) Continuum flux density for this component; (8) \hi redshift; (9) Peak optical depth; (10) Integrated optical depth; (11) Linewidth; (12) ln\,(B); (13) Optical redshift; (14) Closest radio source. These spectra were combined into a single spectrum per source as listed in Table~\ref{tab:finder2}.}
\label{tab:finder}

\begin{tabular}[ht]{clcl rrcr rrrr cl}
\toprule					
\multicolumn{1}{c}{ID}	&	\multicolumn{1}{c}{Selavy ID}		&	\multicolumn{1}{c}{Component name}	&	\multicolumn{1}{c}{FLASH field}		&	\multicolumn{1}{c}{RA} &	\multicolumn{1}{c}{DEC} &	\multicolumn{1}{c}{S$_{\rm cont.}$}	& \multicolumn{1}{c}{$z_{\rm HI}$}	&	\multicolumn{1}{c}{$\tau_{\rm peak}$}	&	\multicolumn{1}{c}{$\tau_{\rm int}$}	&	\multicolumn{1}{c}{Linewidth} &	\multicolumn{1}{c}{ln\,(B)}	& \multicolumn{1}{c}{$z_{\rm opt}$}	&	\multicolumn{1}{c}{Closest radio source} \\

\multicolumn{1}{c}{}	&	\multicolumn{1}{c}{}		&	\multicolumn{1}{c}{}	&	\multicolumn{1}{c}{}	& \multicolumn{1}{c}{(deg)}	&	\multicolumn{1}{c}{(deg)} & \multicolumn{1}{c}{(Jy)}	& \multicolumn{1}{c}{}	&	\multicolumn{1}{c}{}	& \multicolumn{1}{c}{}	&	\multicolumn{1}{c}{(km s$^{-1}$)}	&	\multicolumn{1}{c}{(km s$^{-1}$)}	& \\

\multicolumn{1}{c}{(1)}	&	\multicolumn{1}{c}{(2)} &	\multicolumn{1}{c}{(3)}	&	\multicolumn{1}{c}{(4)}	&	\multicolumn{1}{c}{(5)}	&	\multicolumn{1}{c}{(6)}	&	\multicolumn{1}{c}{(7)}	&	\multicolumn{1}{c}{(8)}	&	\multicolumn{1}{c}{(9)}	&	\multicolumn{1}{c}{(10)} & \multicolumn{1}{c}{(11)} & \multicolumn{1}{c}{(12)} & \multicolumn{1}{c}{(13)} & \multicolumn{1}{c}{(14)} \\
\hline

1	&	SB34581	9a	&	J001425-020556	&	525	&3.606475	&	-2.099110&	0.694	& 0.6785	&	0.216	&	7.90	&	36.6	&	1237.4 & $-$ & PKS\,0011-023 \\ 
\hline				
2	&	SB34581	149a	&	J002331+010114	&	525	&5.881687	&	1.020562&	0.067	& 0.5159	&	0.432	&	32.4	&	74.7	&	76.9 & $-$ & NVSS\,J002331+010114 \\ 
\hline
3	&	SB13290	19a	&	J002604-475618	&	123P	& 6.519942	&	-47.938337 &	0.400	&	0.6745	&	$-$	&	$-$	&	$-$	&	$-$	& $-$	&	MRC\,0023-482\\ 
3	&	SB37448	19a	&	J002604-475617	&	123	&	6.519433	&	-47.938252 & 0.358	& 0.6746	&	0.081	&	4.07	&	50.3	&	44.4 & $-$	&	MRC\,0023-482\\	
\hline					
4	&	SB13281	64a	&	J014141-231510	&	306P	&	25.423772	&	-23.253037 & 0.138	&	0.6707	&	0.161	&	20.02	&	124.8	&	77.5 & $-$ & NVSS\,J014141-231511 \\
4	&	SB37475	67a	&	J014141-231508	&	306	&25.422827	&	-23.252426&	0.136	& 0.6706	&	0.100	&	14.27	&	143.0	&	34.0 & $-$ & NVSS\,J014141-231511\\	
\hline
5	&	SB13281	103a	&	J015516-251423	&	306P &	28.820245 & -25.239974 & 0.099	& 0.7251	&	1.802	&	95.66	&	53.1 &	1174.7 &  $-$ & NVSS\,J015516-251423 \\	
5	&	SB37475	95a	&	J015516-251422	&	306	&28.818926	&	-25.239542&	0.106	& 0.7251	&	1.891	&	95.39	&	50.5	&	1065.4	& $-$	& NVSS\,J015516-251423 \\	
\hline
6	&	SB13269	2a	&	J025544-254741	&	309P	&43.935136 & -25.794825 &	0.604	& 0.6564	&	0.052	&	1.08	&	20.7	&	32.4 & $-$	& PKS\,0253-259 \\	
6	&	SB15212	9a	&	J025544-254739	&	308P	& 43.934931 & -25.794442 &	0.805	& 0.6564	&	0.051	&	1.07	&	21.3	&	74.7 & $-$	& PKS\,0253-259 \\	
\hline
7	&	SB37453	1a	&	J040757-275705	&	311	&61.991078	&	-27.951503&	1.278	& 0.7280	&	0.032	&	2.97	&	94.0	&	254.5 & $-$ & PKS 0405-280 \\ 
\hline
8	&	SB15215	27a	&	J045501-423858	&	170P	& 73.755747 & -42.649620 &	0.288	& 0.6525	&	0.051	&	7.38	&	146.4	&	83.3 & $-$ & SUMSS\,J045501-423858 \\	
\hline
9	&	SB34570	50a	&	J051805-245502	&	314	&79.524788	&	-24.917494&	0.195	& 0.5538	&	0.141	&	16.04	&	114.0	&	272.3	& $-$ & NVSS J051806-245502 \\ 
9	&	SB41061	50a	&	J051805-245502	&	314	&79.525186	&	-24.917212&	0.180	& 0.5538	&	0.156	&	18.61	&	119.0	&	451.8	& $-$ & NVSS J051806-245502 \\
9	&	SB41065	51a	&	J051805-245502	&	314	&79.525202	&	-24.917451&	0.184	& 0.5538	&	0.169	&	18.45	&	109.1	&	500.7	& $-$ & NVSS J051806-245502 \\
\hline
10	&	SB41061	1a	&	J053354-234430	&	314	&83.477481	&	-23.741721&	1.742	& 0.8507	&	0.415	&	81.7	&	197.0	&	36242.7	 & $-$ & MRC 0531-237 \\ 
10	&	SB41065	1a	&	J053354-234430	&	314	&83.477565	&	-23.741920&	1.751	& 0.8508	&	0.359	&	71.48	&	199.1	&	37146.9	& $-$ & MRC 0531-237 \\
\hline 	
11	&	SB13283	228a	&	J090331+010847	&	G09B	& 135.881305 & 1.146627 &	0.053	& 0.5218	&	3.079	&	187.78	&	60.9	&	466.2 & $-$ &  NVSS\,J090331+010846 \\	
11	&	SB34559	191a	&	J090331+010846	&	546	&135.881691	&	1.146269&	0.070	& 0.5218	&	1.344	&	82.19	&	61.1	&	485.4	& $-$ &  NVSS\,J090331+010846 \\	
\hline
12	&	SB13283	91a	&	J090425+012015	&	G09B	& 136.106406 & 1.337547 &	0.087	& 0.8004	&	0.924	&	36.80 &	39.9 &	303.6 & $-$ & NVSS\,J090425+012015 \\		
12	&	SB34549	120a	&	J090425+012013	&	547	&136.107156	&	1.337083&	0.095	& 0.8004	&	1.421	&	28.07	&	19.4	&	35.6	& $-$	& NVSS\,J090425+012015 \\
12	&	SB34559	117a	&	J090425+012013	&	546	&136.106701	&	1.337055&	0.080	& 0.8004	&	1.588	&	46.98	&	29.5	&	238.7	& $-$ &  NVSS\,J090425+012015 \\	
\hline					
13	&	SB11068	403a	&	J091256+030020	&	G09B\_long	& 138.233859 & 3.005576 &	0.033	& 0.8590	&	1.253	&	64.97	&	51.7	&	103.1 & $-$ & NVSS\,J091256+030021 \\
\hline			
14	&	SB13271	2a	&	J092011+175324	&	719P	& 140.046315 & 17.890218 &	1.752	& 0.9040	&	0.009	&	1.90	&	218.0	&	65.6 & $-$ & PKS\,0917+18 \\ 	
14	&	SB41066	3a	&	J092011+175324	&	719	&140.046507	&	17.890180&	1.800	& 0.9044	&	0.01	&	1.14	&	119.6	&	53.7	& $-$ & PKS\,0917+18 \\	
\hline					
15	&	SB13271	57a	&	J092012+161239	&	719P	& 140.051498 & 16.211010 &	0.105	& 0.4362	&	0.082	&	7.37	&	89.2	&	46.9 & $-$ & NVSS\,J092012+161238 \\ 	
15	&	SB34560	55a	&	J092012+161236	&	719	&140.052614	&	16.210097&	0.206	& 0.4362	&	0.139	&	3.28	&	23.8	&	14.9	& $-$ & NVSS\,J092012+161238 \\	
15	&	SB41066	56a	&	J092012+161238	&	719	&140.052235	&	16.210709&	0.202	& 0.4362	&	0.169	&	4.88	&	29.1	&	63.3	& $-$ & NVSS\,J092012+161238 \\	
\hline
16	&	SB34571	8a	&	J094650-202045	&	377	&146.709164	&	-20.345918&	0.550	& 0.9134	&	0.135	&	10.6	&	78.2	&	249.2 & $-$ & NVSS J094650-202044 \\ 
\hline		
17	&	SB34561	188a	&	J100238-195919	&	378	&150.660845	&	-19.988637&	0.057	& 0.4815	&	0.5	&	17.16	&	34.2	&	35.2 & $-$ & NVSS J100238-195917 \\ 
\hline	
18	&	SB13306	65a	&	J113622+004851	&	G12A\_long	& 174.091747 & 0.814267 &	0.165	& 0.5630	&	0.142	&	7.52	&	52.9	&	158.9	& $-$ & NVSS\,J113622+004850 \\
18	&	SB13334	64a	&	J113622+004850	&	G12A	& 174.091646 & 0.814056 &	0.130	& 0.5630	&	0.130 &	6.71	&	51.8	&	28.9 & $-$ & NVSS\,J113622+004850 \\ 
18	&	SB34572	65a	&	J113622+004850	&	553	&174.091546	&	0.814248&	0.166	& 0.5630	&	0.083	&	3.74	&	45.0	&	13.3 & $-$ & NVSS\,J113622+004850 \\	
\hline
19	&	SB34576	62a	&	J150506+022928	&	561	&226.278737	&	2.491188&	0.150	& 0.8085	&	0.112	&	11.96	&	107.4	&	35.4 & $-$ & NVSS J150506+022927 \\ 
\hline	
20	&	SB34552	19a	&	J170135-294917	&	287	&255.398288	&	-29.821407&	0.403	& 0.6299	&	0.149	&	4.44 &	29.9 &	144.0 & $-$ & NVSS J170135-294918 \\ 
\hline
21	&	SB13372	1a	&	J201045-242545	&	J2022-2507	& 302.688009 & -24.429319 &	1.827	&	0.6778	&	0.007	&	1.36	&	19.0	&	358.5 & $-$ & PKS\,2007-245 \\	
\hline
22	&	SB34566	100a	&	J205147+021738	& 575	&312.949592	&	2.294018&	0.115	& 0.8883	&	0.361	& 30.68	&	84.9 &	123.9 & $-$ & NVSS J205147+021740 \\ 
22	&	SB34577	94a	&	J205147+021740	&	575	&312.949333	&	2.294467&	0.107	& 0.8884	&	0.383	&	28.9	&	75.6	&	125.3& $-$ & NVSS J205147+021740 \\
\hline	
23	&	SB10849	8a	&	J215924-241752	&	351P	& 329.853714 & -24.297950 &	0.806	& 0.8680	&	0.028	&	2.68	&	94.5	&	50.3 & $-$ & NVSS J215924-241752 \\ 
\hline
24	&	SB34597	47a	&	J223317-015739	&	579	&338.321991	&	-1.960909&	0.231	& 0.6734	&	0.076	&	5.45	&	71.7	&	43.8 & $-$ & NVSS J223317-015739 \\ 
24	&	SB42298	45a	&	J223317-015739	&	579	&338.322054	&	-1.961156&	0.239	& 0.6733	&	0.073	&	6.48	&	89.2	&	56.3 & $-$ & NVSS J223317-015739 \\
\hline
25	&	SB11051	35a	&	J223605-251918	&	352P	& 339.023937 & -25.321700 &	0.215	& 0.4974	&	0.415	&	11.27	&	27.2	&	512.6 & $-$ & NVSS\,J223605-251919 \\ 
\hline 			
26	&	SB11051	38a	&	J223619-222429	&	352P	& 339.082984 & -22.408156 &	0.207	& 0.7846	&	0.158	&	15.35	&	97.4	&	113.6 & $-$ & NVSS\,J223620-222430 \\ 
\hline		
27	&	SB11051	17a	&	J223722-251003	&	352P	& 339.344983	&	-25.167529 &	0.343	& 0.4641	&	0.072	&	3.06	&	42.7	&	58.4 & $-$ & MRC\,2234-254 \\ 
\hline		
28	&	SB15873	5a	&	J231351-472911	&	160P	&348.466233	&	-47.486505&	1.044	& 0.5811	& 0.078	&	8.80 &	113.0 & 1706.2 & $-$ & PKS\,2311-477 \\ 
28	&	SB34939	5a	&	J231351-472911	&	160	&	348.465800	&	-47.486392&0.952	& 0.5811	&	0.072	&	7.89	&	110.3	&	945.6	& $-$ & PKS\,2311-477 \\	
\hline			
29	&	SB13296	158a	&	J233432-585646	&	121P	&353.634345	&	-58.946112&	0.074	& 0.5769	&	0.366	&	37.43	&	101.6	&	65.4 & $-$ & SUMSS\,J233432-585646 \\ 	
\hline			
30	&	SB34556	46a	&	J233703-015210	&	582	&354.262692	&	-1.869491&	0.232	& 0.7645	&	0.072	&	9.24	&	128.4	&	34.3 & $-$ & NVSS J233702-015209 \\
\bottomrule
\end{tabular}
\end{table}

\clearpage
\section{Observation logs with validation flags}
\label{sec:obslogs}

\clearpage
\onecolumn 
\begin{table}[!ht]
\footnotesize
\tabcolsep 4pt
\renewcommand{\arraystretch}{0.8} 
\caption{Log of observations for FLASH Pilot Survey 1.}
\begin{tabular}{lllrcclccl} 
\toprule
\multicolumn{1}{c}{Field} &  \multicolumn{1}{c}{RA} & \multicolumn{1}{c}{Dec} &  \multicolumn{2}{c}{Observation} & \multicolumn{1}{c}{BeamInt} & \multicolumn{1}{c}{CASDA} & \multicolumn{2}{c}{rms noise} & \multicolumn{1}{c}{Notes} \\
 & \multicolumn{2}{c}{(J2000)} & \multicolumn{1}{c}{Date} & \multicolumn{1}{c}{t (h)} & \multicolumn{1}{c}{(MHz)}  & \multicolumn{1}{c}{SBID} & \multicolumn{1}{c}{Cont} &\multicolumn{1}{c}{Line} & \multicolumn{1}{c}{CASDA validation flag}\\
&&&&&&& (mJy/beam) & (mJy/beam/ch) \\
\multicolumn{1}{c}{(1)} & \multicolumn{1}{c}{(2)} & \multicolumn{1}{c}{(3)} & \multicolumn{1}{c}{(4)} & \multicolumn{1}{c}{(5)} & \multicolumn{1}{c}{(6)} & \multicolumn{1}{c}{(7)} & \multicolumn{1}{c}{(8)} & \multicolumn{1}{c}{(9)} & \multicolumn{1}{c}{(10)} \\ 
\hline
\multicolumn{6}{l}{\bf FLASH survey fields observed and released in CASDA(31 fields)} \\
F302P & 00 00 00.00 & $-25$ 07 47.2 & 17-Dec-2019 &  2 & 5 & 10850 & 0.099 & 5.70 & U \\
F088P & 00 21 10.59 & $-56$ 18 22.7 & 20-Apr-2020 & 2 & 9 & 13299 & 0.083 & 5.33 & U \\ 
F303P  & 00 27 10.19  & $-25$ 07 47.2  & 05-Jan-2020  & 2 & 5 & 11053 & 0.110 & 5.70 & U \\
F123P  & 00 36 55.38  & $-50$ 05 45.3  & 19-Apr-2020  & 2 & 9 & 13290 & 0.079 & 5.32 & U \\
F304P	& 00 54 20.38 & $-$25 07 47.2  & 19-Apr-2020 & 2 & 9 & 13291 & 0.088 & 5.42 & U \\
&& \\
F089P & 01 03 31.76 & $-$56 18 22.7 & 20-Apr-2020 & 2 & 9 & 13298 & 0.082 & 5.60 & U \\
F306P & 01 48 40.75 & $-$25 07 47.2 & 18-Apr-2020	& 2 & 9 & 13281 & 0.078 & 5.36 & U \\
F165P	& 02 13 57.21 & $-$43 52 21.4 & 04-Jul-2020 & 2 & 9 & 15213 & 0.079 & 5.15 & U \\	
F307P	& 02 15 50.94 & $-$25 07 47.2 & 17-Apr-2020	& 2 & 9 & 13268 & 0.078 & 5.13 & U \\
F308P	& 02 43 01.13 & $-$25 07 47.2 & 04-Jul-2020 & 2 & 9 & 15212 & 0.082 & 5.67 & U \\
&& \\
F309P	& 03 10 11.32 & $-$25 07 47.2 & 17-Apr-2020 & 2 & 9 & 13269 & 0.089 & 5.42 & U \\
F170P   & 05 01 23.72 &	$-$43 52 21.4 & 05-Jul-2020 & 2 & 9 & 15215 & 0.083 & 5.09 & U\\
F718P	& 08 43 38.18 & $+$18 51 28.9 & 17-Apr-2020 & 2	& 9 & 13270 & 0.103 & 5.53 & U \\
F719P	& 09 09 49.09 & $+$18 51 28.9 & 17-Apr-2020 & 2	& 9 & 13271 & 0.102 & 5.31 & U\\
F607P	& 10 08 16.55 & $+$06 17 42.1 & 18-Apr-2020  & 2 & 9 & 13284 & 0.133 & 5.21 &  U\\	
&& \\
F608P	& 10 33 06.21 & $+$06 17 42.1 & 17-Apr-2020 & 2 & 9 & 13272 & 0.109 & 5.71 & U \\
F550P	& 10 33 06.21 & $+$00 00 00.0 & 20-Apr-2020 & 2 & 9 & 13305 & 0.102 & 5.48 & U \\
F194P	& 18 25 06.98 & $-$43 52 21.4 & 04-Jul-2020	& 2 & 9 & 15208 & 0.084 & 5.90 & U \\
F195P	& 18 58 36.28 & $-$43 52 21.4 & 05-Jul-2020 & 2 & 9 & 15229 & 0.083 & 5.83 & U \\
F197P	& 20 05 34.88 & $-$43 52 21.4 & 04-Jul-2020 & 2 & 9 & 15209 & 0.085 & 5.71 & U \\
&& \\
F198P	& 20 39 04.19 & $-$43 52 21.4 & 05-Jul-2020 & 2 & 9 & 15230 & 0.081 & 5.68 & U \\ 
F199P	& 21 12 33.49 & $-$43 52 21.4 & 04-Jul-2020 & 2 & 9 & 15211  & 0.081 & 5.28 & U \\
F351P	& 22 11 19.25 & $-$25 07 47.2 & 17-Dec-2019 & 2 & 5 & 10849 & 0.097 & 5.58 & U \\
F352P & 22 38 29.43 &  $-$25 07 47.2	& 05-Jan-2020 & 2 & 5 & 11051 & 0.085 & 5.69 & U\\
F159P & 22 46 09.23	& $-$50 05 45.3	 & 17-Apr-2020 &  2 & 9 & 13278 & 0.082 & 5.71 & U\\
&&\\
F120P & 22 56 28.24	& $-$56 18 22.7	& 20-Apr-2020 &  2	& 9 & 13297 & 0.080 & 5.38 & U \\
F353P & 23 05 39.62	& $-$25 07 47.2	& 05-Jan-2020 &  2 & 5 & 11052 & 0.086 & 5.97 & U \\
F160P & 23 23 04.62	& $-$50 05 45.3	& 04-Sep-2020 &  2 & 9 & 15873 & 0.079 & 5.24 & U \\
F354P	& 23 32 49.81	& $-$25 07 47.2	& 18-Apr-2020 &  2 & 9 & 13279 & 0.080 & 6.14 & U \\
F121P	& 23 38 49.41	& $-$56 18 22.7	& 19-Apr-2020	& 2 & 9 & 13296 & 0.083 & 5.60 & U \\
&& \\
J2022-2507P\footnote{FLASH 305P was observed at an incorrect RA due to a scheduling error and will be reobserved during the Full Survey.} & 20 22 38.4  & $-$25 07 47.2 & 22-Apr-2020 & 2 & 9 & 13372 & 0.085 & 5.66 & U: 305P observed at wrong RA\\
&& \\
\multicolumn{4}{l}{\bf Fields in the GAMA galaxy survey area (10 fields)} \\
FG9A & 08 47 35.59 & $+$00 30 00.0 & 18-Apr-2020 & 2	& 9 & 13285 & 0.093 & 5.61 & U  \\
FG9A\_long & 08 47 35.59 & $+$00 30 00.0 & 19-Apr-2020 & 6	& 9 & 13293 & 0.060 & 3.18 & U \\
FG9B	& 09 12 25 24 & $+00$ 30 00.0 & 18-Apr-2020 & 2	& 9 & 13283 & 0.095 & 5.26 & U \\
FG9B\_long & 09 12 25.24 & $+$00 30 00.0 & 07-Jan-2020 & 6 & 5 & 11068 & 0.076 & 3.39 & U \\
&& \\
FG12A & 11 47 35.17 & $-$00 30 00.0 & 21-Apr-2020	& 2	 & 9 & 13334 & 0.112 & 5.47 & U \\
FG12A\_long & 11 47 35.17 & $-$00 30 00.0 & 20-Apr-2020 & 6	& 9 & 13306 & 0.075 & 3.22 & U \\
FG12B & 12 12 24.83 & $-$00 30 00.0 & 21-Apr-2020 &  2	& 9 & 13335 & 0.101 & 5.58 & U \\
&& \\
FG15A & 14 16 33.10 & $+$00 30 00.0 & 21-Apr-2020 & 2	& 9 & 13336 & 0.104 & 5.42 & U \\
FG15A\_long & 14 16 33.10 & $+$00 30 00.0 & 19-Apr-2020 & 6	& 9 & 13294 & 0.075 & 3.16 & U \\
FG15B & 14 41 22.76 & $+$00 30 00.0 & 17-Apr-2020 &   2 & 9 & 13273 & 0.106 & 5.51 & U \\
\bottomrule
\end{tabular}
\label{tab:obs}
\end{table}

\clearpage
\onecolumn  
\begin{table}[!ht]
\footnotesize
\tabcolsep 6pt
\renewcommand{\arraystretch}{0.8} 
\caption{Log of observations for FLASH Pilot Survey 2.}
\begin{tabular}{lllrcclccl} 
\toprule
\multicolumn{1}{c}{Field} &  \multicolumn{1}{c}{RA} & \multicolumn{1}{c}{Dec} &  \multicolumn{2}{c}{Observation} & \multicolumn{1}{c}{BeamInt} & \multicolumn{1}{c}{CASDA} & \multicolumn{2}{c}{rms noise} & \multicolumn{1}{c}{Note} \\
 & \multicolumn{2}{c}{(J2000)} & \multicolumn{1}{c}{Date} & \multicolumn{1}{c}{t (h) } & \multicolumn{1}{c}{(MHz)}  & \multicolumn{1}{c}{SBID} & \multicolumn{1}{c}{Cont} &\multicolumn{1}{c}{Line} & \multicolumn{1}{c}{CASDA validation flag}\\
&&&&&&& (mJy/beam) & (mJy/beam/ch) \\
\multicolumn{1}{c}{(1)} & \multicolumn{1}{c}{(2)} & \multicolumn{1}{c}{(3)} & \multicolumn{1}{c}{(4)} & \multicolumn{1}{c}{(5)} & \multicolumn{1}{c}{(6)} & \multicolumn{1}{c}{(7)} & \multicolumn{1}{c}{(8)} & \multicolumn{1}{c}{(9)} & \multicolumn{1}{c}{(10)} \\ 
\hline
\multicolumn{6}{l}{\bf FLASH survey fields observed and released in CASDA (51 fields)} \\
F398$^{*}$	&	18	45	49.00	&	$-$18	51	45.4	&	17-Nov-2021	&	2	&	5	&	33616	& 0.082		& 6.08		&	Quality Gate, G	\\
&&	\\																						
F122	&	00 00 00.00	&	$-$50	05	45.2	&	29-Dec-2021	&	2	&	5	&	34941	& 0.079		& 5.44		&	U: ripples, re-obs	\\
F525$^{*}$	&	00	12	24.80	&	+00 00 00.0	&	20-Dec-2021	&	2	&	5	&	34581	& 0.105		& 5.68		&	U: ripples, re-obs	\\
F123	&	00	36	55.30	&	$-$50	05	45.2	&	18-Feb-2022	&	2	&	5	&	37448	& 0.087		& 5.67		&	G	\\
F526$^{*}$	&	00	37	14.40	&	+00 00 00.0	&	26-Dec-2021	&	2	&	5	&	34783	& 0.138		& 5.80		&	U: ripples, ducting, re-obs	\\
F527$^{*}$	&	01	02	04.00	&	+00 00 00.0	&	19-Dec-2021	&	2	&	5	&	34568	& 0.138		& 5.86		&	U*: ducting	\\
&&	\\																						
F305	&	01	21	30.50	&	$-$25	07	47.2	&	18-Feb-2022	&	2	&	5	&	37449	& 0.084		&	5.59	&	G	\\
F528	&	01	26	53.70	&	+00 00 00.0	&	18-Feb-2022	&	2	&	5	&	37450	& 0.128		& 5.52		&	G*	\\
F306	&	01	48	40.70	&	$-$25	07	47.2	&	19-Feb-2022	&	2	&	5	&	37475	& 0.114		& 5.39		&	G	\\
F529	&	01	51	43.40	&	+00 00 00.0	&	18-Dec-2021	&	2	&	5	&	34557	& 0.098		& 5.57		&	U: ripples, re-obs	\\
F255$^{*}$	&	01	52	56.40	&	$-$31	23	14.7	&	15-Jan-2022	&	2	&	5	&	35939	& 0.075		& 5.23		&	U: ripples, re-obs	\\
&&	\\																						
F530	&	02	16	33.10	&	+00 00 00.0	&	17-Feb-2022	&	2	&	5	&	37431	& 0.105		& 5.35		&	G	\\
F256	&	02	21	10.50	&	$-$31	23	14.7	&	20-Dec-2021	&	2	&	5	&	34584	& 0.090		& 5.62		&	U: ripples, re-obs	\\
F531	&	02	41	22.70	&	+00 00 00.0	&	17-Dec-2021	&	2	&	5	&	34546	& 0.098		& 5.75		&	U: ripples, re-obs	\\
F257	&	02	49	24.70	&	$-$31	23	14.7	&	18-Feb-2022	&	2	&	5	&	37451	& 0.082		& 5.66		&	G	\\
F532	&	03	06	12.40	&	+00 00 00.0	&	19-Dec-2021	&	2	&	5	&	34569	& 0.113		& 6.31		&	U: ripples, re-obs	\\
&&	\\																						
F258	&	03	17	38.80	&	$-$31	23	14.7	&	18-Feb-2022	&	2	&	5	&	37452	& 0.074		& 5.23		&	G	\\
F533$^{*}$	&	03	31	02.00	&	+00 00 00.0	&	15-Jan-2022	&	2	&	5	&	35943	& 0.094		& 5.68		&	U: ripples, re-obs	\\
F310	&	03	37	21.50	&	$-$25	07	47.2	&	17-Feb-2022	&	2	&	5	&	37432	& 0.105		& 5.61		&	G	\\
F534	&	03	55	51.70	&	+00 00 00.0	&	18-Dec-2021	&	2	&	5	&	34558	& 0.097		& 5.55		&	U: ripples, re-obs	\\
F311	&	04	04	31.70	&	$-$25	07	47.2	&	18-Feb-2022	&	2	&	5	&	37453	& 0.080		& 5.28		&	G	\\
&&	\\																						
F312	&	04	31	41.80	&	$-$25	07	47.2	&	02-Mar-2022	&	2	&	5	&	37797	& 0.104		& 5.83		&	G*	\\
F313	&	04	58	52.00	&	$-$25	07	47.2	&	17-Dec-2021	&	2	&	5	&	34547	& 0.102		& 5.35		&	U*: ducting	\\
F314$^{*}$	&	05	26	02.00	&	$-$25	07	47.2	&	19-Dec-2021	&	2	&	5	&	34570	& 0.099		& 5.38		&	U: ripples, ducting, re-obs	\\
F545	&	08	28	57.90	&	+00 00 00.0	&	17-Dec-2021	&	2	&	5	&	34548	& 0.105		& 6.08		&	U: ripples, ducting, re-obs	\\
F546	&	08	53	47.50	&	+00 00 00.0	&	18-Dec-2021	&	2	&	5	&	34559	& 0.095		& 5.37		&	U: ripples, re-obs	\\
&&	\\																						
F719$^{*}$	&	09	09	49.00	&	+18	51	28.8	&	18-Dec-2021	&	2	&	5	&	34560	& 0.112		& 5.88		&	U: ripples, ducting, re-obs	\\
F547	&	09	18	37.20	&	+00 00 00.0	&	17-Dec-2021	&	2	&	5	&	34549	& 0.102		& 5.39		&	U: ripples, ducting, re-obs	\\
F377	&	09	36	00.00	&	$-$18	51	45.4	&	19-Dec-2021	&	2	&	5	&	34571	& 0.085		& 5.45		&	U: ripples, ducting, re-obs	\\
F378	&	10	02	10.90	&	$-$18	51	45.4	&	18-Dec-2021	&	2	&	5	&	34561	& 0.096		& 5.55		&	U: ripples, re-obs	\\
F553	&	11	47	35.10	&	+00 00 00.0	&	19-Dec-2021	&	2	&	5	&	34572	& 0.125		& 5.92		&	U: ripples, re-obs	\\
&&	\\																						
F554	&	12	12	24.80	&	+00 00 00.0	&	28-Dec-2021	&	2	&	5	&	34917	& 0.103		&	5.59	&	G	\\
F555	&	12	37	14.40	&	+00 00 00.0	&	19-Dec-2021	&	2	&	5	&	34562	& 0.133		& 6.79		&	U*	\\
F559$^{*}$	&	14	16	33.10	&	+00 00 00.0	&	18-Dec-2021	&	2	&	5	&	34551	& 0.102		& 5.77		&	U: ripples, re-obs	\\
F560	&	14	41	22.70	&	+00 00 00.0	&	19-Dec-2021	&	2	&	5	&	34563	& 0.104		& 5.48		&	U: ripples, re-obs	\\
F561	&	15	06	12.40	&	+00 00 00.0	&	20-Dec-2021	&	2	&	5	&	34576	& 0.105		& 5.81		&	U: ripples, re-obs	\\
&&	\\																						
F011	&	16	45	52.80	&	$-$80	02	36.7	&	19-Dec-2021	&	2	&	5	&	34564	& 0.098		& 5.93		&	U: ripples, re-obs	\\
F287	&	16	56	28.20	&	$-$31	23	14.7	&	18-Dec-2021	&	2	&	5	&	34552	& 0.132		& 6.38		&	U*: ducting	\\
F151	&	17	50	46.10	&	$-$50	05	45.2	&	18-Dec-2021	&	2	&	5	&	34553	& 0.090		& 5.91		&	U: ripples, re-obs	\\
F573	&	20	04	08.00	&	+00 00 00.0	&	18-Dec-2021	&	2	&	5	&	34554	& 0.101		& 5.46		&	U*	\\
F574	&	20	28	57.90	&	+00 00 00.0	&	19-Dec-2021	&	2	&	5	&	34565	& 0.098		& 6.05		&	U: ripples, re-obs	\\
&&	\\																						
F575*	&	20	53	47.50	&	+00 00 00.0	&	19-Dec-2021	&	2	&	5	&	34566	& 0.097		& 5.80		&	U*	\\
F575*	&	20	53	47.50	&	+00 00 00.0	&	20-Dec-2021	&	2	&	5	&	34577	& 0.094		& 5.49		&	U*	\\
F576	&	21	18	37.20	&	+00 00 00.0	&	20-Dec-2021	&	2	&	5	&	34578	& 0.098		& 5.50		&	U: ripples	\\
F577$^{*}$	&	21	43	26.90	&	+00 00 00.0	&	18-Dec-2021	&	2	&	5	&	34555	& 0.098		& 5.56		&	U: ripples, re-obs	\\
F578$^{*}$	&	22	08	16.50	&	+00 00 00.0	&	20-Dec-2021	&	2	&	5	&	34580	& 0.097		& 5.56		&	U: ripples, re-obs	\\
&&	\\																						
F579$^{*}$	&	22	33	06.00	&	+00 00 00.0	&	21-Dec-2021	&	2	&	5	&	34597	& 0.099		& 5.70		&	U: ripples, re-obs	\\
F580	&	22	57	55.80	&	+00 00 00.0	&	19-Dec-2021	&	2	&	5	&	34567	& 0.100		& 6.20		&	U: ripples, re-obs	\\
F581$^{*}$	&	23	22	45.50	&	+00 00 00.0	&	26-Dec-2021	&	2	&	5	&	34781	& 0.120		& 5.65		&	U: ripples, re-obs	\\
F160$^{*}$	&	23	23	05.00	&	$-$50	05	45.2	&	29-Dec-2021	&	2	&	5	&	34939	& 0.078		& 5.42		&	U: ripples, re-obs	\\
F582	&	23	47	35.10	&	+00 00 00.0	&	18-Dec-2021	&	2	&	5	&	34556	& 0.098		& 5.52		&	U: ripples, re-obs	\\
\bottomrule
\end{tabular}
\label{tab:obs2}
\end{table}

\clearpage
\onecolumn 
\begin{table}[!ht]
\footnotesize
\tabcolsep 8pt
\renewcommand{\arraystretch}{0.8} 
\caption{Log of observations for FLASH Pilot Survey 2 re-observed fields.}
\begin{tabular}{lllrcclccl}
\toprule
\multicolumn{1}{c}{Field} &  \multicolumn{1}{c}{RA} & \multicolumn{1}{c}{Dec} &  \multicolumn{2}{c}{Observation} & \multicolumn{1}{c}{BeamInt} & \multicolumn{1}{c}{CASDA} & \multicolumn{2}{c}{rms noise} & \multicolumn{1}{c}{Note} \\
 & \multicolumn{2}{c}{(J2000)} & \multicolumn{1}{c}{Date} & \multicolumn{1}{c}{t (h) } & \multicolumn{1}{c}{(MHz)}  & \multicolumn{1}{c}{SBID} & \multicolumn{1}{c}{Cont} &\multicolumn{1}{c}{Line} & \multicolumn{1}{c}{CASDA validation flag} \\
&&&&&&& (mJy/beam) & (mJy/beam/ch) \\
\multicolumn{1}{c}{(1)} & \multicolumn{1}{c}{(2)} & \multicolumn{1}{c}{(3)} & \multicolumn{1}{c}{(4)} & \multicolumn{1}{c}{(5)} & \multicolumn{1}{c}{(6)} & \multicolumn{1}{c}{(7)} & \multicolumn{1}{c}{(8)} & \multicolumn{1}{c}{(9)} & \multicolumn{1}{c}{(10)} \\ 
\hline
\multicolumn{6}{l}{\bf FLASH survey fields re-observed and released in CASDA (24 fields)} \\
F525	&	00	12	24.80	&	+00	00	00.0	&	08-Jul-2022	&	2	&	5	&	42299	& 0.130		& 5.52		&	G 	\\
F526	&	00	37	14.40	&	$+$00	00	00.0	&	07-Jul-2022	&	2	&	5	&	42275	& 0.137		& 5.52		&	U*: ducting 	\\
F527	&	01	02	04.14	&	$+$00	00	00.0	& 08-Jul-2022		&	2	&	5	&	42300	& 0.124		& 5.56		&	U*: ducting 	\\
F255	&	01	52	56.40	&	$-$31 23 14.7		&	03-Jun-2022	&	2	&	5	&	41226	& 0.080		&	5.30	&	G 	\\
F255	&	01	52	56.40	&	$-$31 23 14.7	&	01-Jun-2022	&	2	&	5	&	41148	& 0.101		& 5.21		&	U*:  	\\
	\\																						
F533	&	03	31	02.07	& +00 00 00.0	&	08-Jul-2022	&	2	&	5	&	42278	& 0.112		& 5.76		&	U*:  	\\
F314	&	05	26	02.26	&	$-$25	07	47.2	&	29-May-2022	&	2	&	5	&	41065	& 0.080		& 4.90		&	U*: ducting 	\\
F314	&	05	26	02.26	&	$-$25	07	47.2	&	29-May-2022	&	2	&	5	&	41061	& 0.073		& 5.05		&	U*:  	\\
F719	&	09	09	49.00	&	+18	51	28.8	& 30-May-2022 &	2	&	5	&	41084	& 0.101		& 4.78		&	U*: ducting 	\\
F719	&	09	09	49.00	&	+18	51	28.8	&	29-May-2022	&	2	&	5	&	41066	& 0.094		& 4.87		&	U*: ducting 	\\
	\\																						
F559	&	14	16	33.10	&	+00	00	00.0	&	30-May-2022	&	2	&	5	&	41085	& 0.096		& 5.13		&	U*: ducting 	\\
F559	&	14	16	33.10	&	+00	00	00.0	&	29-May-2022	&	2	&	5	&	41068	& 0.092		& 5.06		&	U*: ducting 	\\
F398	&	18	45	49.00	&	$-$18	51	45.4	&	29-May-2022	&	2	&	5	&	41071	& 0.082		& 5.99		&	G 	\\
F398	&	18	45	49.00	&	$-$18	51	45.4	&	28-May-2022	&	2	&	5	&	41050	& 0.084		& 6.21		&	U*: ducting 	\\
F577	&	21	43	26.90	&	+00	00	00.0	&	08-Jul-2022	&	2	&	5	&	42296	& 0.109		& 5.55		&	U*: ducting 	\\
	\\																						
F578	&	22	08	16.50	&	+00	00	00.0	&	08-Jul-2022	&	2	&	5	&	42297	& 0.122		& 5.47		&	U*: ducting 	\\
F578	&	22	08	16.50	&	+00	00	00.0	&	14-Aug-2022	&	2	&	5	&	43424	& 0.100		& 5.66		&	G 	\\
F579	&	22	33	06.00	&	+00 00 00.0	&	08-Jul-2022	&	2	&	5	&	42298	& 0.118		& 5.55		&	U*: ducting 	\\
F581	&	23	22	45.50	&	+00	00	00.0	&	30-May-2022	&	2	&	5	&	41105	& 0.350		& 6.36		&	U*:  	\\
F581	&	23	22	45.50	&	+00	00	00.0	&	29-May-2022	&	2	&	5	&	41072	& 0.094		& 5.22		&	U*: ducting 	\\
	\\																						
F581	&	23	22	45.50	&	+00	00	00.0	&	03-Jun-2022	&	2	&	5	&	41225	& 0.112		& 5.75		&	U*:  	\\
F581	&	23	22	45.50	&	+00	00	00.0	&	01-Jun-2022	&	2	&	5	&	41184	& 0.106		& 5.80		&	U* 	\\
F160	&	23	23	04.62	&	$-$50 05 45.3	&	09-Jul-2022	&	2	&	5	&	42323	& 0.100		& 5.66		&	U*: ducting 	\\
F160	&	23	23	04.62	&	$-$50 05 45.3	&	14-Aug-2022	&	2	&	5	&	43426	& 0.075		& 5.18		&	G 	\\
\bottomrule
\end{tabular}
\end{table}

\end{document}